\documentclass[11pt,a4paper]{article}%
\usepackage{jheppub}
\usepackage{amsmath}
\usepackage{graphicx}
\usepackage{amsfonts}
\usepackage{amssymb}%
\newtheorem{theorem}{Theorem}

\newtheorem{lemma}[theorem]{Lemma}
\newtheorem{remark}[theorem]{Remark}

\newenvironment{proof}[1][Proof]{\textbf{#1.} }{\ \rule{0.5em}{0.5em}}

\title{Exact form factors of the \boldmath $O(N)$ $\sigma$-model}

\author[a]{Hrachya M. Babujian}
\author[b]{Angela Foerster}
\author[c]{and Michael Karowski}

\affiliation[a]{Yerevan Physics Institute,\\Alikhanian Brothers 2, Yerevan, 375036 Armenia}
\affiliation[b]{Instituto de F\'{\i}sica da UFRGS,\\Av. Bento Gon\c{c}alves 9500, Porto Alegre, RS - Brazil}
\affiliation[c]{Institut f\"{u}r Theoretische Physik,\\Arnimallee 14, 14195 Berlin, Germany}

\emailAdd{babujian@yerphi.am}
\emailAdd{angela@if.ufrgs.br}
\emailAdd{karowski@physik.fu-berlin.de}

\abstract{
A general form factor formula for the $O(N)~\sigma$-model is constructed and
applied to several operators. The large $N$ limits of these form factors are
computed and compared with the $1/N$ expansion of the $O(N)~\sigma$-model in
terms of Feynman graphs and full agreement is found. In particular, $O(3)$ and
$O(4)$ form factors are discussed. For the $O(3)~\sigma$-model several low
particle form factors are calculated explicitly.
}

\begin{document}

\maketitle

\section{Introduction}

The $O(N)~\sigma$-model is an asymptotically free quantum field theory which
has been attracting high interest, since it exhibits some common features with
Quantum Chromodynamics (QCD), the theory of strong interactions. Although very
powerful, the 4-dimensional QCD still requires an adequate machinery to handle
with the confinement problem. From this point of view, 2-dimensional
integrable models are very useful since they serve as laboratories for
investigations of those properties of quantum field theories which can not be
described via standard methods, such as perturbation theory. We should also
note that exact results and methods which have been developed last decades are
now flourishing and finding applications in the $AdS_{5}\bigotimes S^{5}$
theory \cite{KZar}. It has also been proposed a set of consistency conditions
for the worldsheet form factors for the set of off-shell operators in the
$AdS_{5}\bigotimes S^{5}$ \cite{Kl}. The most important result on the AdS/CFT
correspondence is the remarkable conjecture of Maldacena \cite{Mal}, which
establishes that 10-dimensional string theory on $AdS_{5}\bigotimes S^{5}$
could be equivalent to a 4-dimensional Super Yang Mills (SYM) theory. In order
to establish this conjecture there are ongoing developments on this topic
\cite{KZar} employing exact methods such as the Bethe ansatz. Clearly, such
new developments, among others, place to a higher level the status of the
approaches employed in 2-dimensional integrable Quantum Field Theories (QFT).
Consequently, integrable models in 2-dimensions are now being considered not
only isolated mathematical objects; in opposite, they are universal and also
coming out in higher dimensions, where enough integrals of motion necessary
for integrability are being found. The nonlinear $O(N)~\sigma$-model is
defined by the Lagrangian and the constraint
\begin{equation}
\mathcal{L}=\frac{1}{2}\sum_{\alpha=1}^{N}\left(  \partial_{\mu}%
\varphi_{\alpha}\right)  ^{2}\qquad\text{with}\quad g\sum_{\alpha=1}%
^{N}\varphi_{\alpha}^{2}=1 \label{L}%
\end{equation}
where $\varphi_{\alpha}(x)$ is an isovector $N$-plet set of bosonic fields and
$g$ the coupling constant. This model is integrable, there exist an infinite
set of conservation laws \cite{Po}. In the quantum model the infrared charge
singularity leads to the disintegration of the Goldstone vacuum and to mass
transmutation of particles, which form an $O(N)$ multiplet (see \cite{ZZ3}).

In this article we construct the form factors of the model by using the
solution of the $O(N)$ difference equation, derived previously \cite{BFK5} by
generalizing Tarasov's methods \cite{Tar} (see also \cite{MR}) of the
algebraic Bethe ansatz. Exact form factors for the energy-momentum, the
spin-field and the current are computed and compared with the $1/N$ expansion
of the $O(N)$ $\sigma$- model. We should note that the form factors in $O(3)$
and $O(4)$ sigma models first were calculated by Smirnov \cite{Sm,Sm3} (see
also \cite{BN,Ba,BH,BaW}). In the framework of 2-dimensional integrable QFTs
the central problem is still the computation of the correlation functions or
Weightman functions and the form factor program is exactly devoted to this
purpose. The concept of a generalized form factor was introduced in
\cite{KW,BKW}, where several consistency equations were formulated.
Subsequently this approach was developed further and investigated in different
models by Smirnov \cite{Sm}. Generalized form factors are matrix elements of
fields with many particle states. To construct these objects explicitly one
has to solve generalized Watson's equations which are matrix difference
equations. To solve these equations the so called \textquotedblleft off-shell
Bethe ansatz" is applied \cite{BKZ2,BFKZ,BFK1,BFK5}. The conventional Bethe
ansatz introduced by Bethe \cite{Bethe} is used to solve eigenvalue problems
and its algebraic formulation was developed by Faddeev and coworkers (see e.g.
\cite{TF}). The off-shell Bethe ansatz has been introduced in \cite{B3} to
solve the Knizhnik-Zamolodchikov equations which are differential equations.
For other approches to form factors in integrable quantum field theories see
also \cite{CM2,FMS,YZ,Lu0,Lu,Lu1,BL,LuZ,Or}. The main result of this paper is
the general form factor formula, written as an integral representation, which
provides the solution of all form factors equations and whose main idea is
briefly explained below. In the $O(N)~\sigma$-model the particles form an
isovector $N$-plet of $O(N)$. For a state of $n$ particles of kind $\alpha
_{i}$ with rapidities $\theta_{i}$ and a local operator $\mathcal{O}(x)$ the
matrix element%
\[
\langle\,0\,|\,\mathcal{O}(x)\,|\,\theta_{1},\dots,\theta_{n}\,\rangle
_{\underline{\alpha}}^{in}=e^{-ix(p_{1}+\cdots+p_{n})}F_{\underline{\alpha}%
}^{\mathcal{O}}(\underline{\theta})
\]
defines a form factor which we write as (see \cite{KW})%
\begin{equation}
F_{1\dots n}^{\mathcal{O}}(\underline{\theta})=K_{1\dots n}^{\mathcal{O}%
}(\underline{\theta})\prod_{1\leq i<j\leq n}F(\theta_{ij})\label{2.10}%
\end{equation}
where $F(\theta)$ is the minimal form factor function. We propose the
following ansatz for the K-function in terms of a nested `off-shell' Bethe
ansatz written as a multiple contour integral%
\begin{equation}
\fbox{$\rule[-0.2in]{0in}{0.5in}\displaystyle~K_{1\dots n}^{\mathcal{O}%
}(\underline{\theta})=N_{n}^{\mathcal{O}}\int_{\mathcal{C}_{\underline{\theta
}}^{1}}dz_{1}\cdots\int_{\mathcal{C}_{\underline{\theta}}^{m}}dz_{m}%
\,\tilde{h}(\underline{\theta},\underline{z})\,p^{\mathcal{O}}(\underline
{\theta},\underline{z})\,\,\tilde{\Psi}_{1\dots n}(\underline{\theta
},\underline{z})$\thinspace.}\label{2.16}%
\end{equation}
Here $\tilde{h}(\underline{\theta},\underline{z})$ is a scalar function which
depends only on the S-matrix. The dependence on the specific operator
$\mathcal{O}(x)$ is encoded in the scalar p-function $p^{\mathcal{O}%
}(\underline{\theta},\underline{z})$ which is in general a simple function of
$e^{\theta_{i}}$ and $e^{z_{j}}$. The state $\tilde{\Psi}_{\underline{\alpha}%
}$ in (\ref{2.16}) is a linear combination of the basic Bethe ansatz
co-vectors (see (\ref{BS}))%
\begin{equation}
\tilde{\Psi}_{\underline{\alpha}}(\underline{\theta},\underline{z}%
)=L_{\underline{\mathring{\beta}}}(\underline{z})\,\tilde{\Phi}_{\underline
{\alpha}}^{\underline{\mathring{\beta}}}(\underline{\theta},\underline
{z})\label{PSI}%
\end{equation}
where summation over all $\underline{\mathring{\beta}}=(\mathring{\beta}%
_{1},\dots,\mathring{\beta}_{m})$ is assumed. The $\mathring{\beta}$ form an
$\left(  N-2\right)  $-plet of $O(N-2)$. For $L_{\underline{\mathring{\beta}}%
}(\underline{z})$ we make again an ansatz like (\ref{2.16}). The nested
off-shell Bethe ansatz is obtained by iterating this procedure.

The article is organized as follows. In Section \ref{s2} we recall some
results and fix the notation concerning the $O(N)$ S-matrix, the monodromy
matrix, etc. In Section \ref{s3} we discuss the generalized form factors
formula for the $O(N)~\sigma$-model. In Section \ref{s4} we apply the nested
off-shell Bethe ansatz to solve the $O(N)$ form factor equations. Section
\ref{s5} is devoted to the computation of some examples. The appendices
provide the more complicated proofs of the results we have obtained and
further explicit calculations.

\section{General settings\label{s2}}

\subsection{The $O(N)$ S-matrix}

The 2-particle $O(N)$ S-matrix is of the form \cite{ZZ3}
\begin{equation}
S=b(\theta)\mathbf{1}+c(\theta)\mathbf{P}+d(\theta)\mathbf{K} \label{S}%
\end{equation}
or in terms of the components%
\[
S_{\alpha\beta}^{\delta\gamma}(\theta)=b(\theta)\delta_{\alpha}^{\gamma}%
\delta_{\beta}^{\delta}+c(\theta)\delta_{\alpha}^{\delta}\delta_{\beta
}^{\gamma}+d(\theta)\delta^{\delta\gamma}\delta_{\alpha\beta}%
\]
where $\theta$ is the rapidity difference of the particles. Crossing means
\begin{equation}
S_{\alpha\beta}^{\delta\gamma}(\theta)=\mathbf{C}_{\alpha\alpha^{\prime}%
}S_{\beta\gamma^{\prime}}^{\alpha^{\prime}\delta}(i\pi-\theta)\mathbf{C}%
^{\gamma^{\prime}\gamma} \label{2.2}%
\end{equation}
or in terms of the amplitudes%
\[
b(\theta)=b(i\pi-\theta),\;d(\theta)=c(i\pi-\theta)
\]
if we define the \textquotedblleft charge conjugation
matrices\textquotedblright\ as
\begin{equation}
\mathbf{C}_{\alpha\beta}=\delta_{\alpha\beta}~\text{and }\mathbf{C}%
^{\alpha\beta}=\delta^{\alpha\beta}\,. \label{C}%
\end{equation}
The Yang-Baxter relation%
\begin{equation}
S_{12}(\theta_{12})S_{13}(\theta_{13})S_{23}(\theta_{23})=S_{23}(\theta
_{23})S_{13}(\theta_{13})S_{12}(\theta_{12}) \label{1.3}%
\end{equation}
implies \cite{ZZ3}
\begin{equation}
c(\theta)=-\frac{i\pi\nu}{\theta}b(\theta),\;d(\theta)=-\frac{i\pi\nu}%
{i\pi-\theta}b(\theta) \label{yb}%
\end{equation}
where $\nu=2/(N-2)$. The minimal solution is $b(\theta)=Q(\theta)Q(i\pi
-\theta)$ with%
\begin{equation}
Q(\theta)=\frac{\Gamma\left(  \frac{1}{2}\nu+\frac{1}{2\pi i}\theta\right)
\Gamma\left(  \frac{1}{2}+\frac{1}{2\pi i}\theta\right)  }{\Gamma\left(
\frac{1}{2}+\frac{1}{2}\nu+\frac{1}{2\pi i}\theta\right)  \Gamma\left(
\frac{1}{2\pi i}\theta\right)  }\,. \label{Smin}%
\end{equation}
This minimal solution was first constructed by Zamolodchikov and Zamolodchikov
\cite{ZZ3} and they gave arguments that it provides the $O(N)$ $\sigma$-model
S-matrix
\[
S^{\sigma\text{-model}}(\theta)=S^{\text{min}}(\theta)\,.
\]
The three S-matrix eigenvalues are $S_{\pm}=b\pm c$ and $S_{0}=b+c+Nd$ with%
\begin{equation}
\left(  S_{0},S_{+},S_{-}\right)  =\left(  \frac{\theta+i\pi}{\theta-i\pi
},\frac{\theta-i\pi\nu}{\theta+i\pi\nu},1\right)  S_{-}\,. \label{EV}%
\end{equation}
For later convenience we introduce
\[
\tilde{S}(\theta)=S(\theta)/S_{+}(\theta)=\tilde{b}(\theta)\mathbf{1}%
+\tilde{c}(\theta)\mathbf{P}+\tilde{d}(\theta)\mathbf{K}%
\]
with%
\begin{align}
\tilde{b}(\theta)  &  =\frac{\theta}{\theta-i\pi\nu}\nonumber\\
\tilde{c}(\theta)  &  =\mathbf{-}\frac{i\pi\nu}{\theta-i\pi\nu}\label{1.2}\\
\tilde{d}(\theta)  &  =\mathbf{-}\frac{\theta}{\theta-i\pi\nu}\frac{i\pi\nu
}{i\pi-\theta}\nonumber
\end{align}
We will also need $\mathring{S}(z)$ the S-matrix for $O(N-2)$
\begin{equation}
\tilde{\mathring{S}}(\theta)=\mathring{S}(\theta)/\mathring{S}_{+}%
(\theta)=\tilde{\mathring{b}}(\theta)\mathbf{1}+\tilde{\mathring{c}}%
(\theta)\mathbf{P}+\tilde{\mathring{d}}(\theta)\mathbf{K} \label{1.1a}%
\end{equation}
where $\nu$ is replaced by $\mathring{\nu}=2/(N-4)$.

\subsubsection{Complex basis\label{scb}}

For the Bethe ansatz it is convenient to use instead of the real basis
$|\alpha\rangle_{r}\,,~\left(  \alpha=1,2,\dots,N\right)  $ the complex basis
\[
\left.
\begin{array}
[c]{ccc}%
|\alpha\rangle & = & \frac{1}{\sqrt{2}}\left(  |2\alpha-1\rangle_{r}%
+i|2\alpha\rangle_{r}\right) \\
|\bar{\alpha}\rangle & = & \frac{1}{\sqrt{2}}\left(  |2\alpha-1\rangle
_{r}-i|2\alpha\rangle_{r}\right)
\end{array}
\right\}  ~,~~\alpha=1,2,\dots,\left[  N/2\right]
\]
and in addition $|0\rangle=|\bar{0}\rangle=|N\rangle_{r}$ for N odd. Below we
will use the notation%
\[
|\theta\rangle_{\alpha}=|\alpha(\theta)\rangle,~|\theta\rangle_{\bar{\alpha}%
}=|\bar{\alpha}(\theta)\rangle
\]
for one particle and one antiparticle states with rapidity $\theta$. The
weight vectors%
\[
w=\left(  w_{1},\dots,w_{\left[  N/2\right]  }\right)
\]
of the one-particle states are given by%
\[%
\begin{array}
[c]{ccccl}%
w_{k} & = & \delta_{k\alpha} &  & \text{for }|\alpha\rangle\\
w_{k} & = & -\delta_{k\alpha} &  & \text{for }|\bar{\alpha}\rangle\\
w_{k} & = & 0 &  & \text{for }|0\rangle\,.
\end{array}
\]

\begin{remark}
For even $N$ this means that we consider $O(N)$ as a subgroup of $U(N/2)$. For
$N=3$ we may identify the particles $1,\bar{1},0$ with the pions $\pi_{\pm
},\pi_{0}$.
\end{remark}

The highest weight S-matrix eigenvalue is $a(\theta)=S_{11}^{11}(\theta
)=S_{+}(\theta)$ with%
\begin{align}
a(\theta)  &  =-\frac{\Gamma\left(  \frac{1}{2}+\frac{1}{2\pi i}\theta\right)
\Gamma\left(  \frac{1}{2}+\frac{1}{2}\nu-\frac{1}{2\pi i}\theta\right)
}{\Gamma\left(  \frac{1}{2}-\frac{1}{2\pi i}\theta\right)  \Gamma\left(
\frac{1}{2}+\frac{1}{2}\nu+\frac{1}{2\pi i}\theta\right)  }\frac{\Gamma\left(
1-\frac{1}{2\pi i}\theta\right)  \Gamma\left(  \frac{1}{2}\nu+\frac{1}{2\pi
i}\theta\right)  }{\Gamma\left(  1+\frac{1}{2\pi i}\theta\right)
\Gamma\left(  \frac{1}{2}\nu-\frac{\theta}{2\pi i}\right)  }\label{1.6}\\
&  =-\exp\left(  -2\int_{0}^{\infty}\frac{dt}{t}\frac{e^{-t\nu}+e^{-t}%
}{1+e^{-t}}\sinh t\frac{\theta}{i\pi}\right)  \label{1.7}%
\end{align}
We order the states as: $1,2,\dots,0,\dots,\bar{2},\bar{1}$. Then the charge
conjugation matrix in the complex basis is of the form%
\begin{align}
\mathbf{C}^{\delta\gamma}  &  =\delta^{\delta\bar{\gamma}}\,,~\mathbf{C}%
_{\alpha\beta}=\delta_{\alpha\bar{\beta}}\label{Cc}\\
\mathbf{C}  &  \mathbf{=}\left(
\begin{array}
[c]{ccccc}%
0 & \cdots & 0 & \cdots & 1\\
\vdots & \ddots & \vdots & \cdot & \vdots\\
0 & \cdots & 1 & \cdots & 0\\
\vdots & \cdot & \vdots & \ddots & \vdots\\
1 & \cdots & 0 & \cdots & 0
\end{array}
\right) \nonumber
\end{align}
The annihilation-creation matrix in (\ref{S}) may be written as%
\[
\mathbf{K}_{\alpha\beta}^{\delta\gamma}=\mathbf{C}^{\delta\gamma}%
\mathbf{C}_{\alpha\beta}\,.
\]

\subsection{Nested \textquotedblleft off-shell\textquotedblright\ Bethe
ansatz}

The \textquotedblleft off-shell\textquotedblright\ Bethe ansatz is used to
construct vector valued functions which have symmetry properties according to
a representation of the permutation group generated by a factorizing S-matrix.
In addition they satisfy matrix differential \cite{B1} or difference
\cite{BKZ2} equations. For the application to form factors we use the
co-vector version $K_{_{1\dots n}}(\underline{\theta})\in V_{1\dots n}=\left(
\bigotimes_{i=1}^{n}V\right)  ^{\dag},~(\theta_{i}\in\mathbb{C},\mathbb{~}%
i=1,\dots,n)$. We write the components of the co-vector $K_{_{1\dots n}}$ as
$K_{\underline{\alpha}}$ where $\underline{\alpha}=(\alpha_{1},\dots
,\alpha_{n})$ is a state of $n$ particles. Solutions of the $O(N)$ equations
\begin{align*}
K_{\dots ij\dots}(\dots,\theta_{i},\theta_{j},\dots)  &  =K_{\dots ji\dots
}(\dots,\theta_{j},\theta_{i},\dots)\,\tilde{S}_{ij}(\theta_{ij})\\
K_{\alpha_{1}\alpha_{2}\dots\alpha_{n}}(\theta_{1}+2\pi i,\theta_{2}%
,\dots,\theta_{n})  &  =K_{\alpha_{2}\dots\alpha_{n}\alpha_{1}}(\theta
_{2},\dots,\theta_{n},\theta_{1})
\end{align*}
where constructed in \cite{BFK5,BFK6} (see also \cite{BKZ2,BFK1}). These
equations are equivalent to the form factor equations (i) and (ii) (see
(\ref{1.10}) and (\ref{1.12})). The solutions have been constructed in terms
of a nested $O(N)$ \textquotedblleft off-shell\textquotedblright\ Bethe ansatz
in \cite{BFK5,BFK6}. Here we need special solutions which satisfy in addition
the form factor equation (iii) (see (\ref{1.14})).

\paragraph{Nested \textquotedblleft off-shell\textquotedblright\ Bethe
ansatz:}

We consider a state with $n$ particles and write the off-shell Bethe ansatz
co-vector valued function as
\begin{equation}
\fbox{$\rule[-0.2in]{0in}{0.5in}\displaystyle~K_{\underline{\alpha}%
}(\underline{\theta})=\int_{\mathcal{C}_{\underline{\theta}}}dz_{1}\cdots
\int_{\mathcal{C}_{\underline{\theta}}}dz_{m}\,\tilde{k}(\underline{\theta
},\underline{z})\,\tilde{\Psi}_{\underline{\alpha}}(\underline{\theta
},\underline{z})$~} \label{BA}%
\end{equation}
where $\underline{\alpha}=(\alpha_{1},\dots,\alpha_{n})$, $\underline{\theta
}=\left(  \theta_{1},\dots,\theta_{n}\right)  $ and $\underline{z}=\left(
z_{1},\dots,z_{m}\right)  $. This ansatz transforms the complicated matrix
equations (\ref{1.10})-(\ref{1.14}) to simple equations for the scalar
function $\tilde{k}(\underline{\theta},\underline{z})$ (see \cite{BFK5} and
below). The integration contour $\mathcal{C}_{\underline{\theta}}$ will be
specified in section \ref{s4}. The state $\tilde{\Psi}_{\underline{\alpha}}$
in (\ref{BA}) is the linear combination (\ref{PSI}) of the basic Bethe ansatz
co-vectors (\ref{BS}). For the co-vector valued function $L_{\underline
{\mathring{\beta}}}(\underline{z})$ (which lies in a tensor product of smaller
spaces of dimension $N-2$) we make again an ansatz like (\ref{BA}). Iterating
this procedure we obtain the nested off-shell\ Bethe ansatz. This iteration
ends up at the $O(3)$ or $O(4)$ cases which will be discussed separately.

As usual in the context of the algebraic Bethe ansatz \cite{FST,TF} the basic
Bethe ansatz co-vectors $\tilde{\Phi}_{\underline{\alpha}}^{\underline
{\mathring{\beta}}}$ are obtained from the monodromy matrix. We consider a
state with $n$ particles and as is usual in the context of the algebraic Bethe
Ansatz we define \cite{FST,TF} the monodromy matrix by%
\begin{equation}
\tilde{T}_{1\dots n,0}(\underline{\theta},\theta_{0})=\tilde{S}_{10}%
(\theta_{1}-\theta_{0})\,\cdots\tilde{S}_{n0}(\theta_{n}-\theta_{0}).
\label{T}%
\end{equation}
It is a matrix acting in the tensor product of the \textquotedblleft quantum
space\textquotedblright\ $V^{1\dots n}=V_{1}\otimes\cdots\otimes V_{n}$ and
the \textquotedblleft auxiliary space\textquotedblright\ $V_{0}$. All vector
spaces $V_{i}$ are isomorphic to a space $V$ whose basis vectors label all
kinds of particles. Here $V\cong\mathbb{C}^{N}$ is the space of the vector
representation of $O(N)$.

Suppressing the indices $1\ldots n$ we write the monodromy matrix in the
complex basis as (following the notation of Tarasov \cite{Tar})
\begin{equation}
\tilde{T}_{\alpha}^{\alpha^{\prime}}=\left(
\begin{array}
[c]{ccc}%
\tilde{A}_{1} & (\tilde{B}_{1})_{\mathring{\alpha}} & \tilde{B}_{2}\\
(\tilde{C}_{1})^{\mathring{\alpha}^{\prime}} & \big(\tilde{A}_{2}%
\big)_{\mathring{\alpha}}^{\mathring{\alpha}^{\prime}} & (\tilde{B}%
_{3})^{\mathring{\alpha}^{\prime}}\\
\tilde{C}_{2} & (\tilde{C}_{3})_{\mathring{\alpha}} & \tilde{A}_{3}%
\end{array}
\right)  \label{T1}%
\end{equation}
where $\alpha,\alpha^{\prime}$ assume the values $1,2,\dots,(0),\dots,\bar
{2},\bar{1}$ corresponding to the basis vectors of the auxiliary space
$V\cong\mathbb{C}^{N}$ and $\mathring{\alpha},\mathring{\alpha}^{\prime}$
assume the values $2,\dots,(0),\dots,\bar{2}$ corresponding to the basis
vectors of $\mathring{V}\cong\mathbb{C}^{N-2}$. We will also use the notation
$\tilde{A}=\tilde{A}_{1},~\tilde{B}=\tilde{B}_{1},~\tilde{C}=\tilde{C}_{1}$
and $\tilde{D}=\tilde{A}_{2}$ which is an $\left(  N-2\right)  \times\left(
N-2\right)  $ matrix in the auxiliary space. As usual the Yang-Baxter algebra
relation for the S-matrix yields the typical $TTS$-relation which implies the
basic algebraic properties of the sub-matrices $\tilde{A}_{i},\tilde{B}%
_{i},\tilde{C}_{i}$.

The reference co-vector is defined as usual by%
\[
\Omega\tilde{B}_{i}=0\,
\]
with the solution
\begin{equation}
\Omega_{\underline{\alpha}}=\delta_{\alpha_{1}}^{1}\cdots\delta_{\alpha_{n}%
}^{1}\,. \label{omega}%
\end{equation}
It satisfies
\begin{gather*}
\Omega\tilde{T}({\underline{\theta},z})=\Omega\left(
\begin{array}
[c]{ccc}%
a_{1}(\underline{\theta},z) & 0 & 0\\
\ast & a_{2}(\underline{\theta},z)\mathbf{1} & 0\\
\ast & \ast & a_{3}(\underline{\theta},z)
\end{array}
\right)  \,,\\
a_{1}(\underline{\theta},z)=1\,,~a_{2}(\underline{\theta},z)=\prod_{k=1}%
^{n}\tilde{b}(\theta_{i}-z),~a_{3}(\underline{\theta},z)=\prod_{k=1}%
^{n}\left(  \tilde{b}(\theta_{i}-z)+\tilde{d}(\theta_{i}-z)\right)  \,.
\end{gather*}

The basic Bethe ansatz co-vectors in (\ref{BA}) are defined as (for more
details see \cite{BFK5})%
\begin{equation}
\tilde{\Phi}_{\underline{\alpha}}^{\underline{\mathring{\beta}}}%
(\underline{\theta},\underline{z})=\left(  \Pi_{\underline{\beta}}%
^{\underline{\mathring{\beta}}}(\underline{z})\Omega\tilde{T}_{1}^{\beta_{m}%
}(\underline{\theta},z_{m})\dots\tilde{T}_{1}^{\beta_{1}}(\underline{\theta
},z_{1})\right)  _{\underline{\alpha}}=~~~%
\begin{array}
[c]{c}%
\unitlength4mm\begin{picture}(10,9) \thicklines\put(3.8,1){$\theta_1$} \put(8.3,1){$\theta_n$} \put(6,3.8){$z_i$} \put(5.0,6.2){1} \put(8.0,6.2){1} \put(9.5,1.8){1} \put(9.5,3.2){1} \put(9.5,4.8){1} \put(5.8,5.5){$\dots$} \put(.3,5.5){$_{\dots}$} \put(1.6,5.5){$_{\dots}$} \put(5,1){\line(0,1){5}} \put(8,1){\line(0,1){5}} \put(9,6){\oval(15.5,5)[lb]} \put(9,6){\oval(18,8)[lb]} \put(9,6){\oval(13,2)[lb]} \put(-.2,6){$\framebox(3,1){$\Pi$}$} \put(0,7){\line(0,1){1}} \put(1.25,7){\line(0,1){1}} \put(2.5,7){\line(0,1){1}} \put(4.4,0){$\alpha_1$} \put(7.8,0){$\alpha_n$} \put(-.2,8.3){$\mathring{\beta}_1$} \put(2.2,8.3){$\mathring{\beta}_m$} \end{picture}
\end{array}
\label{BS}%
\end{equation}
The matrix $\Pi_{\underline{\beta}}^{\underline{\mathring{\beta}}}%
(\underline{z})$ intertwines between the S-matrix $S$ of $O(N)$ and
$\mathring{S}$ of $O(N-2)$%
\begin{equation}
\tilde{\mathring{S}}_{ij}(z_{ij}\mathring{\nu}/\nu)\Pi_{\dots ij\dots
}(\underline{z})=\Pi_{\dots ji\dots}(\underline{z})\tilde{S}_{ij}(z_{ij})\,.
\label{PiS}%
\end{equation}
This matrix $\Pi$ is necessary\footnote{This matrix $\Pi$ is trivial for the
$SU(N)$ Bethe ansatz because the $SU(N)$ S-matrix amplitudes do not depend on
$N$ for a suitable normalization and parameterization.} because for the next
level Bethe ansatz the S-matrix $\mathring{S}(\theta)$ for $O(N-2)$ has to be
used. The co-vectors (\ref{BS}) are generalizations of vectors introduced by
Tarasov \cite{Tar} for a 3-state model, the Korepin-Izergin model. The
following relations for special components of $\Pi$ will be used below (for
more details see \cite{BFK5,BFK6})%
\begin{equation}
\Pi_{\underline{\beta}}^{\underline{\mathring{\beta}}}=\left\{
\begin{array}
[c]{lll}%
0 & \text{for} & \beta_{1}=1\\
0 & \text{for} & \beta_{m}=\bar{1}\\
\delta_{\beta_{1}}^{\mathring{\beta}_{1}}\,\Pi_{\beta_{2}\dots\beta_{m}%
}^{\mathring{\beta}_{2}\dots\mathring{\beta}_{m}} & \text{for} & \beta_{1}%
\neq\bar{1}\\[1mm]%
\Pi_{\beta_{1}\dots\beta_{m-1}}^{\mathring{\beta}_{1}\dots\mathring{\beta
}_{m-1}}\,\delta_{\beta_{m}}^{\mathring{\beta}_{m}} & \text{for} & \beta
_{m}\neq1\,.
\end{array}
\right.  \label{Pi}%
\end{equation}
In particular for $n=2$%
\begin{equation}
\Pi_{\beta_{1}\beta_{2}}^{\mathring{\beta}_{1}\mathring{\beta}_{2}}%
(\underline{z})=\delta_{\beta_{1}}^{\mathring{\beta}_{1}}\delta_{\beta_{2}%
}^{\mathring{\beta}_{2}}+f(z_{12})\mathbf{\mathring{C}}^{\mathring{\beta}%
_{1}\mathring{\beta}_{2}}\delta_{\beta_{1}}^{\bar{1}}\delta_{\beta2}%
^{1}\,,~~f(z)=\frac{i\pi\nu}{z+i\pi\left(  1-\nu\right)  } \label{Pi2}%
\end{equation}

\begin{remark}
\label{r1}The $\Pi$-matrix is responsible for the fact that the Bethe state
$\tilde{\Psi}_{\underline{\alpha}}(\underline{\theta},\underline{z})$ is a
symmetric function of the $z_{i}$, if the co-vector valued function
$L_{\underline{\mathring{\beta}}}(\underline{z})$ in $(\ref{PSI})$ satisfies
equation $(\ref{ik})$ for level $k=1$.
\end{remark}

It is well known (see \cite{BFK5}) that the `off-shell' Bethe ansatz states
are highest weight states if they satisfy certain matrix difference equations.
If there are $n$ particles, the $O(N)$ weights are%
\[
(w_{1},\dots,w_{\left[  N/2\right]  })=\left\{
\begin{array}
[c]{lll}%
\left(  n-n_{1},\dots,n_{\left[  N/2\right]  -1}-n_{\left[  N/2\right]
}\right)  & \text{for} & N~\text{odd}\\
\left(  n-n_{1},\dots,n_{\left[  N/2\right]  -2}-n_{-}-n_{+},n_{-}%
-n_{+}\right)  & \text{for} & N~\text{even}%
\end{array}
\right.
\]
where $n_{1}=m,n_{2},\dots$ are the numbers of $\tilde{T}$ operators in
(\ref{BS}) and the higher levels of the nesting. In particular $n_{\pm}$ are
the numbers of positive/negative chirality spinor $C$-operators. For the
on-shell Bethe ansatz for $N$ even see also \cite{dVK}. As is well known (see
e.g. \cite{OWR,Re1,RW} and references therein), the various levels of the
nested Bethe ansatz correspond to the nodes of the Dynkin diagrams of the
corresponding Lie algebras $D_{N/2}$ for $N=$ even and $B_{[N/2]}$ for $N=$
odd:
\[
D_{N/2}:~~%
\begin{array}
[t]{c}%
\unitlength3mm\begin{picture}(16,2.6)(0,.4) \put(.6,-.5){$n_1$} \put(1,1){\circle{1}} \put(1.5,1){\line(1,0){2}} \put(3.6,-.5){$n_2$} \put(4,1){\circle{1}} \put(4.5,1){\line(1,0){2}} \put(6.8,.9){$\dots$} \put(8.5,1){\line(1,0){2}} \put(11,1){\circle{1}} \put(11.25,1.45){\line(2,1){2}} \put(11.25,.55){\line(2,-1){2}} \put(15,2.3){$n_+$} \put(15,-1){$n_-$} \put(13.8,2.6){\circle{1}} \put(13.8,-.6){\circle{1}} \end{picture}\\
~
\end{array}
~~\qquad B_{[N/2]}:~~%
\begin{array}
[t]{c}%
\unitlength3mm\begin{picture}(16,2)(0,.4) \put(.6,-.5){$n_1$} \put(1,1){\circle{1}} \put(1.5,1){\line(1,0){2}} \put(3.6,-.5){$n_2$} \put(4,1){\circle{1}} \put(4.5,1){\line(1,0){2}} \put(6.8,.9){$\dots$} \put(8.5,1){\line(1,0){2}} \put(11,1){\circle{1}} \put(11.5,1.2){\line(1,0){2}} \put(11.5,.8){\line(1,0){2}} \put(13,-.5){$n_{[N/2]}$} \put(14,1){\circle{1}} \bezier{77}(12,1.4)(12.5,1.2)(13,1) \bezier{77}(12,.6)(12.5,.8)(13,1) \end{picture}\\
~
\end{array}
\]

\section{Generalized form factors}

\label{s3}

For a state of $n$ particles of kind $\alpha_{i}$ with rapidities $\theta_{i}$
and a local operator $\mathcal{O}(x)$ we define the form factor functions
$F_{\alpha_{1}\dots\alpha_{n}}^{\mathcal{O}}(\theta_{1},\dots,\theta_{n})$, or
using a short hand notation $F_{\underline{\alpha}}^{\mathcal{O}}%
(\underline{\theta})$, by
\begin{equation}
\langle\,0\,|\,\mathcal{O}(x)\,|\,\theta_{1},\dots,\theta_{n}\,\rangle
_{\underline{\alpha}}^{in}=e^{-ix(p_{1}+\cdots+p_{n})}F_{\underline{\alpha}%
}^{\mathcal{O}}(\underline{\theta})~,~~\text{for}~\theta_{1}>\dots>\theta_{n}.
\label{1.8}%
\end{equation}
where $\underline{\alpha}=(\alpha_{1},\dots,\alpha_{n})$ and $\underline
{\theta}=(\theta_{1},\dots,\theta_{n})$. For all other arrangements of the
rapidities the functions $F_{\underline{\alpha}}^{\mathcal{O}}(\underline
{\theta})$ are given by analytic continuation. Note that the physical value of
the form factor, i.e. the left hand side of (\ref{1.8}), is given for ordered
rapidities as indicated above and the statistics of the particles. The
$F_{\underline{\alpha}}^{\mathcal{O}}(\underline{\theta})$ are considered as
the components of a co-vector valued function $F_{1\dots n}^{\mathcal{O}%
}(\underline{\theta})\in V_{1\dots n}=\left(  V^{1\dots n}\right)  ^{\dagger}$.

Now we formulate the main properties of form factors in terms of the functions
$F_{1\dots n}^{\mathcal{O}}$.

\subsection{Form factor equations}

The co-vector valued function $F_{1\dots n}^{\mathcal{O}}({\underline{\theta}%
})$ defined by (\ref{1.8}) is meromorphic in all variables $\theta_{1}%
,\dots,\theta_{n}$ and satisfies the following relations:

\begin{itemize}
\item[(i)] The Watson's equations describe the symmetry property under the
permutation of both, the variables $\theta_{i},\theta_{j}$ and the spaces
$i,j=i+1$ at the same time
\begin{equation}
F_{\dots ij\dots}^{\mathcal{O}}(\dots,\theta_{i},\theta_{j},\dots)=F_{\dots
ji\dots}^{\mathcal{O}}(\dots,\theta_{j},\theta_{i},\dots)\,S_{ij}(\theta_{ij})
\label{1.10}%
\end{equation}
for all possible arrangements of the $\theta$'s.

\item[(ii)] The crossing relation implies a periodicity property under the
cyclic permutation of the rapidity variables and spaces
\begin{multline}
^{~\text{out,}\bar{1}}\langle\,p_{1}\,|\,\mathcal{O}(0)\,|\,p_{2},\dots
,p_{n}\,\rangle_{2\dots n}^{\text{in,conn.}}\\
=F_{1\ldots n}^{\mathcal{O}}(\theta_{1}+i\pi,\theta_{2},\dots,\theta
_{n})\mathbf{C}^{\bar{1}1}=F_{2\ldots n1}^{\mathcal{O}}(\theta_{2}%
,\dots,\theta_{n},\theta_{1}-i\pi)\mathbf{C}^{1\bar{1}} \label{1.12}%
\end{multline}
The charge conjugation matrix $\mathbf{C}^{\bar{1}1}$ is given by (\ref{Cc}).

\item[(iii)] There are poles determined by one-particle states in each
sub-channel. In particular the function $F_{\underline{\alpha}}^{\mathcal{O}%
}({\underline{\theta}})$ has a pole at $\theta_{12}=i\pi$ such that
\begin{equation}
\operatorname*{Res}_{\theta_{12}=i\pi}F_{1\dots n}^{\mathcal{O}}(\theta
_{1},\dots,\theta_{n})=2i\,\mathbf{C}_{12}\,F_{3\dots n}^{\mathcal{O}}%
(\theta_{3},\dots,\theta_{n})\left(  \mathbf{1}-S_{2n}\dots S_{23}\right)  \,.
\label{1.14}%
\end{equation}

\item[(v)] Naturally, since we are dealing with relativistic quantum field
theories we finally have
\begin{equation}
F_{1\dots n}^{\mathcal{O}}(\theta_{1}+\mu,\dots,\theta_{n}+\mu)=e^{s\mu
}\,F_{1\dots n}^{\mathcal{O}}(\theta_{1},\dots,\theta_{n}) \label{1.18}%
\end{equation}
if the local operator transforms under Lorentz transformations as
$\mathcal{O}\rightarrow e^{s\mu}\mathcal{O}$ where $s$ is the
\textquotedblleft spin\textquotedblright\ of $\mathcal{O}$.
\end{itemize}

As was shown in \cite{BFKZ} the properties (i) - (iii) follow from general
LSZ-assumptions and \textquotedblleft maximal analyticity\textquotedblright,
which means that $F_{1\dots n}^{\mathcal{O}}(\underline{\theta})$ is a
meromorphic function with respect to all $\theta$'s, and in the `physical'
strips $0<\operatorname{Im}\theta_{ij}<\pi$~$(\theta_{ij}=\theta_{i}%
-\theta_{j},\,i<j)$ there are only poles of physical origin as for example
bound state poles. In general there is also the form factor equation (iv)
referring to bound states. Since there are no bound states in the
$O(N)~\sigma$-model this equation is empty.

We will now provide a constructive and systematic way of how to solve the form
factor equations for the co-vector valued function $F_{1\dots n}^{\mathcal{O}%
}$, once the scattering matrix is given.

\paragraph{Minimal form factors:}

The solutions of Watson's and the crossing equations (i) and (ii) for two
particles
\begin{align*}
\text{(i)}  &  :F\left(  \theta\right)  =S\left(  \theta\right)  F\left(
-\theta\right) \\
\text{(ii)}  &  :F\left(  i\pi-\theta\right)  =F\left(  i\pi-\theta\right)
\end{align*}
with no poles in the physical strip $0\leq\operatorname{Im}\theta\leq\pi$ and
at most a simple zero at $\theta=0$ are the minimal form factors. For the
construction of the off-shell Bethe ansatz the minimal form factor of highest
weight eigenvalue of the $O(N)$ S-matrix $a(\theta)=S_{+}(\theta)$ of
(\ref{1.6}) is essential
\begin{equation}
F\left(  \theta\right)  =c\exp\left(  \int_{0}^{\infty}\frac{dt}{t\sinh
t}\frac{1-e^{-t\nu}}{1+e^{-t}}\left(  1-\cosh t\left(  1-\frac{\theta}{i\pi
}\right)  \right)  \right)  \,. \label{minF}%
\end{equation}
For convenience we have introduced the constant $c$, which is defined below
(\ref{c}). The two other minimal form factors belonging to the S-matrix
eigenvalues $S_{-}\left(  \theta\right)  $ and $S_{0}\left(  \theta\right)  $
(see (\ref{EV})) are \cite{KW}
\begin{align}
F_{-}\left(  \theta\right)   &  =\frac{i}{\sinh\frac{1}{2}\theta}\frac
{\Gamma^{2}\left(  \frac{1}{2}+\frac{1}{2}\nu\right)  }{\Gamma\left(
1+\frac{1}{2}\nu-\frac{1}{2\pi i}\theta\right)  \Gamma\left(  \frac{1}{2}%
\nu+\frac{1}{2\pi i}\theta\right)  }F_{+}\left(  \theta\right)  \label{minF-}%
\\
F_{0}\left(  \theta\right)   &  =\frac{\sinh\theta}{i\pi-\theta}F_{-}\left(
\theta\right) \label{minF0}\\
F_{+}\left(  \theta\right)   &  =\frac{1}{c}F\left(  \theta\right)  .\nonumber
\end{align}

\section{Nested \textquotedblleft off-shell\textquotedblright\ Bethe ansatz
for $O(N)$}

\label{s4}

\subsection{The fundamental theorem}

We write the general form factor $F_{1\dots n}^{\mathcal{O}}(\underline
{\theta})$ for n-particles following \cite{KW} as in (\ref{2.10}) where
$F(\theta)$ is the minimal form factor function (\ref{minF}). The K-function
$K_{1\dots n}^{\mathcal{O}}(\underline{\theta})$ contains the entire pole
structure and is determined by the form factor equations (i) - (iii). We
propose the ansatz (\ref{2.16}) for the K-function in terms of a nested
`off-shell' Bethe ansatz (\ref{BA})
\[
\fbox{$\rule[-0.2in]{0in}{0.5in}\displaystyle~K_{1\dots n}^{\mathcal{O}%
}(\underline{\theta})=N_{n}^{\mathcal{O}}\int_{\mathcal{C}_{\underline{\theta
}}^{1}}dz_{1}\cdots\int_{\mathcal{C}_{\underline{\theta}}^{m}}dz_{m}%
\,\tilde{h}(\underline{\theta},\underline{z})\,p^{\mathcal{O}}(\underline
{\theta},\underline{z})\,\,\tilde{\Psi}_{1\dots n}(\underline{\theta
},\underline{z})$}%
\]
written as a multiple contour integral. The scalar function $\tilde
{h}(\underline{\theta},\underline{z})$ depends only on the S-matrix and not on
the specific operator $\mathcal{O}(x)$%
\begin{equation}
\tilde{h}(\underline{\theta},\underline{z})=\prod_{i=1}^{n}\prod_{j=1}%
^{m}\tilde{\phi}_{j}(\theta_{i}-z_{j})\prod_{1\leq i<j\leq m}\tau_{ij}%
(z_{i}-z_{j})\,. \label{h}%
\end{equation}
\begin{figure}[h]%
\[
\unitlength4.5mm\begin{picture}(26,13)(0,-2)
\thicklines
\put(1,0){
\put(.15,-2){$^{\hbox{\circle{.3}}}~\theta_n-i\pi $}
\put(-.12,8){${\times}~\theta_n+4i\pi $}
\put(-.12,4){${\times}~\theta_n+2i\pi $}
\put(.15,-1){$^{\hbox{\circle{.3}}}~\theta_n-i\pi\nu$}
\put(0,3){$\bullet~~\theta_n+i\pi(2-\nu)$}
\put(0,7){$\bullet~~\theta_n+i\pi(4-\nu)$}
\put(.15,8.3){\oval(.8,19)[b]}
\put(-.25,5){\vector(0,-1){0}}
}
\put(9,3){\dots}
\put(12,0){
\put(.15,-2){$^{\hbox{\circle{.3}}}~\theta_2-i\pi $}
\put(-.12,8){${\times}~\theta_2+4i\pi $}
\put(-.12,4){${\times}~\theta_2+2i\pi $}
\put(.15,-1){$^{\hbox{\circle{.3}}}~\theta_2-i\pi\nu$}
\put(0,3){$\bullet~~\theta_2+i\pi(2-\nu)$}
\put(0,7){$\bullet~~\theta_2+i\pi(4-\nu)$}
\put(.15,8.3){\oval(.8,19)[b]}
\put(-.25,5){\vector(0,-1){0}}
}
\put(20,2){
\put(.15,-2){$^{\hbox{\circle{.3}}}~\theta_1-i\pi $}
\put(-.12,8){${\times}~\theta_1+4i\pi $}
\put(-.12,4){${\times}~\theta_1+2i\pi $}
\put(.15,-1){$^{\hbox{\circle{.3}}}~\theta_1-i\pi\nu$}
\put(0,3){$\bullet~~\theta_1+i\pi(2-\nu)$}
\put(0,7){$\bullet~~\theta_1+i\pi(4-\nu)$}
\put(.15,8.3){\oval(.8,19)[b]}
\put(-.25,5){\vector(0,-1){0}}
}
\end{picture}
\]
\caption{The integration contour $\mathcal{C}_{\underline{\theta}}^{o}$. The
bullets and the crosses refer to poles and zeroes of the integrand resulting
from $\,\tilde{\psi}(\theta_{i}-z_{j})$ and the small open circles refer to
poles originating from $\tilde{S}(\theta_{i}-z_{j})$.}%
\label{f5.1}%
\end{figure}\begin{figure}[h]%
\[
\unitlength4.5mm\begin{picture}(26,12)(0,-3)
\thicklines
\put(1,0){
\put(-.17,-3){${\times}~~\theta_n-i\pi(4+\nu)$}
\put(-.17,1){${\times}~~\theta_n-i\pi(2+\nu)$}
\put(-.19,5.1){${\times}$}
\put(0,-2){$\bullet~~\theta_n-4\pi i$}
\put(0,2){$\bullet~~\theta_n-2\pi i$}
\put(.15,3.9){$^{\hbox{\circle{.3}}}~\theta_n-i\pi $}
\put(.15,5){$^{\hbox{\circle{.3}}}~\theta_n-i\pi\nu$}
\put(0,6){$\bullet~~\theta_n$}
\put(.15,6.2){\oval(1,1)}
\put(.5,6.68){\vector(1,0){0}}
\put(.15,-3){\oval(.8,12)[t]}
\put(-.25,0){\vector(0,1){0}}
}
\put(8,3){\dots}
\put(12,0){
\put(-.17,-3){${\times}~~\theta_2-i\pi(4+\nu)$}
\put(-.17,1){${\times}~~\theta_2-i\pi(2+\nu)$}
\put(-.19,5.1){${\times}$}
\put(0,-2){$\bullet~~\theta_2-4\pi i$}
\put(0,2){$\bullet~~\theta_2-2\pi i$}
\put(.15,3.9){$^{\hbox{\circle{.3}}}~\theta_2-i\pi $}
\put(.15,5){$^{\hbox{\circle{.3}}}~\theta_2-i\pi\nu$}
\put(0,6){$\bullet~~\theta_2$}
\put(.15,6.2){\oval(1,1)}
\put(.5,6.68){\vector(1,0){0}}
\put(.15,-3){\oval(.8,12)[t]}
\put(-.25,0){\vector(0,1){0}}
}
\put(20,2){
\put(-.17,-3){${\times}~~\theta_1-i\pi(4+\nu)$}
\put(-.17,1){${\times}~~\theta_1-i\pi(2+\nu)$}
\put(-.19,5.1){${\times}$}
\put(0,-2){$\bullet~~\theta_1-4\pi i$}
\put(0,2){$\bullet~~\theta_1-2\pi i$}
\put(.15,3.9){$^{\hbox{\circle{.3}}}~\theta_1-i\pi $}
\put(.15,5){$^{\hbox{\circle{.3}}}~\theta_1-i\pi\nu$}
\put(0,6){$\bullet~~\theta_1$}
\put(.15,6.2){\oval(1,1)}
\put(.5,6.68){\vector(1,0){0}}
\put(.15,-3){\oval(.8,12)[t]}
\put(-.25,0){\vector(0,1){0}}
}
\end{picture}
\]
\caption{The integration contour $\mathcal{C}_{\underline{\theta}}^{e}$. The
bullets and the crosses refer to poles and zeroes of the integrand resulting
from $\,\tilde{\chi}(\theta_{i}-z_{j})$ and the small open circles refer to
poles originating from $\tilde{S}(\theta_{i}-z_{j})$.}%
\label{f5.2}%
\end{figure}The functions $\tilde{\phi}_{j}$ and $\tau_{ij}$ have to satisfy
the shift equations%
\begin{align}
\tilde{\phi}_{j}(\theta-2\pi i)  &  =\tilde{b}(\theta)\tilde{\phi}_{j}%
(\theta)\label{shiftphi}\\
\tau_{ij}(z-2\pi i)/\tilde{b}(2\pi i-z)  &  =\tau_{ij}(z)/\tilde{b}(z)
\label{shifttau}%
\end{align}
which follow from the form factor equation (ii) or (\ref{1.12})
\cite{BFK5,BFK6}. Here, for the $O(N)$ form factors, they depend on whether
$i,j$ are even or odd%
\begin{gather}
\tilde{\phi}_{e}(\theta)=\tilde{\chi}(\theta),~\tilde{\phi}_{o}(\theta
)=\tilde{\psi}(\theta)\label{2.17}\\
\tau_{ee}(z)=\tau_{oo}(z)=\frac{1}{\tilde{\chi}(-z)\tilde{\chi}(z)},~\tau
_{eo}(z)=\tau_{oe}(-z)=\frac{1}{\tilde{\chi}(-z)\tilde{\psi}(z)} \label{2.19}%
\end{gather}
where $\tilde{\psi}(z)$ and $\tilde{\chi}(z)$ are
\begin{equation}
\tilde{\psi}(\theta)=\frac{\Gamma\left(  1-\frac{1}{2}\nu+\frac{1}{2\pi
i}\theta\right)  }{\Gamma\left(  1+\frac{1}{2\pi i}\theta\right)  }%
\,,~~\tilde{\chi}(\theta)=\frac{\Gamma(-\frac{1}{2\pi i}\theta)}{\Gamma
(\frac{1}{2}\nu-\frac{1}{2\pi i}\theta)}\,. \label{psichi}%
\end{equation}
In addition the equation
\begin{equation}
F(\theta)F(\theta+i\pi)\tilde{\psi}(-\theta-i\pi+i\pi\nu)\tilde{\chi}%
(-\theta)=1 \label{FF}%
\end{equation}
is satisfied. It follows from the form factor equation (iii) or (\ref{1.14})
as will be discussed in appendix \ref{sd}.

Notice that the equations (\ref{psichi}) and (\ref{FF}) also determine the
normalization constant $c$ in (\ref{minF}) as%
\begin{equation}
c=\frac{1}{\sqrt{2}\pi}\Gamma\left(  \frac{3}{4}\right)  \Gamma\left(
\frac{1}{4}+\frac{1}{2}\nu\right)  \exp\left(  \int_{0}^{\infty}\frac{dt}%
{t}\frac{1-e^{-t\nu}}{1+e^{-t}}\frac{1-\cosh\frac{1}{2}t}{\sinh t}\right)  \,.
\label{c}%
\end{equation}

The dependence on the specific operator $\mathcal{O}(x)$ is encoded in the
scalar p-function $p^{\mathcal{O}}(\underline{\theta},\underline{z})$ which is
in general a simple function of $e^{\theta_{i}}$ and $e^{z_{j}}$ (see below).
By means of the ansatz (\ref{2.10}) and (\ref{2.16}) we have transformed the
complicated form factor equations (i) - (v) (which are in general matrix
equations) into much simpler scalar equations for the scalar p-function (see
(\ref{4.25})).

The integration contours (corresponding to the functions $\tilde{\phi}%
_{e}(\theta)=\tilde{\chi}(\theta)$ and $\tilde{\phi}_{o}(\theta)=\tilde{\psi
}(\theta)$) $\mathcal{C}_{\underline{\theta}}^{o}$ and $\mathcal{C}%
_{\underline{\theta}}^{e}$ are depicted in Fig.~\ref{f5.1} and Fig.~\ref{f5.2}.

\begin{theorem}
\label{TN}We make the following assumptions:

\begin{enumerate}
\item The p-function $p(\underline{\theta},\underline{z})$ satisfies the
equations
\begin{equation}
\left.
\begin{array}
[c]{rl}%
(\mathrm{i}^{\prime}): & p(\underline{\theta},\underline{z})~\text{is
symmetric under }\theta_{i}\leftrightarrow\theta_{j}\\
(\mathrm{ii}^{\prime}): & p(\underline{\theta},\underline{z})=p(\theta
_{1}+2\pi i,\theta_{2},\dots,\underline{z})=p(\underline{\theta},z_{1}+2\pi
i,z_{2},\dots)\\
(\mathrm{iii}^{\prime}): & p(\underline{\theta},\underline{z})=\,p(\underline
{\check{\theta}},\underline{\check{z}})~\text{for }\theta_{12}=i\pi
,\,z_{1}=\theta_{1}-i\pi\nu\text{ and }z_{2}=\theta_{2}%
\end{array}
\right\}  \label{4.25}%
\end{equation}
where the short notations $\underline{\check{\theta}}=(\theta_{3},\dots
,\theta_{n})$ and $\underline{\check{z}}=(z_{3},\dots,z_{m})$ are used.

\item The higher level function $L_{\underline{\beta}}(\underline{z})$ in
$(\ref{PSI})$ satisfies $(\mathrm{i})^{(k)}$ - $(\mathrm{iii})^{(k)}$ of
$(\ref{ik})$ - $(\ref{iiik})$ for $k=1$

\item The normalization constants in $(\ref{2.16})$ satisfy (for $N>4$)%
\begin{equation}
N_{m}^{\mathcal{O}}=\frac{1}{\left[  \frac{1}{2}m\right]  \left[  \frac{1}%
{2}m+\frac{1}{2}\right]  }\frac{\nu F(i\pi)}{8\pi}\frac{\tilde{\chi}%
(-i\pi\left(  1+\nu\right)  )}{\tilde{\psi}^{2}(i\pi\nu)\tilde{\chi}(-i\pi
)}N_{m-2}^{\mathcal{O}} \label{norm}%
\end{equation}

\end{enumerate}

\noindent then the co-vector valued function $F_{\underline{\alpha}%
}(\underline{\theta})$ given by the ansatz $(\ref{2.10})$ and the integral
representation $(\ref{2.16})$ satisfies the form factor equations
$(\mathrm{i}),(\mathrm{ii})$ and $(\mathrm{iii})$ of $(\ref{1.10})$ -
$(\ref{1.14})$.
\end{theorem}

\noindent The proof of this\ theorem can be found in appendix \ref{sd}. The
normalization relation (\ref{norm}), of course, depends on how the higher
level K-functions are normalized. This will be discussed in subsection
\ref{s2.4}.

\subsection{O(3) form factors}

\label{s42}

In the complex basis the three one-particle states are $1,0,\bar{1}$. The
S-matrix for $\nu=2$ is%
\[
S^{O(3)}(\theta)=\frac{\theta-i\pi}{\theta+i\pi}\left(  \frac{\theta}%
{\theta-2i\pi}\mathbf{1-}\frac{2i\pi}{\theta-2i\pi}\mathbf{P-}\frac{\theta
}{\theta-2i\pi}\frac{2i\pi}{i\pi-\theta}\mathbf{K}\right)
\]
and the eigenvalues are
\begin{align*}
S_{+}^{O(3)}\left(  \theta\right)   &  =\frac{\theta-i\pi}{\theta+i\pi}\\
S_{-}^{O(3)}\left(  \theta\right)   &  =\frac{\theta-i\pi}{\theta+i\pi}%
\frac{\theta+2\pi i}{\theta-2\pi i}\\
S_{0}^{O(3)}\left(  \theta\right)   &  =\frac{\theta+2\pi i}{\theta-2\pi i}\,.
\end{align*}
The minimal form factors for these S-matrix eigenvalues are
\begin{align*}
F_{+}\left(  \theta\right)   &  =\tfrac{1}{2}\left(  \theta-i\pi\right)
\tanh\tfrac{1}{2}\theta\\
F_{-}\left(  \theta\right)   &  =\tfrac{1}{2}\pi^{2}\frac{\left(  \theta
-i\pi\right)  }{\theta\left(  \theta-2\pi i\right)  }\tanh\tfrac{1}{2}\theta\\
F_{0}\left(  \theta\right)   &  =-\pi^{2}\frac{1}{\theta\left(  \theta-2\pi
i\right)  }\sinh^{2}\tfrac{1}{2}\theta\,.
\end{align*}
The general form factors%
\[
F_{\underline{\alpha}}^{\mathcal{O}}(\underline{\theta})=K_{\underline{\alpha
}}^{\mathcal{O}}(\underline{\theta})\prod_{1\leq i<j\leq n}F(\theta_{ij})
\]
are given by (\ref{2.10}) with $F\left(  \theta\right)  =2F_{+}\left(
\theta\right)  =\left(  \theta-i\pi\right)  \tanh\tfrac{1}{2}\theta$ and the
(one level) `off-shell' Bethe ansatz (\ref{2.16})
\begin{equation}
K_{\underline{\alpha}}^{\mathcal{O}}(\underline{\theta})=N_{m}^{\mathcal{O}%
}\int_{\mathcal{C}_{\underline{\theta}}^{1}}dz_{1}\cdots\int_{\mathcal{C}%
_{\underline{\theta}}^{m}}dz_{m}\,\tilde{h}(\underline{\theta},\underline
{z})\,p^{\mathcal{O}}(\underline{\theta},\underline{z})\,\,\tilde{\Psi
}_{\underline{\alpha}}(\underline{\theta},\underline{z}) \label{BA3}%
\end{equation}
with $\tilde{h}(\underline{\theta},\underline{z})$ given by (\ref{h}). The
functions $\tilde{\phi}_{j}(\theta)$ and $\tau_{ij}(z)$ we get from
(\ref{psichi}) (up to inessential constants) as
\begin{align*}
\tilde{\psi}(\theta)  &  =\tilde{\chi}(\theta)=\frac{1}{\theta}\\
\tau(z)  &  =z^{2}%
\end{align*}
such that (\ref{FF}) holds. It turns out that in (\ref{BA3}) we have to
calculate only some residues because for $\nu=2$ many zeroes cancel poles such
that we may replace the contour integrals $\int_{\mathcal{C}_{\underline
{\theta}}^{j}}dz\dots$ for even and odd $j$
\[
\int_{\mathcal{C}_{\underline{\theta}}}dz\dots\rightarrow\sum_{i=1}^{n}\left(
{\oint_{\theta_{i}}}+{\oint_{\theta_{i}-2\pi i}}\right)  dz\dots
\]
where ${\oint_{\theta}}dz\dots$ means an integral along a small circle around
$\theta$. The state $\tilde{\Psi}_{\underline{\alpha}}$ in (\ref{BA3}) is here
proportional to Bethe ansatz co-vectors (\ref{BS})%
\[
\tilde{\Psi}_{\underline{\alpha}}(\underline{\theta},\underline{z}%
)=L(\underline{z})\,\tilde{\Phi}_{\underline{\alpha}}(\underline{\theta
},\underline{z})
\]
where the scalar function
\[
L(\underline{z})=\prod\limits_{1\leq i<j\leq m}L(z_{ij})\,,~~L(z)=\frac
{\left(  z-i\pi\right)  }{z\left(  z-2\pi i\right)  }\tanh\tfrac{1}{2}z
\]
is the minimal solution of the equations (\ref{ik}), (\ref{iik}) and
(\ref{iiik}) with the scalar S-matrix%
\[
\tilde{\mathring{S}}(z\mathring{\nu}/\nu)=\tilde{\mathring{S}}(-z)=\frac
{z-i\pi}{z+i\pi}\frac{z+2i\pi}{z-2i\pi}\,.
\]
The $O(3)$ weight of the state is%
\[
w=n-m\,.
\]
For explicit examples see section \ref{s5} and appendix \ref{O3}.

\subsection{O(4) form factors}

\label{s2.3}

In the complex basis the four one-particle states are $1,2,\bar{2},\bar{1}$.
The S-matrix for $\nu=1$ is%
\[
S^{O(4)}(\theta)=a^{O(4)}(\theta)\left(  \frac{\theta}{\theta-i\pi}%
\mathbf{1-}\frac{i\pi}{\theta-i\pi}\mathbf{P-}\frac{\theta}{\theta-i\pi}%
\frac{i\pi}{i\pi-\theta}\mathbf{K}\right)
\]
and the S-matrix eigenvalues are%
\begin{align}
a^{O(4)}(\theta)  &  =S_{+}^{O(4)}(\theta)=-\left(  \frac{\Gamma\left(
1-\frac{\theta}{2\pi i}\right)  \Gamma\left(  \frac{1}{2}+\frac{\theta}{2\pi
i}\right)  }{\Gamma\left(  1+\frac{\theta}{2\pi i}\right)  \Gamma\left(
\frac{1}{2}-\frac{\theta}{2\pi i}\right)  }\right)  ^{2}\nonumber\\
S_{-}^{O(4)}(\theta)  &  =\frac{i\pi+\theta}{i\pi-\theta}\left(  \frac
{\Gamma\left(  1-\frac{\theta}{2\pi i}\right)  \Gamma\left(  \frac{1}{2}%
+\frac{\theta}{2\pi i}\right)  }{\Gamma\left(  1+\frac{\theta}{2\pi i}\right)
\Gamma\left(  \frac{1}{2}-\frac{\theta}{2\pi i}\right)  }\right)
^{2}\label{EW4}\\
S_{0}^{O(4)}(\theta)  &  =-\left(  \frac{\Gamma\left(  1-\frac{\theta}{2\pi
i}\right)  \Gamma\left(  \frac{3}{2}+\frac{\theta}{2\pi i}\right)  }%
{\Gamma\left(  1+\frac{\theta}{2\pi i}\right)  \Gamma\left(  \frac{3}{2}%
-\frac{\theta}{2\pi i}\right)  }\right)  ^{2}.\nonumber
\end{align}
The group isomorphy $O(4)\simeq SU(2)\otimes SU(2)$ reflects in terms of the
S-matrices. The $O(4)$ S-matrix can be written as a tensor product of two
$SU(2)$ S-matrices \cite{BKKW,BW,BFK0}%
\begin{align*}
S^{SU(2)}(\theta)  &  =a^{SU(2)}(\theta)\left(  \frac{\theta}{\theta-i\pi
}\mathbf{1-}\frac{i\pi}{\theta-i\pi}\mathbf{P}\right) \\
a^{SU(2)}(\theta)  &  =S_{+}^{SU(2)}(\theta)=-\frac{\Gamma\left(
1-\frac{\theta}{2\pi i}\right)  \Gamma\left(  1-\frac{1}{N}+\frac{\theta}{2\pi
i}\right)  }{\Gamma\left(  1+\frac{\theta}{2\pi i}\right)  \Gamma\left(
1-\frac{1}{N}-\frac{\theta}{2\pi i}\right)  }%
\end{align*}
or more precisely%
\begin{align}
\Gamma_{\alpha}^{AB}\Gamma_{\beta}^{CD}\left(  S^{O(4)}\right)  _{\delta
\gamma}^{\alpha\beta}  &  =-\left(  ^{+\!}S^{SU(2)}\right)  _{C^{\prime
}A^{\prime}}^{AC}\left(  ^{-\!}S^{SU(2)}\right)  _{D^{\prime}B^{\prime}}%
^{BD}\Gamma_{\delta}^{C^{\prime}D^{\prime}}\Gamma_{\gamma}^{A^{\prime
}B^{\prime}}\,.\label{SSS}\\%
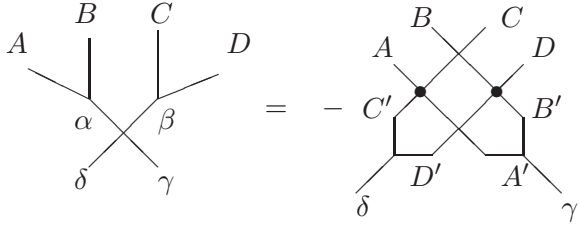
\begin{array}
[c]{c}%
\unitlength1mm\begin{picture}(31,25)(2,-5) \put(13,8){\line(1,-1){9}} \put(22,8){\line(-1,-1){9}} \put(13,8){\line(0,1){8}} \put(13,8){\line(-2,1){8}} \put(22,8){\line(0,1){9}} \put(22,8){\line(5,2){8}} \put(2,14){$A$} \put(11,18){$B$} \put(31,14){$D$} \put(21,18){$C$} \put(11,4){$\alpha$} \put(22,4){$\beta$} \put(11,-4){$\delta$} \put(22,-4){$\gamma$} \end{picture}
\end{array}
~  &  =~~-%
\begin{array}
[c]{c}%
\unitlength1mm\begin{picture}(27,30)(0,-2) \put(5,18){\line(1,-1){12}} \put(17,6){\line(1,0){5}} \put(22,6){\line(0,1){5}} \put(22,11){\line(-1,1){12}} \put(17,23){\line(-1,-1){12}} \put(5,11){\line(0,-1){5}} \put(5,6){\line(1,0){5}} \put(10,6){\line(1,1){12}} \put(22,6){\line(1,-1){5}} \put(5,6){\line(-1,-1){5}} \put(2,19){$A$} \put(7,23){$B$} \put(23,19){$D$} \put(19,23){$C$} \put(7,2){$D'$} \put(19,2){$A'$} \put(.8,11){$C'$} \put(23,11){$B'$} \put(0,-2){$\delta$} \put(27,-2){$\gamma$} \put(7.5,13.5){$\bullet$} \put(17.5,13.5){$\bullet$} \end{picture}
\end{array}
\nonumber
\end{align}
The $SU(2)$ S-matrices $^{\pm}S^{SU(2)}(\theta)$ correspond to the spinor
representations of $O(4)$ with positive (negative) chirality (see
\cite{ShWi,KT1}). In particular%
\begin{equation}
a^{O(4)}(\theta)=-\left(  \frac{\Gamma\left(  1-\frac{\theta}{2\pi i}\right)
\Gamma\left(  \frac{1}{2}+\frac{\theta}{2\pi i}\right)  }{\Gamma\left(
1+\frac{\theta}{2\pi i}\right)  \Gamma\left(  \frac{1}{2}-\frac{\theta}{2\pi
i}\right)  }\right)  ^{2}=-\left(  a^{SU(2)}(\theta)\right)  ^{2}\,.
\label{aa}%
\end{equation}
The relative S-matrix for states of different chirality is trivial
$S=\mathbf{1}$. The intertwiners $\Gamma_{\alpha}^{AB}$ have been discussed in
\cite{KT1,BFK5}. In the complex basis of the $O(4)$ states and the fundamental
$SU(2)$ representations the intertwiner matrix is diagonal and%
\begin{equation}
\left(
\begin{array}
[c]{cccc}%
\Gamma_{1}^{\uparrow_{+}\uparrow_{-}}, & \Gamma_{2}^{\uparrow_{+}%
\downarrow_{-}}, & \Gamma_{\bar{2}}^{\downarrow_{+}\uparrow_{-}}, &
\Gamma_{\bar{1}}^{\downarrow_{+}\downarrow_{-}}%
\end{array}
\right)  =\left(
\begin{array}
[c]{cccc}%
-1, & 1, & 1, & 1
\end{array}
\right)  . \label{inter}%
\end{equation}

The minimal form factors for the S-matrix eigenvalues (\ref{EW4}) are%
\begin{align}
F^{O(4)}\left(  \theta\right)   &  =F_{+}^{O(4)}\left(  \theta\right)
=\exp\left(  \int_{0}^{\infty}\frac{dt}{t\sinh t}\frac{1-e^{-t}}{1+e^{-t}%
}\left(  1-\cosh t\left(  1-\frac{\theta}{i\pi}\right)  \right)  \right)
\nonumber\\
F_{-}^{O(4)}\left(  \theta\right)   &  =2\frac{1}{\theta-i\pi}\coth\tfrac
{1}{2}\theta\,F_{+}^{O(4)}\left(  \theta\right) \label{Fmin4}\\
F_{0}^{O(4)}\left(  \theta\right)   &  =\frac{1}{i\pi-\theta}\sinh
\theta\,F_{-}^{O(4)}\left(  \theta\right)  .\nonumber
\end{align}
Equation (\ref{aa}) for the highest weight S-matrix amplitudes means that the
highest weight minimal form factors $F=F_{+}$ are related by
\begin{equation}
F^{O(4)}\left(  \theta\right)  =\frac{i}{\sinh\frac{1}{2}\theta}\left(
F^{SU(2)}(\theta)\right)  ^{2}. \label{F4F2}%
\end{equation}

Similarly, as for S-matrices (\ref{SSS}) the group isomorphy $O(4)\simeq
SU(2)\otimes SU(2)$ reflects in terms of the form factors. The co-vector
valued function%
\begin{equation}
F_{\underline{\alpha}}^{\mathcal{O}}(\underline{\theta})=c_{n}\sum_{l}%
\prod_{i<j}\coth\tfrac{1}{2}\theta_{ij}\,F_{\underline{A}}^{\,^{+\!}%
\mathcal{O}_{l}}(\underline{\theta})F_{\underline{B}}^{\,^{-\!}\mathcal{O}%
_{l}}(\underline{\theta})\Gamma_{\underline{\alpha}}^{\underline{A}%
\underline{B}} \label{FFF}%
\end{equation}
$\allowbreak$is a candidate for an $O(4)$ form factor if $F_{\underline{A}%
}^{\,^{+\!}\mathcal{O}_{l}}$ and $F_{\underline{B}}^{\,^{-\!}\mathcal{O}_{l}}$
are $SU(2)$ form factors. The $SU(2)$ form factor equations (i) and (ii) for
$F_{\underline{A}}^{\,^{+\!}\mathcal{O}_{l}}$ and $F_{\underline{B}}%
^{\,^{-\!}\mathcal{O}_{l}}$ imply the $O(4)$ form factor equations (i) and
(ii) for $F_{\underline{\alpha}}^{\mathcal{O}}$. Moreover the double poles of
$F_{\underline{A}}^{\,^{+\!}\mathcal{O}_{l}}(\underline{\theta})F_{\underline
{B}}^{\,^{-\!}\mathcal{O}_{l}}(\underline{\theta})$ at $\theta_{ij}=i\pi$ are
made to simple poles by the $\coth\tfrac{1}{2}\theta_{ij}$. However, the
$SU(2)$ form factor equations (iii) for $F_{\underline{A}}^{\,^{+\!}%
\mathcal{O}_{l}}$ and $F_{\underline{B}}^{\,^{-\!}\mathcal{O}_{l}}$ will in
general not imply the $O(4)$ form factor equation (iii) for $F_{\underline
{\alpha}}^{\mathcal{O}}$. This problem was discussed in \cite{Sm3,BaW} and
will discussed in this paper in appendix \ref{so4} and in terms of some
examples in section \ref{s5}. We write formally%
\begin{equation}
\mathcal{O}\equiv\sum_{l}\,^{\,+\!}\mathcal{O}_{l}\,\times\,^{-\!}%
\mathcal{O}_{l}\,. \label{OOO}%
\end{equation}
This equation is to be understood as the relation (\ref{FFF}) of the form factors.

\subsection{Higher level off-shell Bethe ansatz}

\label{s2.4}

For convenience we use the variables $u$ and $v$ with $\theta=i\pi\nu
_{k}u,~z=i\pi\nu_{k}v$ and $\nu_{k}=2/(N-2k-2)$. Let $S^{(k)}(\theta)$ be the
$O(N-2k)$ S-matrix with%
\begin{align}
\tilde{S}^{(k)}(u)  &  =S^{(k)}/S_{+}^{(k)}=\tilde{b}(u)\mathbf{1}+\tilde
{c}(u)\mathbf{P}+\tilde{d}_{k}(u)\mathbf{K}\nonumber\\
\tilde{b}(u)  &  =\frac{u}{u-1},~\tilde{c}(u)=\frac{-1}{u-1},~\tilde{d}%
_{k}(u)=\frac{u}{u-1}\frac{1}{u-1/\nu_{k}} \label{Su}%
\end{align}
We define%
\begin{align}
K_{\underline{\alpha}}^{(k)}(\underline{u})  &  =\tilde{N}_{m_{k}}^{(k)}%
\int_{\mathcal{C}_{\underline{u}}^{1}}dv_{1}\cdots\int_{\mathcal{C}%
_{\underline{u}}^{m_{k}}}dv_{m_{k}}\,\tilde{h}(\underline{u},\underline
{v})p^{(k)}(\underline{u},\underline{v})\,\,\tilde{\Psi}_{\underline{\alpha}%
}^{(k)}(\underline{u},\underline{v})\label{Kk}\\
\tilde{\Psi}_{\underline{\alpha}}^{(k)}(\underline{u},\underline{v})  &
=L_{\underline{\mathring{\beta}}}^{(k)}(\underline{v})\,\big(\tilde{\Phi
}^{(k)}\big)_{\underline{\alpha}}^{\underline{\mathring{\beta}}}(\underline
{u},\underline{v}),\quad L_{\underline{\mathring{\beta}}}^{(k)}(\underline
{v})=K_{\underline{\mathring{\beta}}}^{(k+1)}(\underline{v}).\nonumber
\end{align}
with $\underline{u}=u_{1},\dots,u_{n_{k}},~\underline{v}=v_{1},\dots,v_{m_{k}%
}$ and $m_{k}=n_{k+1}$.

The equations (i)$^{(k)}$ - (iii)$^{(k)}$ for $k>0$ read in terms of these
variables as

\begin{itemize}
\item[(i)$^{(k)}$] The symmetry property under the permutation of both, the
variables $u_{i},u_{j}$ and the spaces $i,j=i+1$ at the same time
\begin{equation}
K_{\dots ij\dots}^{(k)}(\dots,u_{i},u_{j},\dots)=K_{\dots ji\dots}^{(k)}%
(\dots,u_{j},u_{i},\dots)\,\tilde{S}_{ij}^{(k)}(u_{ij}) \label{ik}%
\end{equation}
for all possible arrangements of the $u$'s.

\item[(ii)$^{(k)}$] The periodicity property under the cyclic permutation of
the rapidity variables and spaces
\begin{equation}
K_{1\ldots n_{k}}^{(k)}(u_{1}+2/\nu,u_{2},\dots,u_{n_{k}})\mathbf{C}^{\bar
{1}1}=K_{2\ldots n_{k}1}^{(k)}(u_{2},\dots,u_{n_{k}},u_{1})\mathbf{C}%
^{1\bar{1}} \label{iik}%
\end{equation}
with the charge conjugation matrix $\mathbf{C}^{\bar{1}1}$.

\item[(iii)$^{(k)}$] The function $K_{1\ldots n}^{(k)}(\underline{u})$ has a
pole at $u_{12}=1/\nu_{k}$ such that%
\begin{equation}
\operatorname*{Res}_{u_{12}=1/\nu_{k}}K_{1\dots n_{k}}^{(k)}(u_{1}%
,\dots,u_{n_{k}})=\prod_{j=3}^{n_{k}}\tilde{\psi}(u_{i1}+1)\tilde{\chi}%
(u_{i2})\mathbf{C}_{12}K_{3\dots n_{k}}^{(k)}(u_{3},\dots,u_{n_{k}})\,.
\label{iiik}%
\end{equation}

\end{itemize}

These equations are similar to the form factor equations (i) - (iii) of
(\ref{1.10}) - (\ref{1.14}) for $O(N-2k)$. However, there are two differences:

\begin{enumerate}
\item the shift in (ii)$^{(k)}$ is not the one of $O(N-2k)$ but that of $O(N)$,

\item in (iii)$^{(k)}$ there is only one term on the right hand side.
\end{enumerate}

The p-function $p^{(k)}(\underline{u},\underline{v})$ satisfies the equations
\begin{equation}
\left.
\begin{array}
[c]{rl}%
(\mathrm{i}^{\prime}): & p^{(k)}(\underline{u},\underline{v})~\text{is
symmetric under }u_{i}\leftrightarrow u_{j},~v_{i}\leftrightarrow v_{j}\\
(\mathrm{ii}^{\prime}): & p^{(k)}(\underline{u},\underline{v})=p^{(k)}%
(u_{1}+2/\nu,u_{2},\dots,\underline{v})=p^{(k)}(\underline{u},v_{1}%
+2/\nu,v_{2},\dots)\\
(\mathrm{iii}^{\prime}): & p^{(k)}(\underline{u},\underline{z})=\,p^{(k)}%
(\underline{\check{u}},\underline{\check{v}})~\text{for }u_{12}=1/\nu
_{k},~v_{1}=u_{1}-1~\text{and }v_{2}=u_{2}\,.
\end{array}
\right\}  \label{pk}%
\end{equation}
The short notations $\underline{\check{u}}=(u_{3},\dots,u_{n_{k}})$ and
$\underline{\check{v}}=(v_{3},\dots,v_{m_{k}})$ are used. Below we will
replace $p^{(k)}(\underline{u},\underline{v})$ by $1$ which will not change
the results, if the $p^{(k)}$ satisfy the conditions (\ref{pk}).

\begin{lemma}
\label{L1}The vector valued function $K_{\underline{\alpha}}^{(k)}%
(\underline{u})$ of $(\ref{Kk})$ for $0<k<\left[  \frac{1}{2}\left(
N-3\right)  \right]  $ satisfies the equations $(\mathrm{i})^{(k)}$ -
$(\mathrm{iii})^{(k)}$, if the corresponding relations are satisfied for
$K^{(k+1)}$ and the normalizations satisfy%
\begin{equation}
\tilde{N}_{m_{k}}^{(k)}=\frac{1}{\left[  \frac{1}{2}m_{k}\right]  \left[
\frac{1}{2}m_{k}+\frac{1}{2}\right]  }\frac{\Gamma^{2}\left(  1+\frac{1}{2}%
\nu\right)  }{4\pi^{2}}\frac{\tilde{\chi}(1/\nu_{k+1}-2/\nu)}{\tilde{\chi
}(1/\nu_{k}-2/\nu)}\tilde{N}_{m_{k}-2}^{(k)}\,, \label{NN}%
\end{equation}
where $\left[  x\right]  $ is the largest integer $\leq x$. The numbers
$m_{k}=n_{k+1}$ are given by the numbers of particles $n=n_{0}$ and the
weights of the operator $\mathcal{O}$%
\begin{equation}
w^{\mathcal{O}}=(w_{1},\dots,w_{\left[  N/2\right]  })=\left\{
\begin{array}
[c]{lll}%
\left(  n_{0}-n_{1},\dots,n_{\left[  N/2\right]  -1}-n_{\left[  N/2\right]
}\right)  & \text{for} & N~\text{odd}\\
\left(  n_{0}-n_{1},\dots,n_{\left[  N/2\right]  -2}-n_{-}-n_{+},n_{-}%
-n_{+}\right)  & \text{for} & N~\text{even\thinspace.}%
\end{array}
\right.  \label{w}%
\end{equation}

\end{lemma}

\noindent The proof of this\ lemma can be found in appendix \ref{se}. The
cases $k=M=\left[  \frac{1}{2}\left(  N-3\right)  \right]  $ have to be
considered separately.

\begin{lemma}
\label{L3}For $N=$ odd the level $k=M=(N-3)/2$ means an $O(3)$ problem with
$\nu_{M}=2$ and $K_{\underline{\alpha}}^{(M)}(\underline{u})$ of $(\ref{Kk})$
satisfies the equations $(\mathrm{i})^{(k)}$ - $(\mathrm{iii})^{(k)},$ if
\begin{equation}
\tilde{N}_{m_{M}}^{(M)}=\frac{1}{\left[  \frac{1}{2}m_{M}\right]  \left[
\frac{1}{2}m_{M}+\frac{1}{2}\right]  }\frac{\Gamma^{3}\left(  1+\frac{1}{4}%
\nu\right)  \Gamma\left(  1+\frac{1}{2}\nu\right)  }{2\pi^{2}\Gamma\left(
1-\frac{1}{4}\nu\right)  }\tilde{N}_{m_{M}-2}^{(M)}\,.\label{NM}%
\end{equation}
In particular for $N=3$
\begin{equation}
\tilde{N}_{m}=\frac{1}{m\left(  m-1\right)  }\frac{1}{16\pi}\tilde{N}%
_{m_{M}-2}\,.\label{N3}%
\end{equation}

\end{lemma}

\noindent The proof of this\ lemma can be found in appendix \ref{se1}. Note
that the shift in (\ref{iik}) is not that of $O(3)$, but that of $O(N)$.

For $N$ even and $k=M=(N-4)/2$ we have $\nu_{M}=1$ as for $O(4)$, however, the
shift in (\ref{iik}) is not that of $O(4)$ but that of $O(N)$. We use the
technique of subsection \ref{s2.3} and set analogously to (\ref{FFF})%
\begin{equation}
K_{\underline{\alpha}}^{(M)}(\underline{u})=d_{n_{M}}\prod_{1\leq i<j\leq
n_{M}}\sin\tfrac{1}{2}\pi\nu\left(  u_{ij}-1\right)  \,^{+}K_{\underline{A}%
}^{SU(2)}(\underline{u})\,^{-}K_{\underline{B}}^{SU(2)}(\underline{u}%
)\Gamma_{\underline{\alpha}}^{\underline{A}\underline{B}}. \label{KKK}%
\end{equation}
with%
\begin{align}
\,^{\pm}K_{\underline{A}}^{SU(2)}(\underline{u})  &  =\,^{\pm}\tilde
{N}_{m_{\pm}}\int_{\mathcal{C}_{\underline{u}}^{1}}dv_{1}\cdots\int
_{\mathcal{C}_{\underline{u}}^{m_{\pm}}}dv_{m_{\pm}}\,\tilde{h}(\underline
{u},\underline{v})p_{\pm}(\underline{u},\underline{v})\,\,\tilde{\Psi
}_{\underline{A}}^{SU(2)}(\underline{u},\underline{v})\label{KSU2}\\
\tilde{h}(\underline{u},\underline{v})  &  =\prod_{i=1}^{n}\prod_{j=1}%
^{m}\tilde{\phi}_{\nu}(u_{i}-v_{j})\prod_{1\leq i<j\leq m}\tau_{\nu}%
(v_{ij})\,\nonumber\\
\tilde{\phi}_{\nu}(u)  &  =\Gamma\left(  -\tfrac{1}{2}\nu u\right)
\Gamma\left(  1-\tfrac{1}{2}\nu+\tfrac{1}{2}\nu u\right)  ,~\tau_{\nu
}(u)=\frac{1}{\tilde{\phi}_{\nu}(u)\tilde{\phi}_{\nu}(-u)}\,. \label{phinu}%
\end{align}
The p-functions $p_{\pm}$ satisfy the conditions of e.g. \cite{BFK1}. Note
that $\tilde{\phi}_{\nu}(u)$ satisfies%
\[
\tilde{\phi}_{\nu}(u-2/\nu)=-\tilde{b}(u)\tilde{\phi}_{\nu}(u)
\]
which implies the shift relation (\ref{iik}). For $N=4$ i.e. $\nu=1$ we obtain
the $\tilde{\phi}$-function of $SU(2)$ (see e.g. \cite{BFK1}).

\begin{lemma}
\label{L4}For $N=$ even and $k=M=(N-4)/2$ the K-function of $(\ref{KKK})$
satisfies the equations $(\mathrm{i})^{(k)}$ - $(\mathrm{iii})^{(k)}$ if%
\begin{align*}
d_{n_{M}}  &  =-\frac{2}{\nu\pi^{3}}d_{n_{M}-2}\\
\,^{\pm}\tilde{N}_{m_{\pm}}  &  =\frac{1}{m_{\pm}}\frac{(-1)^{m_{\pm}}}{2\pi
i\Gamma^{2}\left(  -\tfrac{1}{2}\nu\right)  }\,^{\pm}\tilde{N}_{m_{\pm}-1}.
\end{align*}

\end{lemma}

\noindent The proof of this\ lemma can be found in appendix \ref{se2}.

\section{Examples}

\label{s5}

\subsection{Field}

\label{s5.3}

The fundamental field $\varphi^{\alpha}(x)$ in the Lagrangian (\ref{L})
transforms as the vector representation of $O(N)$ and has therefore the
weights $w=(w_{1},\dots,w_{\left[  N/2\right]  })=(1,0,\dots,0)$ (see
\cite{BFK5,BFK6}) which implies with (\ref{w}) that the numbers $n_{i}$ of
integrations in the various levels of the off-shell Bethe ansatz satisfy%
\[
\left\{
\begin{array}
[c]{lll}%
n-1=n_{1}=n_{2}=\dots=n_{\left[  N/2\right]  } & \text{for} & N~\text{odd}\\
n-1=n_{1}=n_{2}=\dots=n_{\left[  N/2\right]  -2}=n_{-}+n_{+},n_{-}=n_{+} &
\text{for} & N~\text{even\thinspace.}%
\end{array}
\right.
\]

Because the Bethe ansatz yields highest weight states we obtain the matrix
elements of the highest weight component of $\varphi^{\alpha}$ which means in
the complex basis $\alpha=1$. We use the short notation $\varphi=\varphi^{1}$
and propose for the $n$-particle form factors ($n=m+1$ odd)%
\begin{align*}
\langle0|\varphi(0)|\underline{\theta}\rangle_{\underline{\alpha}}  &
=F_{\underline{\alpha}}^{\varphi}(\underline{\theta})=\prod_{i<j}F(\theta
_{ij})K_{\underline{\alpha}}^{\varphi}(\underline{\theta})\\
K_{\underline{\alpha}}^{\varphi}(\underline{\theta})  &  =N_{n}^{\varphi}%
\int_{\mathcal{C}_{\underline{\theta}}^{o}}dz_{1}\dots\int_{\mathcal{C}%
_{\underline{\theta}}^{e}}dz_{m}\,\tilde{h}(\underline{\theta},\underline
{z})p^{\varphi}(\underline{\theta},\underline{z})\,\tilde{\Psi}_{\underline
{\alpha}}(\underline{\theta},\underline{z})
\end{align*}
with the p-function for $n=m+1=$ odd $>1$
\begin{equation}
p^{\varphi}(\underline{\theta},\underline{z})=\frac{\Bigg({\textstyle}%
\sum\limits_{\underset{j\,\text{odd}}{j=1}}^{m-1}e^{z_{j}+i\pi\nu
}+{\textstyle}\sum\limits_{\underset{j\,\text{even}}{j=2}}^{m}e^{z_{j}%
}\Bigg)\Bigg({\textstyle}\sum\limits_{\underset{j\,\text{odd}}{j=1}}%
^{m-1}e^{-z_{j}-i\pi\nu}+{\textstyle}\sum\limits_{\underset{j\,\text{even}%
}{j=2}}^{m}e^{-z_{j}}\Bigg)}{\bigg({\textstyle\sum\limits_{j=1}^{n}}%
e^{\theta_{j}}\bigg)\bigg({\textstyle\sum\limits_{j=1}^{n}}e^{-\theta_{j}%
}\bigg)-1} \label{pphi}%
\end{equation}
which satisfies (\ref{4.25}). The scalar function $\tilde{h}(\underline
{\theta},\underline{z})$ is given by (\ref{h}) and the Bethe ansatz
$\tilde{\Psi}_{\underline{\alpha}}(\underline{\theta},\underline{z})$ state by
(\ref{PSI}) and (\ref{BS}).

The one particle form factor is trivial%
\[
\langle0|\varphi(0)|\theta\rangle_{\alpha}=F_{\alpha}^{\varphi}(\theta
)=\delta_{_{\alpha}}^{1}\,.
\]
The three particle form factor is obtained by the ansatz (\ref{2.10}), the
integral representation (\ref{2.16}) and the state (\ref{PSI}) for $n=3,~m=2$%
\begin{align}
\langle0|\varphi(0)|\underline{\theta}\rangle_{\underline{\alpha}} &
=F_{\underline{\alpha}}^{\varphi}(\underline{\theta})=F(\theta_{12}%
)F(\theta_{13})F(\theta_{23})K_{\underline{\alpha}}^{\varphi}(\underline
{\theta})\label{phi3}\\
K_{\underline{\alpha}}^{\varphi}(\underline{\theta}) &  =N_{3}^{\varphi}%
\int_{\mathcal{C}_{\underline{\theta}}^{o}}dz_{1}\int_{\mathcal{C}%
_{\underline{\theta}}^{e}}dz_{2}\,\tilde{h}(\underline{\theta},\underline
{z})p^{\varphi}(\underline{\theta},\underline{z})\,\tilde{\Psi}_{\underline
{\alpha}}(\underline{\theta},\underline{z})\nonumber
\end{align}
with%
\begin{align*}
\tilde{h}(\underline{\theta},\underline{z}) &  =\prod_{i=1}^{3}\left(
\tilde{\psi}(\theta_{i}-z_{1})\tilde{\chi}(\theta_{i}-z_{2})\right)  \frac
{1}{\tilde{\chi}(z_{12})\tilde{\psi}(-z_{12})},\\
p^{\varphi}(\underline{\theta},\underline{z}) &  =\frac{\left(  e^{z_{1}%
+i\pi\nu}+e^{z_{2}}\right)  \left(  e^{-z_{1}-i\pi\nu}+e^{-z_{2}}\right)
}{\left(  e^{\theta_{1}}+e^{\theta_{2}}+e^{\theta_{3}}\right)  \left(
e^{-\theta_{1}}+e^{-\theta_{2}}+e^{-\theta_{3}}\right)  -1},\\
\tilde{\Psi}_{\underline{\alpha}}(\underline{\theta},\underline{z}) &
=L_{\underline{\mathring{\beta}}}(\underline{z})\,\left(  \Pi_{\underline
{\beta}}^{\underline{\mathring{\beta}}}(\underline{z})\Omega\tilde{T}%
_{1}^{\beta_{2}}(\underline{\theta},z_{2})\tilde{T}_{1}^{\beta_{1}}%
(\underline{\theta},z_{1})\right)  _{\underline{\alpha}}.
\end{align*}
The higher level function $L_{\mathring{\beta}_{1}\mathring{\beta}_{2}%
}(\underline{z})=\mathbf{\mathring{C}}_{\mathring{\beta}_{1}\mathring{\beta
}_{2}}L(z_{12})$ is given by the solution of lemma \ref{L1} for the $O(N-2)$
weights $w=(0,\dots,0)$. In appendix \ref{sf} we get the result%
\begin{equation}
L(z)=\frac{\Gamma\left(  \frac{1}{2}+\frac{1}{2}\nu-\frac{1}{2\pi i}z\right)
\Gamma\left(  -\frac{1}{2}+\frac{1}{2}\nu+\frac{1}{2\pi i}z\right)  }%
{\Gamma\left(  1+\frac{1}{2}\nu-\frac{1}{2\pi i}z\right)  \Gamma\left(
\frac{1}{2}\nu+\frac{1}{2\pi i}z\right)  }\,.\label{Lem}%
\end{equation}
We could not perform the integrations\footnote{Doing one integral we obtain a
generalization of Meijer's G-functions. The second integration does not yield
known functions (to our knowledge). One could, of course, apply numerical
integration techniques and one could determine the asymptotic behavior for
large $\theta^{\prime}$s which is under investigation \cite{BFK9}.} in
(\ref{phi3}) for general $N$, but we calculate the 3-particle form factor for
$O(3)$ and $O(4)$. In addition we expand the exact expression in
$1/N$-expansion to compare the result with the $1/N$-expansion of the $\sigma
$-model in terms of Feynman graphs.

\subparagraph{$O(3)$-form factors of the field $\varphi(x):$}

As explained in subsection \ref{s42} we perform the integrations in
(\ref{phi3}) for $O(3)$ by calculating a finite number of residues. We obtain
(see appendix \ref{sh})%
\[
F_{\alpha\beta\gamma}^{\varphi}(\underline{\theta})=\pi^{3}g_{\alpha
\beta\gamma}^{\varphi}(\underline{\theta})G(\theta_{12})G(\theta_{13}%
)G(\theta_{23})
\]
with%
\begin{align}
g_{\alpha\beta\gamma}^{\varphi}(\underline{\theta})  &  =\theta_{23}%
\delta_{\alpha}^{1}\mathbf{C}_{\beta\gamma}+\left(  2\pi i-\theta_{13}\right)
\delta_{\beta}^{1}\mathbf{C}_{\alpha\gamma}+\theta_{12}\delta_{\gamma}%
^{1}\mathbf{C}_{\alpha\beta}\nonumber\\
G(\theta)  &  =\frac{1}{\theta\left(  \theta-2\pi i\right)  }\tanh\tfrac{1}%
{2}\theta\,F(\theta)=\frac{\theta-i\pi}{\theta\left(  \theta-2\pi i\right)
}\tanh^{2}\tfrac{1}{2}\theta\label{G}%
\end{align}
which agrees with the result of \cite{BN} obtained by different methods.

The 5-particle form factor of the field for $O(3)$ is determined by the same
technique in appendix \ref{sh}.

\subparagraph{$O(4)$-form factors of the field $\varphi(x):$}

We apply the techniques of subsection \ref{s2.3} where the $O(4)$ form factor
is written in terms of $SU(2)$ ones as (\ref{FFF}). For details see appendix
\ref{so4}. We use the general formula (\ref{OOO}) with%
\begin{align}
^{+}\mathcal{O}_{1}  &  =\,^{-}\mathcal{O}_{2}=\,\psi_{+}^{A}\nonumber\\
^{+}\mathcal{O}_{2}  &  =\,^{-}\mathcal{O}_{1}=\,\psi_{-}^{B} \label{OOOphi}%
\end{align}
and write formally%
\begin{equation}
\varphi^{\alpha}\equiv-\tfrac{1}{2}\left(  \psi_{+}^{A}\times\psi_{-}^{B}%
+\psi_{-}^{A}\times\psi_{+}^{B}\right)  \,\Gamma_{AB}^{\alpha}%
\end{equation}
where
\[
\psi^{A}(x)=\binom{\psi_{+}^{A}(x)}{\psi_{-}^{A}(x)}%
\]
is the fundamental field of the $SU(2)$ chiral Gross-Neveu model (see
\cite{BFK1,BFK3}) with statistics factors (see appendix \ref{so4})%
\[
\sigma_{\pm}=\pm i
\]
and \textquotedblleft spin\textquotedblright\ $s=1/4$ such that
\[
\langle\,0\,|\,\psi^{A}(0)\,|\,\theta\,\rangle_{B}=\delta_{B}^{A}%
\binom{e^{-\frac{1}{4}\theta}}{e^{+\frac{1}{4}\theta}}\,.
\]
Because the Bethe ansatz yields highest weight states we obtain the matrix
elements of the field $\varphi(x)=\varphi^{1}(x)$ with%
\[
\varphi\equiv\tfrac{1}{2}\left(  \psi_{+}\times\psi_{-}+\psi_{-}\times\psi
_{+}\right)
\]
where $\psi_{\pm}=\psi_{\pm}^{\uparrow}$ are the highest weight components of
the $SU(2)$ fields. The $O(4)$-weights of $\varphi(x)$ are $w_{O(4)}^{\varphi
}=w_{O(4)}^{\psi_{+}}+w_{O(4)}^{\psi_{-}}=(1/2,1/2)+(1/2,-1/2)=(1,0)$ as we
need. This follows because $SU(2)$-weights of $\psi_{\pm}$ are $w_{SU(2)}%
^{\psi_{\pm}}=(1,0)$ which mean the $O(4)$-weights $w_{O(4)}^{\psi_{\pm}%
}=(1/2,\pm1/2)$ (for more details see section 3.1.3 of \cite{BFK6}). The
1-particle form factor is%
\begin{align*}
\langle\,0\,|\,\varphi\,|\,\theta\,\rangle_{\alpha}  &  =\tfrac{1}{2}\left(
\langle\,0\,|\times\langle\,0\,|\right)  \,\left(  \psi_{+}\times\psi_{-}%
+\psi_{-}\times\psi_{+}\right)  \,\left(  |\,\theta\,\rangle\times
|\,\theta\,\rangle\right) \\
&  =\tfrac{1}{2}\delta_{\alpha}^{1}\left(  e^{-\frac{1}{4}\theta}e^{\frac
{1}{4}\theta}+e^{\frac{1}{4}\theta}e^{-\frac{1}{4}\theta}\right)
=\delta_{\alpha}^{1}\,.
\end{align*}
The $n=3$ particle form factor (\ref{FFF}) is (up to const.)%
\[
F_{\alpha\beta\gamma}^{\varphi}(\theta_{1},\theta_{2},\theta_{3})=\prod
_{i<j}\coth\tfrac{1}{2}\theta_{ij}\left(  F_{ABC}^{\psi_{+}}(\underline
{\theta})F_{A^{\prime}B^{\prime}C^{\prime}}^{\psi_{-}}(\underline{\theta
})+F_{ABC}^{\psi_{-}}(\underline{\theta})F_{A^{\prime}B^{\prime}C^{\prime}%
}^{\psi_{+}}(\underline{\theta})\right)  \Gamma_{\alpha}^{AA^{\prime}}%
\Gamma_{\beta}^{BB^{\prime}}\Gamma_{\gamma}^{CC^{\prime}}%
\]
(see also (8.5) in \cite{BaW}). This $O(4)$ form factor satisfies the form
factor equations (i), (ii) and (iii). The three-particle $SU(2)$ form factors
$F_{\underline{A}}^{\psi_{\pm}}(\underline{\theta})$ have been discussed in
\cite{BFK1,BFK3}, they can be expressed in terms of Meijer's G-functions.

\subparagraph{1/N expansion}

For convenience we multiply the field with the Klein-Gordon operator and take
\[
{\mathcal{O}(}x{)}=i(\square+m^{2})\varphi(x)\,.
\]
We obtain (see appendix \ref{sg})%
\begin{equation}
F^{\mathcal{O}}_{\alpha\beta\gamma}(\underline{\theta})=-\frac{8\pi i}{N}%
m^{2}\,\left(  \delta_{\alpha}^{1}\mathbf{C}_{\beta\gamma}\frac{\sinh
\theta_{23}}{i\pi-\theta_{23}}+\delta_{\beta}^{1}\mathbf{C}_{\alpha\gamma
}\frac{\sinh\theta_{13}}{i\pi-\theta_{13}}+\delta_{\gamma}^{1}\mathbf{C}%
_{\alpha\beta}\frac{\sinh\theta_{12}}{i\pi-\theta_{12}}\right)  +O(N^{-2})
\label{1overN}%
\end{equation}
which agrees with the $1/N$ expansion using Feynman graphs (see appendix
\ref{s1/N}).

\subsection{Current}

\label{s5.2}

The classical Noether current (real basis)%
\[
J_{\mu}^{\alpha\beta}=\varphi^{\alpha}\partial_{\mu}\varphi^{\beta}%
-\varphi^{\beta}\partial_{\mu}\varphi^{\alpha}%
\]
transforms as the antisymmetric tensor representation of $O(N)$ and has
therefore the weights $w^{J}=(w_{1},\dots,w_{\left[  N/2\right]
})=(1,1,0,\dots,0)$ (see \cite{BFK5,BFK6}) which implies with (\ref{w}) that
the numbers $n_{i}$ of integrations in the various levels of the off-shell
Bethe ansatz satisfy%
\[
\left\{
\begin{array}
[c]{lll}%
n-2=n_{1}-1=n_{2}=\dots=n_{\left[  N/2\right]  } & \text{for} & N~\text{odd}\\
n-2=n_{1}-1=n_{2}=\dots=n_{\left[  N/2\right]  -2}=n_{-}+n_{+},n_{-}=n_{+} &
\text{for} & N~\text{even\thinspace.}%
\end{array}
\right.
\]

Because the Bethe ansatz yields highest weight states we obtain the matrix
elements of the highest weight component of $J_{\mu}^{\alpha\beta}$ which
means in the complex basis
\[
J_{\mu}=J_{\mu}^{12}=\varphi^{1}\partial_{\mu}\varphi^{2}-\varphi^{2}%
\partial_{\mu}\varphi^{1}.
\]
The conservation law $\partial_{\mu}J^{\mu}=0$ implies that there exists a
pseudo-potential $J(x)$ with%
\[
J^{\mu}(x)=\epsilon^{\mu\nu}\partial_{\nu}J(x).
\]
For the form factors we have%
\[
F_{\underline{\alpha}}^{J^{\mu}}(\underline{\theta})=-i\epsilon^{\mu\nu}%
P_{\nu}F_{\underline{\alpha}}^{J}(\underline{\theta}),~P={\textstyle\sum}%
p_{i}\,.
\]
We propose for $n=m+1$ even%
\begin{align}
\langle0|J(0)|\underline{\theta}\rangle_{\underline{\alpha}}  &
=F_{\underline{\alpha}}^{J}(\underline{\theta})=\prod_{i<j}F(\theta
_{ij})K_{\underline{\alpha}}^{J}(\underline{\theta})\nonumber\\
K_{\underline{\alpha}}^{J}(\underline{\theta})  &  =N_{n}^{J}\int
_{\mathcal{C}_{\underline{\theta}}^{o}}dz_{1}\dots\int_{\mathcal{C}%
_{\underline{\theta}}^{e}}dz_{m}\,\tilde{h}(\underline{\theta},\underline
{z})p^{J}(\underline{\theta},\underline{z})\,\tilde{\Psi}_{\underline{\alpha}%
}(\underline{\theta},\underline{z}) \label{FJ}%
\end{align}
with the p-function
\begin{equation}
p^{J}(\underline{\theta},\underline{z})=\exp\bigg({\textstyle\sum
\limits_{i=1}^{n}}\theta_{i}-{\textstyle\sum\limits_{j=1}^{m}}z_{j}-\tfrac
{1}{2}n\,i\pi\nu\bigg)/{\textstyle\sum\limits_{i=1}^{n}}e^{\theta_{i}}
\label{pJ}%
\end{equation}
which satisfies (\ref{4.25}). The scalar function $\tilde{h}(\underline
{\theta},\underline{z})$ is given by (\ref{h}) and the Bethe ansatz state
$\tilde{\Psi}_{\underline{\alpha}}(\underline{\theta},\underline{z})$ by
(\ref{PSI}) and (\ref{BS}).

We calculate the 2-particle form factor%
\begin{align}
K_{\underline{\alpha}}^{J}(\underline{\theta})  &  =N_{1}^{J}\int
_{\mathcal{C}_{\underline{\theta}}^{o}}dz\,\tilde{\psi}(\theta_{1}%
-z)\tilde{\psi}(\theta_{2}-z)\,\tilde{\Psi}_{\underline{\alpha}}%
(\underline{\theta},z)\,e^{\theta_{1}+\theta_{2}-z}/\left(  e^{\theta_{1}%
}+e^{\theta_{2}}\right) \label{J2}\\
\tilde{\Psi}_{\underline{\alpha}}(\underline{\theta},z)  &  =\tilde{b}%
(\theta_{1}-z)\tilde{c}(\theta_{2}-z)\delta_{\alpha_{1}}^{1}\delta_{\alpha
_{2}}^{2}+\tilde{c}(\theta_{1}-z)\delta_{\alpha_{1}}^{2}\delta_{\alpha_{2}%
}^{1}\,.\nonumber
\end{align}
The integration in (\ref{J2}) can be performed using the result of Example 5.3
in \cite{BFK5}. We obtain for $N>4$%
\begin{align}
F_{\underline{\alpha}}^{J}(\underline{\theta})  &  =\left(  \delta_{\alpha
_{1}}^{1}\delta_{\alpha_{2}}^{2}-\delta_{\alpha_{1}}^{2}\delta_{\alpha_{2}%
}^{1}\right)  \tanh\tfrac{1}{2}\theta_{12}F_{-}(\theta_{12})\label{FJ2}\\
F_{\underline{\alpha}}^{J_{\mu}^{\alpha\beta}}(\underline{\theta})  &
=i\left(  \delta_{\alpha_{1}}^{\alpha}\delta_{\alpha_{2}}^{\beta}%
-\delta_{\alpha_{1}}^{\beta}\delta_{\alpha_{2}}^{\alpha}\right)  \left(
p_{1}-p_{2}\right)  _{\mu}F_{-}(\theta_{12})\nonumber
\end{align}
which agrees with the result of \cite{KW} where also the $1/N$-expansion was checked.

\subparagraph{$O(3)$-form factors of the current:}

The 2-particle form factor for $O(3)$ is obtained from Example 5.4 in
\cite{BFK5} as%
\begin{equation}
F_{\underline{\alpha}}^{J}(\underline{\theta})=\left(  \delta_{\alpha_{1}}%
^{1}\delta_{\alpha_{2}}^{0}-\delta_{\alpha_{1}}^{0}\delta_{\alpha_{2}}%
^{1}\right)  \tfrac{1}{2}\pi^{2}G(\theta) \label{FJ23}%
\end{equation}
with $G(\theta)$ given by (\ref{G}).

For the 4-particle form factor for $O(3)$ we use again the techniques of
subsection \ref{s42}. Performing the integrations in (\ref{FJ}) by calculating
a finite number of residues we obtain in appendix \ref{sh} for example%
\begin{align*}
F_{0001}^{J}(\underline{\theta})  &  =\tfrac{1}{2}\pi^{5}\left(  \theta
_{12}\theta_{13}\theta_{23}+2\pi i\theta_{14}\left(  \theta_{34}-2\pi
i\right)  -2i\pi^{3}\right)  \prod_{i<j}G(\theta_{ij})\\
F_{0\bar{1}11}^{J}(\underline{\theta})  &  =\tfrac{1}{2}\pi^{5}\left(
\theta_{34}-2i\pi\right)  \left(  \theta_{23}\theta_{24}-\theta_{12}\left(
\theta_{12}-i\pi\right)  \right)  \prod_{i<j}G(\theta_{ij}).
\end{align*}
The other components are obtained by the form factor equations (i) and (ii).
These results agree with those of \cite{BN} which were obtained by different methods.

\subparagraph{$O(4)$-form factors of the current:}

The $O(4)$ form factor is written again in terms of $SU(2)$ ones as in
(\ref{FFF}). We apply again the techniques of subsection \ref{s2.3} (for
details see appendix \ref{so4}) and use the general formula (\ref{OOO}) with%
\begin{equation}
\mathcal{O}\equiv\mathcal{O}_{1}\times\mathcal{O}_{2}+\mathcal{O}_{3}%
\times\mathcal{O}_{4} \label{OOOJ}%
\end{equation}
The p-functions of these operators are proposed to be%
\begin{align*}
p^{\mathcal{O}_{1}}(\underline{\theta},\underline{z})  &  =\exp\left(
\tfrac{1}{2}{\textstyle\sum\limits_{i=1}^{n}}\theta_{i}-{\textstyle\sum
\limits_{i=1}^{m}}z_{i}\right)  /{\textstyle\sum\limits_{i=1}^{n}}\exp
\theta_{i}\,,~~m=\tfrac{1}{2}n-1\\
p^{\mathcal{O}_{2}}(\underline{\theta},\underline{z})  &  =\exp\left(
-\tfrac{1}{2}{\textstyle\sum\limits_{i=1}^{n}}\theta_{i}+{\textstyle\sum
\limits_{i=1}^{m}}z_{i}\right)  ,~~m=\tfrac{1}{2}n\\
p^{\mathcal{O}_{3}}(\underline{\theta},\underline{z})  &  =\left(
p^{\mathcal{O}_{1}}(\underline{\theta},\underline{z})\right)  ^{2}\\
p^{\mathcal{O}_{4}}(\underline{\theta},\underline{z})  &  =\left(
p^{\mathcal{O}_{2}}(\underline{\theta},\underline{z})\right)  ^{2}%
\end{align*}
where the $SU(2)$ weights are $w^{\mathcal{O}_{1}}=w^{\mathcal{O}_{3}%
}=(2,0),~w^{\mathcal{O}_{2}}=w^{\mathcal{O}_{4}}=(0,0)$, which means that the
$O(4)$ weight vector is $w^{\mathcal{O}}=(1,1)$.

With (\ref{F4F2}) we get for the K-functions%
\begin{equation}
K_{\underline{\alpha}}^{J}(\underline{\theta})=\prod_{i<j}\cosh\tfrac{1}%
{2}\theta_{ij}\left(  \,K_{\underline{A}}^{\mathcal{O}_{1}}(\underline{\theta
})K_{\underline{B}}^{\mathcal{O}_{2}}(\underline{\theta})+K_{\underline{A}%
}^{\mathcal{O}_{3}}(\underline{\theta})K_{\underline{B}}^{\mathcal{O}_{4}%
}(\underline{\theta})\right)  \Gamma_{\underline{\alpha}}^{\underline
{A}\underline{B}} \label{KKKx}%
\end{equation}
The results of \cite{BFK1,BFK3} imply (for $n=2)$
\[
K_{\alpha_{1}\alpha_{2}}^{J}(\underline{\theta})=\left(  \delta_{\alpha_{1}%
}^{1}\delta_{\alpha_{2}}^{2}-\delta_{\alpha_{1}}^{2}\delta_{\alpha_{2}}%
^{1}\right)  \frac{2}{\theta_{12}-i\pi}%
\]
which agrees with (\ref{FJ2})%
\[
F_{\underline{\alpha}}^{J_{\mu}^{\alpha\beta}}(\underline{\theta})=i\left(
\delta_{\alpha_{1}}^{\alpha}\delta_{\alpha_{2}}^{\beta}-\delta_{\alpha_{1}%
}^{\beta}\delta_{\alpha_{2}}^{\alpha}\right)  \left(  p_{1}-p_{2}\right)
_{\mu}F_{-}(\theta_{12})
\]
because of (\ref{Fmin4}).

\subsection{Energy momentum}

\label{s5.1}Because $T^{\mu\nu}$ is an $O(N)$ iso-scalar we have the weights
$w=(w_{1},\dots,w_{\left[  N/2\right]  })=(0,\dots,0)$ (see \cite{BFK5,BFK6})
which implies with (\ref{w}) that%
\[
\left\{
\begin{array}
[c]{lll}%
n=n_{1}=\dots=n_{\left[  N/2\right]  } & \text{for} & N~\text{odd}\\
n=n_{1}=\dots=n_{\left[  N/2\right]  -2}=n_{-}+n_{+},n_{-}=n_{+} & \text{for}
& N~\text{even\thinspace.}%
\end{array}
\right.
\]
Following \cite{BN} we write the energy momentum tensor in terms of an energy
momentum potential%
\begin{align*}
T^{\mu\nu}(x)  &  =R^{\mu\nu}(i\partial_{x})T(x)\\
R^{\mu\nu}(P)  &  =-P^{\mu}P^{\nu}+g^{\mu\nu}P^{2}.
\end{align*}
For the potential we propose the $n$-particle form factor as%
\begin{align}
\langle0|T(0)|\underline{\theta}\rangle_{\underline{\alpha}}  &
=F_{\underline{\alpha}}^{T}(\underline{\theta})=N_{n}^{T}\prod_{i<j}%
F(\theta_{ij})K_{\underline{\alpha}}^{T}(\underline{\theta})\nonumber\\
K_{\underline{\alpha}}^{T}(\underline{\theta})  &  =\int_{\mathcal{C}%
_{\underline{\theta}}^{1}}dz_{1}\dots\int_{\mathcal{C}_{\underline{\theta}%
}^{m}}dz_{m}\,\tilde{h}(\underline{\theta},\underline{z})p^{T}(\underline
{\theta},\underline{z})\,\tilde{\Psi}_{\underline{\alpha}}(\underline{\theta
},\underline{z}) \label{EM}%
\end{align}
with the p-function for $n=m=$ even
\begin{equation}
p^{T}(\underline{\theta},\underline{z})=1 \label{pT}%
\end{equation}
which satisfies (\ref{4.25}). The scalar function $\tilde{h}(\underline
{\theta},\underline{z})$ is given by (\ref{h}) and the Bethe ansatz
$\tilde{\Psi}_{\underline{\alpha}}(\underline{\theta},\underline{z})$ state by
(\ref{PSI}) and (\ref{BS}). The form factor of the energy momentum tensor is
then%
\begin{align}
\langle0|T^{\mu\nu}(0)|\underline{\theta}\rangle_{\underline{\alpha}}  &
=F_{\underline{\alpha}}^{T^{\mu\nu}}(\underline{\theta})=\prod_{i<j}%
F(\theta_{ij})K_{\underline{\alpha}}^{T^{\mu\nu}}(\underline{\theta
})\nonumber\\
K_{\underline{\alpha}}^{T^{\mu\nu}}(\underline{\theta})  &  =\left(  -P^{\mu
}P^{\nu}+g^{\mu\nu}P^{2}\right)  K_{\underline{\alpha}}^{T}(\underline{\theta
})\,,~P=\sum p_{i}\,. \label{EM0}%
\end{align}

We do not calculate the integrals in (\ref{EM}) for general $N$, but we derive
the 2 particle form factor following \cite{KW}. In addition we calculate
integrals explicitly for $N=3,~N=4$ and $N\rightarrow\infty$ in appendix
\ref{sf}. Using the arguments of \cite{KW} we write
\begin{align}
F_{\alpha_{1}\alpha_{2}}^{T^{\mu\nu}}(\underline{\theta})  &  =\left(
-p_{1}^{\mu}p_{2}^{\nu}-p_{2}^{\mu}p_{1}^{\nu}+g^{\mu\nu}\left(  p_{1}%
p_{2}+m^{2}\right)  \right)  \mathbf{C}_{\alpha_{1}\alpha_{2}}F_{0}%
(\theta_{12})\label{EMN}\\
F_{\alpha_{1}\alpha_{2}}^{T}(\underline{\theta})  &  =\frac{1}{2\cosh^{2}%
\frac{1}{2}\theta_{12}}\mathbf{C}_{\alpha_{1}\alpha_{2}}F_{0}(\theta
_{12})\nonumber
\end{align}
where $F_{0}(\theta)$ is the minimal form factor (\ref{minF0}) in the scalar
channel belonging to the S-matrix eigenvalue $S_{0}\left(  \theta\right)  $.
The normalization means that the energy momentum operator satisfies for a one
particle state the eigenvalue equation
\[
P^{\mu}\,|\,\theta\,\rangle_{\alpha}=\int dx^{1}T^{\mu0}(x)\,|\,\theta
\,\rangle_{\alpha}=|\,\theta\,\rangle_{\alpha}\,p^{\mu}(\theta).
\]
Using (\ref{minF0}) for general $N$ we obtain explicitly (with $\theta
=\theta_{12}$)%
\begin{equation}
F_{\underline{\alpha}}^{T}(\underline{\theta})=-\mathbf{C}_{\alpha_{1}%
\alpha_{2}}\frac{\left(  \Gamma\left(  \frac{1}{2}+\frac{1}{2}\nu\right)
\right)  ^{2}}{2\pi^{2}}\frac{\Gamma\left(  -\frac{1}{2}+\frac{1}{2}%
\frac{\theta}{i\pi}\right)  \Gamma\left(  \frac{1}{2}-\frac{1}{2}\frac{\theta
}{i\pi}\right)  }{\Gamma\left(  1+\frac{1}{2}\nu-\frac{1}{2}\frac{\theta}%
{i\pi}\right)  \Gamma\left(  \frac{1}{2}\nu+\frac{1}{2}\frac{\theta}{i\pi
}\right)  }F\left(  \theta\right)  \,. \label{EM2}%
\end{equation}

\subparagraph{$O(3)$-form factors of energy momentum:}

Again we perform the integrations in (\ref{EM}) for $O(3)$ by calculating a
finite number of residues.

It turns out that the leading term in the limit $\nu\rightarrow2$ (i.e.
$N\rightarrow3$) vanishes and we have to calculate the contribution of order
$(\nu-2)$. For the 2-particle form factor we obtain%
\[
F_{\underline{\alpha}}^{T}(\underline{\theta})=-\mathbf{C}_{\alpha_{1}%
\alpha_{2}}\tfrac{1}{2}\pi^{2}\frac{1}{\theta_{12}-i\pi}G(\theta_{12})
\]
which agrees with (\ref{EM2}) for $N=3$.

For the 4-particle form factor for $O(3)$ we obtain in appendix \ref{sh} for
example the component%
\[
F_{\bar{1}\bar{1}11}^{T}(\underline{\theta})=\tfrac{1}{2}\pi^{5}\left(
\theta_{12}-2\pi i\right)  \left(  \theta_{34}-2\pi i\right)  \prod
_{i<j}G(\theta_{ij})\,.
\]
The other components are obtained by the form factor equations (i) and (ii).
These results agree again with those of \cite{BN} which were obtained by
different methods.

\subparagraph{$O(4)$-form factors of energy momentum:}

Applying the results of Appendix \ref{sf} we obtain from the general formula
(\ref{FFF})%
\[
F_{\alpha_{1}\alpha_{2}}^{T}(\underline{\theta})=-2\mathbf{C}_{\alpha
_{1}\alpha_{2}}\left(  \frac{1}{\theta_{12}-i\pi}\right)  ^{2}F\left(
\theta_{12}\right)
\]
which agrees (\ref{EMN}) for $N=4$.

\subparagraph{1/N expansion:}

In appendix \ref{sf} we also calculate the integrals in (\ref{EM}) explicitly
for $n=2$ and $N\rightarrow\infty$ and find
\[
F_{\alpha_{1}\alpha_{2}}^{T}(\underline{\theta})=-\mathbf{C}_{\alpha_{1}%
\alpha_{2}}\frac{1}{\theta_{12}-i\pi}\tanh\tfrac{1}{2}\theta_{12}+O(1/N)
\]
which agrees with (\ref{EMN}) for $N\rightarrow\infty$. This result agrees
also with the one obtained by calculating Feynman graphs. This calculation is
similar to that, which was done in \cite{KW} for the $O(N)$ Gross-Neveu model
up to $O(1/N^{2})$. Note that the leading term for $N\rightarrow\infty$ is not
the free value.

\subsection*{Conclusions:}

\addcontentsline{toc}{section}{Conclusions}

In this paper the general form factor formula for the $O(N)$-sigma model is
constructed. As an application, the general $O(N)$ form factors for the field,
the current and the energy momentum operators are presented in terms of
integral representations. The large $N$ limits of these form factors are
compared with the $1/N$-expansion of the $O(N)$-sigma model in terms of
Feynman graphs and full agreement is found. Using these general results some
examples of $O(3)$ and $O(4)$ form factors for low particle numbers are
computed explicitly and agreement is found with previous results \cite{BN}
obtained by different methods. We believe that our results may be relevant to
understand the behavior of correlation functions in theories with
asymptotically freedom like $4D~QCD$.

\paragraph{Acknowledgment:}

The authors have profited from discussions with A. Fring, R. Schra\-der, F.
Smirnov and A. Belavin. H.B. thanks to A. Tsvelik and P. Wiegmann for valuable
discussions. M.K. thanks J. Balog and P. Weisz for discussions and hospitality
at the Max-Planck Institut f\"{u}r Physik (M\"{u}nchen), where parts of this
work have been performed. H.B. and M.K. were supported by Humboldt Foundation.
H.B. is also supported by the Armenian grant 11-1 c028 and Armenian-Russian
grant-AR-17. H.B. thanks the High Energy section at ICTP for hospitality in
the September of 2012. A.F. acknowledges financial support from CAPES -
Coordena\c{c}\~{a}o de Aperfei\c{c}oamento de Pessoal de Nivel Superior (Proc.
10126-12-0 ).

\appendix

\section*{Appendices}

\addcontentsline{toc}{part}{Appendices}

\renewcommand{\theequation}{\mbox{\Alph{section}.\arabic{equation}}} \setcounter{equation}{0}

\section{Proof of the main theorem
\protect\ref{TN}%
}

\label{sd}

The co-vector valued function $K_{\underline{\alpha}}^{\mathcal{O}}%
(\underline{\theta})$ given by the integral representation (\ref{2.16}) can be
written as a sum of \textquotedblleft Jackson-type Integrals" as investigated
in \cite{BFK5} because of the identity%
\begin{equation}
\int_{\mathcal{C}_{a}}dz\Gamma(a-z)f(z)=2\pi i\,\operatorname*{Res}_{z=a}%
\sum_{l=-\infty}^{\infty}\Gamma(a-z-l)f(z+l) \label{Int}%
\end{equation}
where the $\mathcal{C}_{a}$ encircles the poles of $\Gamma(a-z)$
anti-clockwise. For these expressions symmetry properties and matrix
difference equation have been proved in \cite{BFK5} which imply the form
factor equations (i) and (ii). Therefore we only have to prove, that the
assumptions of theorem \ref{TN} picks those solutions of (i) and (ii), which
in addition satisfy the residue relation (iii)%
\[
\operatorname*{Res}_{\theta_{12}=i\pi}F_{1\dots n}^{\mathcal{O}}(\theta
_{1},\dots,\theta_{n})=2i\,\mathbf{C}_{12}\,F_{3\dots n}^{\mathcal{O}}%
(\theta_{3},\dots,\theta_{n})\left(  \mathbf{1}-S_{2n}\dots S_{23}\right)
\,.
\]

\begin{proof}
The K-function $K_{1\dots n}^{\mathcal{O}}(\underline{\theta})$ defined by
(\ref{2.10}) contains the entire pole structure and is determined by the form
factor equations (i) - (iii) which read in terms of $K_{1\dots n}%
^{\mathcal{O}}(\underline{\theta})$ as%
\begin{gather}
K_{\dots ij\dots}^{\mathcal{O}}(\dots,\theta_{i},\theta_{j},\dots)=K_{\dots
ji\dots}^{\mathcal{O}}(\dots,\theta_{j},\theta_{i},\dots)\,\tilde{S}%
_{ij}(\theta_{ij})\label{2.12}\\
K_{1\ldots n}^{\mathcal{O}}(\theta_{1}+2\pi i,\theta_{2},\dots,\theta
_{n})\mathbf{C}^{\bar{1}1}=K_{2\ldots n1}^{\mathcal{O}}(\theta_{2}%
,\dots,\theta_{n},\theta_{1})\mathbf{C}^{1\bar{1}}\label{2.14}\\
\operatorname*{Res}_{\theta_{12}=i\pi}K_{1\dots n}(\underline{\theta}%
)=\frac{2i\,}{F(i\pi)}\mathbf{C}_{12}\prod_{i=3}^{n}\tilde{\psi}(\theta
_{i1}+i\pi\nu)\tilde{\chi}(\theta_{i2})K_{3\dots n}(\theta_{3},\dots
,\theta_{n})\left(  \mathbf{1}-S_{2n}\dots S_{23}\right)  \label{2.15}%
\end{gather}
where (\ref{FF}) has been used. The residue of $K_{1\dots n}(\underline
{\theta})$ consists of two terms
\[
\operatorname*{Res}_{\theta_{12}=i\pi}K_{1\dots n}(\underline{\theta
})=\bigg(\operatorname*{Res}_{\theta_{12}=i\pi}^{(1)}+\operatorname*{Res}%
_{\theta_{12}=i\pi}^{(2)}\bigg)K_{1\dots n}(\underline{\theta})\,.
\]
This is because for each $z_{j}$ integration with $j$ even the contours will
be \textquotedblleft pinched\textquotedblright\ at two points (see Fig.
\ref{f5.2}):

\begin{itemize}
\item[(1)] $z_{j}=\theta_{2}\approx\theta_{1}-i\pi$

\item[(2)] $z_{j}=\theta_{1}-2\pi i\approx\theta_{2}-i\pi$
\end{itemize}

In appendix \ref{se} we prove for general level $k$ of the off-shell Bethe
ansatz the residue formulas. The general result imply for $k=0$ that the
contribution from the pinching (1) gives%
\[
\operatorname*{Res}_{\theta_{12}=i\pi}^{(1)}K_{1\dots n}(\underline{\theta
})=\frac{2i\,}{F(i\pi)}\mathbf{C}_{12}\prod_{i=3}^{n}\tilde{\psi}(\theta
_{i1}+i\pi\nu)\tilde{\chi}(\theta_{i2})K_{3\dots n}(\theta_{3},\dots
,\theta_{n})
\]
if the normalization relation (\ref{norm}) holds. Therefore we have proved%
\[
\operatorname*{Res}_{\theta_{12}=+i\pi}^{(1)}F_{1\dots n}(\theta_{1}%
,\dots,\theta_{n})=2i\,\mathbf{C}_{12}\,F_{3\dots n}(\theta_{3},\dots
,\theta_{n})\,.
\]
To investigate $\operatorname*{Res}\limits_{\theta_{12}=i\pi}^{(2)}F_{1\dots
n}(\underline{\theta})\,$due to the pinching at $z_{j}=\theta_{1}-2\pi
i\approx\theta_{2}-i\pi$ we use (ii) and (i) to write%
\begin{align*}
F_{1\ldots n}(\underline{\theta}) &  =\mathbf{C}_{1\bar{1}}F_{2\ldots
n1}(\theta_{2},\dots,\theta_{n},\theta_{1}-2\pi i)\mathbf{C}^{1\bar{1}}\\
&  =\mathbf{C}_{1\bar{1}}F_{21\ldots n}(\theta_{2},\theta_{1}-2\pi
i,\dots,\theta_{n})\mathbf{C}^{1\bar{1}}S_{\bar{1}n}\dots S_{\bar{1}3}%
\end{align*}
We use the result for $\operatorname*{Res}\limits_{\theta_{1}=\theta_{2}+i\pi
}^{(1)}$ and obtain%
\begin{align*}
\operatorname*{Res}_{\theta_{1}=\theta_{2}+i\pi}^{(2)}F_{1\ldots n}%
(\underline{\theta}) &  =-\operatorname*{Res}_{\theta_{2}=\left(  \theta
_{1}-2\pi i\right)  +i\pi}^{(1)}\mathbf{C}_{1\bar{1}}F_{21\ldots n}(\theta
_{2},\theta_{1}-2\pi i,\dots,\theta_{n})\mathbf{C}^{1\bar{1}}S_{\bar{1}n}\dots
S_{\bar{1}3}\\
&  =-\mathbf{C}_{1\bar{1}}2i\,\mathbf{C}_{21}\,F_{3\dots n}^{\mathcal{O}%
}(\theta_{3},\dots,\theta_{n})\mathbf{C}^{1\bar{1}}S_{\bar{1}n}\dots
S_{\bar{1}3}\\
&  =-2i\mathbf{C}_{12}\,\,F_{3\dots n}^{\mathcal{O}}(\theta_{3},\dots
,\theta_{n})S_{2n}\dots S_{23}\,.
\end{align*}

\end{proof}

\section{O(4) solutions of (iii)}

\label{so4}

In order to get solutions of the form factor equation (iii) in the form of
(\ref{FFF})%
\begin{equation}
F_{\underline{\alpha}}^{\mathcal{O}}(\underline{\theta})=c_{n}\sum_{l}%
\prod_{i<j}\coth\tfrac{1}{2}\theta_{ij}\,F_{\underline{A}}^{\,^{+\!}%
\mathcal{O}_{l}}(\underline{\theta})F_{\underline{B}}^{\,^{-\!}\mathcal{O}%
_{l}}(\underline{\theta})\Gamma_{\underline{\alpha}}^{\underline{A}%
\underline{B}}%
\end{equation}
we need that
\[
\operatorname*{Res}\limits_{\theta_{12}=i\pi}F_{\underline{\alpha}%
}^{\mathcal{O}}(\underline{\theta})=2i\,F_{\underline{\check{\alpha}}%
}^{\mathcal{O}}(\underline{\check{\theta}})\left(  \mathbf{C}_{\alpha
_{1}\alpha_{2}}\mathbf{1}_{\underline{\check{\alpha}}}^{\underline
{\check{\alpha}}^{\prime}}-\mathbf{C}_{\alpha_{1}\alpha_{2}^{\prime}}\left(
S_{2n}^{O(4)}\dots S_{23}^{O(4)}\right)  _{\alpha_{2}\underline{\check{\alpha
}}}^{\underline{\check{\alpha}^{\prime}}\alpha_{2}^{\prime}}\right)
\]
with $\underline{\check{\theta}}=(\theta_{3},\dots,\theta_{n})$ etc. The fact
that fields of the chiral Gross-Neveu model posses a generalized statistics
implies that the form factor equations contain statistics factors $\sigma$
(see \cite{BFK1,BFK3}). The residue equation (iii) for $SU(2)$ form factors
reads as%
\begin{equation}
\operatorname*{Res}_{\theta_{12}=i\pi}F_{1\dots n}^{\mathcal{O}}(\theta
_{1},\dots,\theta_{n})=2i\,\mathbf{C}_{12}\,F_{3\dots n}^{\mathcal{O}}%
(\theta_{3},\dots,\theta_{n})\left(  \mathbf{1}-\sigma^{\mathcal{O}}\rho
S_{2n}\dots S_{23}\right)  \,. \label{1.14ax}%
\end{equation}
where $S$ is the $SU(2)$ S-matrix. Here $\sigma^{\mathcal{O}}$ is the
statistics factor of the operator $\mathcal{O}(x)$ with respect to the
particle $2$ and $\rho$ is a sign factor due to the unusual crossing relation
of the S-matrix, in \cite{BFK1,BFK3} was shown that%
\[
\sigma^{\mathcal{O}}=e^{i\pi\frac{1}{2}Q^{\mathcal{O}}},~\rho=(-1)^{1+\frac
{1}{2}\left(  n-Q^{\mathcal{O}}\right)  }=\pm1
\]
where $Q^{\mathcal{O}}=n\operatorname{mod}2$ is the charge of the operator
$\mathcal{O}$. By charge conjugation we have for the conjugate operator
$\sigma^{\mathcal{\bar{O}}}=e^{-i\pi\frac{1}{2}Q^{\mathcal{O}}}$. The
condition $(\mathrm{ii}_{1}^{\prime})$ in \cite{BFK1} for the $SU(2)$
p-function also contains the statistics and extra sign factors
\[
p^{\mathcal{O}}(\underline{\theta},\underline{z})=\sigma^{\mathcal{O}}%
\rho(-1)^{m+1}p(\theta_{1}+2\pi i,\theta_{2},\dots,\underline{z})\,.
\]

We calculate (for simplicity we skip here all constants, questions of
normalization will be considered in appendix \ref{se2})%

\begin{multline*}
\operatorname*{Res}\limits_{\theta_{12}=i\pi}F_{\underline{\alpha}%
}^{\mathcal{O}}(\underline{\theta})\\
=\prod_{2<i<j}\coth\tfrac{1}{2}\theta_{ij}\sum_{l}\,F_{\underline{\check{A}%
}^{\prime}}^{\,^{+\!}\mathcal{O}_{l}}(\underline{\check{\theta}}%
)F_{\underline{\check{B}}^{\prime}}^{\,^{-\!}\mathcal{O}_{l}}(\underline
{\check{\theta}})\left(  \mathbf{C}_{A_{1}A_{2}}\mathbf{1}_{\underline
{\check{A}}}^{\underline{\check{A}}^{\prime}}-\mathbf{C}_{A_{1}A_{2}^{\prime}%
}\,^{+\!}\sigma_{l}\rho\left(  S_{2n}\dots S_{23}\right)  _{A_{2}%
\underline{\check{A}}}^{\underline{\check{A}}^{\prime}A_{2}^{\prime}}\right)
\\
\times\left(  \mathbf{C}_{B_{1}B_{2}}\mathbf{1}_{\underline{\check{B}}%
}^{\underline{\check{B}}^{\prime}}-\mathbf{C}_{B_{1}B_{2}^{\prime}}%
\,^{-\!}\sigma_{l}\rho\left(  S_{2n}\dots S_{23}\right)  _{B_{2}%
\underline{\check{B}}}^{\underline{\check{B}}^{\prime}B_{2}^{\prime}}\right)
\Gamma_{\underline{\alpha}}^{\underline{A}\underline{B}}\\
=F_{\underline{\check{\alpha}}}^{\mathcal{O}}(\underline{\check{\theta}%
})\left(  \mathbf{C}_{\alpha_{1}\alpha_{2}}\mathbf{1}_{\underline
{\check{\alpha}}}^{\underline{\check{\alpha}}^{\prime}}-\mathbf{C}_{\alpha
_{1}\alpha_{2}^{\prime}}\left(  S_{2n}^{O(4)}\dots S_{23}^{O(4)}\right)
_{\alpha_{2}\underline{\check{\alpha}}}^{\underline{\check{\alpha}}^{\prime
}\alpha_{2}^{\prime}}\right)
\end{multline*}
if%
\begin{multline}
\sum_{l}\,F_{\underline{\check{A}}^{\prime}}^{\,^{+\!}\mathcal{O}_{l}%
}(\underline{\check{\theta}})F_{\underline{\check{B}}^{\prime}}^{\,^{-\!}%
\mathcal{O}_{l}}(\underline{\check{\theta}})\left(  \mathbf{C}_{A_{1}A_{2}%
}\mathbf{1}_{\underline{\check{A}}}^{\underline{\check{A}}^{\prime}}%
\mathbf{C}_{B_{1}B_{2}^{\prime}}\,^{-\!}\sigma_{l}\rho\left(  S_{2n}\dots
S_{23}\right)  _{B_{2}\underline{\check{B}}}^{\underline{\check{B}}^{\prime
}B_{2}^{\prime}}\right. \\
+\left.  \mathbf{C}_{A_{1}A_{2}^{\prime}}\,^{+\!}\sigma_{l}\rho\left(
S_{2n}\dots S_{23}\right)  _{A_{2}\underline{\check{A}}}^{\underline{\check
{A}}^{\prime}A_{2}^{\prime}}\mathbf{C}_{B_{1}B_{2}}\mathbf{1}_{\underline
{\check{B}}}^{\underline{\check{B}}^{\prime}}\right)  \Gamma_{\underline
{\alpha}}^{\underline{A}\underline{B}}=0 \label{if}%
\end{multline}
and
\[
\,^{+\!}\sigma_{l}\,^{-\!}\sigma_{l}=(-1)^{n-1}\,.
\]
It was used that%
\[
\mathbf{C}_{A_{1}A_{2}}\mathbf{C}_{B_{1}B_{2}}\Gamma_{\alpha_{1}}^{A_{1}B_{1}%
}\Gamma_{\alpha_{2}}^{A_{2}B_{2}}=-\mathbf{C}_{\alpha_{1}\alpha_{2}}%
\]
and
\begin{align*}
&  \,^{+\!}\sigma_{l}\,^{-\!}\sigma_{l}\rho^{2}\mathbf{C}_{A_{1}A_{2}^{\prime
}}\mathbf{C}_{B_{1}B_{2}^{\prime}}\Gamma_{\alpha_{1}}^{A_{1}B_{1}}\left(
S_{2n}\dots S_{23}\right)  _{A_{2}\underline{\check{A}}}^{\underline{\check
{A}}^{\prime}A_{2}^{\prime}}\left(  S_{2n}\dots S_{23}\right)  _{B_{2}%
\underline{\check{B}}}^{\underline{\check{B}}^{\prime}B_{2}^{\prime}}%
\Gamma_{\alpha_{2}}^{A_{2}B_{2}}\Gamma_{\underline{\hat{\alpha}}}%
^{\underline{\check{A}}\underline{\check{B}}}\\
&  =\,^{+\!}\sigma_{l}\,^{-\!}\sigma_{l}(-1)^{n-2}\mathbf{C}_{A_{1}%
A_{2}^{\prime}}\mathbf{C}_{B_{1}B_{2}^{\prime}}\Gamma_{\alpha_{1}}^{A_{1}%
B_{1}}\Gamma_{\alpha_{2}^{\prime}}^{A_{2}^{\prime}B_{2}^{\prime}}%
\Gamma_{\underline{\hat{\alpha}}^{\prime}}^{\underline{\hat{A}}^{\prime
}\underline{\hat{B}}^{\prime}}\left(  S_{2n}^{O(4)}\dots S_{23}^{O(4)}\right)
_{\alpha_{2}\underline{\check{\alpha}}}^{\underline{\check{\alpha}^{\prime}%
}\alpha_{2}^{\prime}}\\
&  =\,^{+\!}\sigma_{l}\,^{-\!}\sigma_{l}(-1)^{n-1}\mathbf{C}_{\alpha_{1}%
\alpha_{2}^{\prime}}\left(  S_{2n}^{O(4)}\dots S_{23}^{O(4)}\right)
_{\alpha_{2}\underline{\check{\alpha}}}^{\underline{\check{\alpha}^{\prime}%
}\alpha_{2}^{\prime}}=\mathbf{C}_{\alpha_{1}\alpha_{2}^{\prime}}\left(
S_{2n}^{O(4)}\dots S_{23}^{O(4)}\right)  _{\alpha_{2}\underline{\check{\alpha
}}}^{\underline{\check{\alpha}^{\prime}}\alpha_{2}^{\prime}}%
\end{align*}
where (\ref{SSS}) has been used. The condition (\ref{if}) holds if%
\begin{equation}
\sum_{l}\,^{\pm\!}\sigma_{l}\,F_{\underline{A}}^{\,^{+\!}\mathcal{O}_{l}%
}(\underline{\theta})\,F_{\underline{B}}^{\,^{-\!}\mathcal{O}_{l}}%
(\underline{\theta})\Gamma_{\underline{\alpha}}^{\underline{A}\underline{B}%
}=0\,. \label{if1}%
\end{equation}

\paragraph{Scalar operators:}

Let us first consider an $O(4)$ iso scalar operator $\mathcal{O}$, then $n$ is
even and $F_{\underline{\alpha}}^{\mathcal{O}}(\underline{\theta})$ is of the
form%
\[
F_{\underline{\alpha}}^{\mathcal{O}}(\underline{\theta})=\sum_{\pi\in
S_{n}^{\prime}}f_{\pi}(\underline{\theta})\mathbf{C}_{\pi\alpha_{1}\pi
\alpha_{2}}\dots\mathbf{C}_{\pi\alpha_{n-1}\pi\alpha_{n}}%
\]
where the set $S_{n}^{\prime}$ contains all $n!/(2^{n/2}(n/2)!)$ permutations
of $\{1,\dots,n\}$ with the restrictions%
\begin{align*}
\pi\alpha_{1}  &  <\pi\alpha_{2},\ldots,\pi\alpha_{n-1}<\pi\alpha_{n}\\
\pi\alpha_{1}  &  <\pi\alpha_{3}<\ldots<\pi\alpha_{n-1}\,.
\end{align*}
Obviously, if the special components $F_{\underline{\alpha}}^{\mathcal{O}%
}(\underline{\theta})$ with $\alpha_{i}\in\{1,\bar{1}\}$ vanish, then all
$f_{\pi}(\underline{\theta})$ vanish and $F_{\underline{\alpha}}^{\mathcal{O}%
}(\underline{\theta})=0$ for all $\underline{\alpha}$.

For the case of two term of the $l$-sum in (\ref{if1}) and
\begin{align*}
^{+\!}\mathcal{O}_{1}  &  =\,^{-\!}\mathcal{O}_{2}=\,\mathcal{O}_{1}\\
^{+\!}\mathcal{O}_{2}  &  =\,^{-\!}\mathcal{O}_{1}=\,\mathcal{O}_{2}\\
\sigma_{1}  &  =-\sigma_{2}\,,~\sigma_{1}\sigma_{2}=-1\Rightarrow\sigma
_{1}=\pm1
\end{align*}
we have in (\ref{if1})%
\[
\sum_{l}\,^{\,\pm}\sigma_{l}\,F_{\underline{A}}^{\,^{+}\mathcal{O}_{l}%
}(\underline{\theta})\,F_{\underline{B}}^{\,^{-}\mathcal{O}_{l}}%
(\underline{\theta})\Gamma_{\underline{\alpha}}^{\underline{A}\underline{B}%
}=\sigma_{1}\left(  F_{\underline{A}}^{\,\mathcal{O}_{1}}(\underline{\theta
})\,F_{\underline{B}}^{\,\mathcal{O}_{2}}(\underline{\theta})-F_{\underline
{A}}^{\,\mathcal{O}_{2}}(\underline{\theta})\,F_{\underline{B}}^{\,\mathcal{O}%
_{1}}(\underline{\theta})\right)  \Gamma_{\underline{\alpha}}^{\underline
{A}\underline{B}}=0
\]
because for $\alpha_{i}\in\{1,\bar{1}\}$ the symmetry $\Gamma_{\underline
{\alpha}}^{\underline{A}\underline{B}}=\Gamma_{\underline{\alpha}}%
^{\underline{B}\underline{A}}$ holds (see (\ref{inter})). As an example of
this construction see that for the energy momentum in subsection \ref{s5.1}
and appendix \ref{sf}.

\paragraph{Vector operators:}

For the highest weight component of an iso vector $O(4)$ operator the number
$n$ is odd and the form factors are of the form%
\[
F_{\underline{\alpha}}^{\mathcal{O}}(\underline{\theta})=\sum_{\pi\in
S_{n}^{\prime}}f_{\pi}(\underline{\theta})\mathbf{C}_{\pi\alpha_{1}\pi
\alpha_{2}}\dots\mathbf{C}_{\pi\alpha_{n-2}\pi\alpha_{n-1}}\delta_{\pi
\alpha_{n}}^{1}%
\]
with the restrictions%
\begin{align*}
\pi\alpha_{1}  &  <\pi\alpha_{2},\ldots,\pi\alpha_{n-2}<\pi\alpha_{n-1}\\
\pi\alpha_{1}  &  <\pi\alpha_{3}<\ldots<\pi\alpha_{n-2}\,.
\end{align*}
As above, if the special components $F_{\underline{\alpha}}^{\mathcal{O}%
}(\underline{\theta})$ with $\alpha_{i}\in\{1,\bar{1}\}$ vanish, then all
$f_{\pi}(\underline{\theta})$ vanish and $F_{\underline{\alpha}}^{\mathcal{O}%
}(\underline{\theta})=0$ for all $\underline{\alpha}$. For the case of two
term of the $l$-sum in (\ref{if1}) and
\begin{align*}
^{+\!}\mathcal{O}_{1}  &  =\,^{-\!}\mathcal{O}_{2}=\,\mathcal{O}_{1}\\
^{+\!}\mathcal{O}_{2}  &  =\,^{-\!}\mathcal{O}_{1}=\,\mathcal{O}_{2}\\
\sigma_{1}  &  =-\sigma_{2}\,,~\sigma_{1}\sigma_{2}=1\Rightarrow\sigma_{1}=\pm
i
\end{align*}
again (\ref{if1}) holds. As an example of this construction see that for field
in subsection \ref{s5.3}.

\paragraph{Anti-symmetric tensor operators:}

The construction is similar as above, however, one needs 4 operators as for
the current in subsection \ref{s5.2}.

\paragraph{The highest level off-shell Bethe ansatz for even $N$:}

For these constructions one has to apply a modification of the $O(4)$
construction above because the shift in equation (\ref{iik}) is not that of
$O(4)$ but that of $O(N)$ (see appendix \ref{se2}).

\section{Higher level K-functions}

\subsection{Proof of lemma
\protect\ref{L1}%
}

\label{se}

Lemma \ref{L1} also holds for $k=0$ in (\ref{iiik}), if $\operatorname*{Res}$
is replaced by $\overset{(1)}{\operatorname*{Res}}$ as explained in appendix
\ref{sd}. For convenience we use here the variables $u$ and $v$ with
$\theta=i\pi\nu_{k}u,~z=i\pi\nu_{k}v$ and $\nu_{k}=2/(N-2k-2)$ (for the
S-matrix see (\ref{Su}).

\begin{proof}
The relations (i)$^{(k)}$ and (ii)$^{(k)}$ follow as above in the proof of
theorem \ref{TN} from the results of \cite{BFK5}. To prove (iii)$^{(k)}$ we
calculate%
\begin{equation}
\operatorname*{Res}_{u_{12}=1/\nu_{k}}K_{1\dots n_{k}}^{(k)}(\underline
{u})=\operatorname*{Res}_{u_{12}=1/\nu_{k}}\tilde{N}_{m_{k}}^{(k)}%
\int_{\mathcal{C}_{\underline{u}}^{1}}dv_{1}\cdots\int_{\mathcal{C}%
_{\underline{u}}^{m_{k}}}dv_{m_{k}}\,\tilde{h}(\underline{u},\underline
{v})\,\tilde{\Psi}_{1\dots n_{k}}^{(k)}(\underline{u},\underline{v})\,.
\label{RKk}%
\end{equation}
For $j$ even contours will be \textquotedblleft pinched\textquotedblright\ at
$v_{j}=u_{2}\approx u_{1}-1/\nu_{k}$.\footnote{For $k=0$ there is a second
pinching point at $v_{j}=u_{1}-2/\nu_{k}\approx u_{2}-1/\nu_{k}$ as explained
in appendix \ref{sd}.} Due to symmetry it is sufficient to determine the
contribution from one of the $v_{j}$ and multiply the result by $\left[
\frac{1}{2}m_{k}\right]  $. We take for convenience $v_{j}=v_{2}$, then the
contribution is given by the $v_{2}$ integration $\oint_{u_{2}}dv_{2}\cdots$
along small circle around $v_{2}=u_{2}$ (see figure \ref{f5.1}). The S-matrix
$\tilde{S}(u_{2}-v_{2})$ yields the permutation operator $\tilde
{S}(0)=\mathbf{P}$ and the S-matrix $\tilde{S}(u_{1}-v_{2})$ yields
$\mathbf{K}$ after taking $\operatorname*{Res}_{u_{12}=1/\nu_{k}}\tilde
{S}(u_{12})$. In the representation of the Bethe state (\ref{BS})
\[
\left(  \tilde{\Phi}^{(k)}\right)  _{\underline{\alpha}}^{\underline
{\mathring{\beta}}}(\underline{u},\underline{v})=\left(  \left(  \Pi
^{(k)}\right)  _{\underline{\beta}}^{\underline{\mathring{\beta}}}%
(\underline{v})\Omega_{k}\left(  \tilde{T}^{(k)}\right)  _{k+1}^{\beta_{m_{k}%
}}(\underline{u},v_{m_{k}})\dots\left(  \tilde{T}^{(k)}\right)  _{k+1}%
^{\beta_{1}}(\underline{u},v_{1})\right)  _{\underline{\alpha}}%
\]
we may move for generic values of the other $v_{j}$ the operator $\left(
\tilde{T}^{(k)}\right)  _{k+1}^{\beta_{2}}(\underline{u},v_{2})$ to the left
by means of the $TTS=STT$ commutation rule (14) of \cite{BFK5} and (\ref{PiS})
(using the short notation $\tilde{T}_{k}(v_{i})=\tilde{T}_{k}(\underline
{u},v_{i})$)%
\[
\Pi_{k}\Omega_{k}\tilde{T}_{k}(v_{m})\dots\tilde{T}_{k}(v_{2})\tilde{T}%
_{k}(v_{1})=\tilde{S}_{k+1}(v_{32})\dots\tilde{S}_{k+1}(v_{m2})\Pi_{k}%
\Omega_{k}\tilde{T}_{k}(v_{2})\tilde{T}_{k}(v_{m})\dots\tilde{T}_{k}(v_{1}).
\]
Because of (\ref{Pi}) $\left(  \Pi^{(k)}\right)  _{\dots\overline{k+1}}%
^{\dots\mathring{\beta}_{m_{k}}}=0$ we find $\operatorname*{Res}_{u_{12}%
=1/\nu_{k}}\Pi_{k}\Omega_{k}\tilde{T}_{k}(\underline{u},v_{2})=0.$ However,
the pole of $\left(  L^{(k)}\right)  _{1\dots m_{k}}(\underline{v})$ (see e.g.
(\ref{iiik})) at $v_{12}=1/\nu_{k+1}$ will produce a singular contribution
from the $v_{i}$-integration $\oint_{u_{1}-1}dv_{i}\cdots$ (which is a part of
$\int_{\mathcal{C}_{u}^{o}}dv_{i}\cdots$) for $i$ odd. We have a $0/0$
situation which we can resolve as follows:

We take $i,j=1,2$ and multiply the result by $\left[  \frac{1}{2}m_{k}\right]
\left[  \frac{1}{2}m_{k}+\frac{1}{2}\right]  $ and shift again $\tilde{T}%
_{k}(v_{1})$ and $\tilde{T}_{k}(v_{2})$ through all the other $\tilde{T}%
_{k}(v_{i})$ as above%
\begin{align*}
&  \Pi_{1\dots m_{k}}^{(k)}\left(  \Omega_{k}\big(\tilde{T}_{k}\big)^{m_{k}%
}\dots\big(\tilde{T}_{k}\big)^{2}\big(\tilde{T}_{k}\big)^{1}\right)
_{\underline{\alpha}}\\
&  =\Pi_{1\dots m_{k}}^{(k)}\tilde{S}_{21}^{(k)}\dots\tilde{S}_{m_{k}1}%
^{(k)}\tilde{S}_{32}^{(k)}\dots\tilde{S}_{m_{k}2}^{(k)}\left(  \Omega
_{k}\big(\tilde{T}_{k}\big)^{1}\big(\tilde{T}_{k}\big)^{2}\big(\tilde{T}%
_{k}\big)^{m_{k}}\dots\big(\tilde{T}_{k}\big)^{3}\right)  _{\underline{\alpha
}}\\
&  =\tilde{S}_{21}^{(k+1)}\dots\tilde{S}_{m_{k}1}^{(k+1)}\tilde{S}%
_{32}^{(k+1)}\dots\tilde{S}_{m_{k}2}^{(k+1)}\Pi_{3\dots m_{k}21}^{(k)}\left(
\Omega_{k}\big(\tilde{T}_{k}\big)^{1}\big(\tilde{T}_{k}\big)^{2}\big(\tilde
{T}_{k}\big)^{m_{k}}\dots\big(\tilde{T}_{k}\big)^{3}\right)  _{\underline
{\alpha}}\,.
\end{align*}
Applying this to $L_{1\dots m_{k}}^{(k)}(\underline{v})$ using (\ref{ik}) for
higher levels we get%
\[
\tilde{\Psi}_{\underline{\alpha}}^{(k)}(\underline{u},\underline{v})=L_{3\dots
m_{k}21}^{(k)}\Pi_{3\dots m_{k}21}^{(k)}\left(  \Omega_{k}\big(\tilde{T}%
_{k}\big)^{1}\big(\tilde{T}_{k}\big)^{2}\big(\tilde{T}_{k}\big)^{m_{k}}%
\dots\big(\tilde{T}_{k}\big)^{3}\right)  _{\underline{\alpha}}%
\]
For $u_{12}\approx1/\nu_{k},~v_{1}\approx u_{1}-1$ and $v_{2}\approx u_{2}$
(i.e. $v_{12}\approx1/\nu_{k}-1=1/\nu_{k+1})$ we may replace inside
$\tilde{\Psi}_{\underline{\alpha}}^{(k)}$ the S-matrices%
\begin{align*}
\tilde{S}_{k}(u_{1}-v_{1}) &  \rightarrow\tilde{c}(u_{1}-v_{1})\mathbf{P}\\
\tilde{S}_{k}(u_{2}-v_{1}) &  \rightarrow1\\
\tilde{S}_{k}(u_{1}-v_{2}) &  \rightarrow\tilde{d}_{k}(u_{12})\mathbf{K}\\
\tilde{S}_{k}(u_{2}-v_{2}) &  \rightarrow\mathbf{P\,.}%
\end{align*}
For the first relation it has been used that $\tilde{b}(u)\approx-\tilde
{c}(u)\approx1/(u-1)$ and the only nonvanishing contribution from $\tilde
{S}_{k}(u_{1}-v_{1})$ is $\tilde{b}(u_{1}-v_{1})\left(  \mathbf{1-P+}%
const\,\mathbf{K}\right)  _{\mathring{\beta}^{\prime}1}^{\mathring{\beta}%
1}=\tilde{c}(u_{1}-v_{1})\mathbf{P}_{\mathring{\beta}^{\prime}1}%
^{\mathring{\beta}1}$ because $\left(  \mathbf{1-P}\right)  _{\alpha\beta
}^{11}=\,\mathbf{K}_{\alpha\beta}^{11}=0$ and moreover $\left(  \Pi
^{(k)}\right)  _{\dots\overline{k+1}}^{\dots\mathring{\beta}_{m_{k}}}=0$ holds
(see (\ref{Pi})) . Therefore we may replace (see Fig. for $k=0$)%
\begin{align}
\tilde{\Psi}_{\underline{\alpha}}^{(k)}(\underline{u},\underline{v}) &
\rightarrow\tilde{c}(u_{1}-v_{1})\tilde{d}_{k}(u_{12})L_{3\dots m_{k}21}%
^{(k)}(\underline{v}^{\prime})\mathbf{\mathring{C}}^{21}\label{Psik'}\\
&  \times\mathbf{C}_{\hat{\alpha}}\prod_{j=3}^{m_{k}}\frac{1}{a_{k}%
(u_{1}-v_{j})a_{k}(u_{2}-v_{j})}\Pi_{3\dots m_{k}}^{(k)}\left(  \Omega
_{k}\left(  \tilde{T}_{k}\right)  ^{m_{k}}\dots\left(  \tilde{T}_{k}\right)
^{3}\right)  _{\underline{\check{\alpha}}}\,.\nonumber
\end{align}%
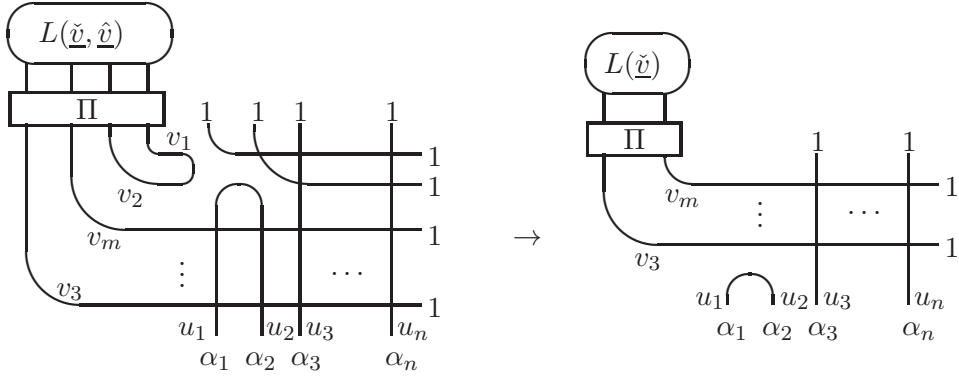
\[
\unitlength4mm\begin{picture}(31,12)(13.5,0)
\thicklines
\put(19,2.6){$\vdots$}
\put(23,1){\line(0,1){7}}
\put(24,3){$\dots$}
\put(26,1){\line(0,1){7}}
\put(18.5,8){\oval(1,2)[lb]}
\put(27,8){\oval(23,7)[lb]}
\put(19.,1){$u_1$}
\put(21.9,1){$u_2$}
\put(23.2,1){$u_3$}
\put(26.2,1){$u_n$}
\put(19.7,0){$\alpha_1$}
\put(21.2,0){$\alpha_2$}
\put(22.7,0){$\alpha_3$}
\put(25.8,0){$\alpha_n$}
\put(15,2.3){$v_3$}
\put(16,4.){$v_{m}$}
\put(17,5.4){$v_{2}$}
\put(18.6,7.3){$v_1$}
\put(27.2,1.6){1}
\put(27.2,4){1}
\put(27.2,5.6){1}
\put(27.2,6.6){1}
\put(19.7,8.1){1}
\put(21.4,8.1){1}
\put(22.8,8.1){1}
\put(25.8,8.1){1}
\put(13.5,8){\framebox(5,1)[]{$\Pi$}}
\put(27,8){\oval(26.,12.)[lb]}
\put(21,1){\oval(1.5,10)[t]}
\put(18.5,8){\oval(3.5,4.)[lb]}
\put(27,8.){\oval(11,4)[lb]}
\put(18.5,6.5){\oval(2,1)[r]}
\put(27,8){\oval(14,2)[lb]}
\put(14,9){\line(0,1){1}}
\put(15.5,9){\line(0,1){1}}
\put(16.75,9){\line(0,1){1}}
\put(18,9){\line(0,1){1}}
\put(16.1,11){\oval(5.4,2)}
\put(14.4,10.7){$L(\underline{\check v},\underline{\hat v})$}
\put(30,4){$\to$}
\put(38,4.6){$\vdots$}
\put(40,2){\line(0,1){5}}
\put(41,5){$\dots$}
\put(43,2){\line(0,1){5}}
\put(44,7){\oval(18,2)[lb]}
\put(36.,2){$u_1$}
\put(38.8,2){$u_2$}
\put(40.2,2){$u_3$}
\put(43.2,2){$u_n$}
\put(36.7,1){$\alpha_1$}
\put(38.2,1){$\alpha_2$}
\put(39.7,1){$\alpha_3$}
\put(42.8,1){$\alpha_n$}
\put(34,3.3){$v_3$}
\put(35,5.4){$v_{m}$}
\put(44.2,3.6){1}
\put(44.2,5.6){1}
\put(39.8,7.1){1}
\put(42.8,7.1){1}
\put(32.5,7){\framebox(3,1)[]{$\Pi$}}
\put(44,7){\oval(22,6)[lb]}
\put(37.8025,2.022){\oval(1.5,2)[t]}
\put(33,8){\line(0,1){1}}
\put(35,8){\line(0,1){1}}
\put(34.1,10){\oval(3.4,2)}
\put(33,9.7){$L(\underline{\check v})$}
\end{picture}
\]
Note that unitarity and crossing imply for $u_{12}=1/\nu_{k}$%
\begin{align*}
&  \mathbf{C}_{12}\tilde{S}^{(k)}(u_{1}-v_{m_{k}})\dots\tilde{S}^{(k)}%
(u_{1}-v_{3})\tilde{S}^{(k)}(u_{2}-v_{m_{k}})\dots\tilde{S}^{(k)}(u_{2}%
-v_{3})\\
&  =\frac{\mathbf{C}_{12}S^{(k)}(u_{1}-v_{m_{k}})\dots S^{(k)}(u_{1}%
-v_{3})S^{(k)}(u_{2}-v_{m_{k}})\dots S^{(k)}(u_{2}-v_{3})}{a_{k}%
(u_{1}-v_{m_{k}})\dots a_{k}(u_{1}-v_{3})a_{k}(u_{2}-v_{mk})\dots a_{k}%
(u_{2}-v_{3}))}\\
&  =\mathbf{C}_{12}\prod_{j=3}^{m_{k}}\frac{\mathbf{1}}{a_{k}(u_{1}%
-v_{j})a_{k}(u_{2}-v_{j})}\,.
\end{align*}
We calculate for $v_{12}\rightarrow1/\nu_{k+1}$%
\[
L_{3\dots m_{k}21}^{(k)}(\underline{v}^{\prime})\mathbf{\mathring{C}}%
^{21}=\left(  \operatorname*{Res}_{v=1/\nu_{k+1}}\tilde{d}_{k+1}(v)\right)
^{-1}\prod_{j=3}^{m_{k}}a_{k+1}(v_{1j})a_{k+1}(v_{2j})\tilde{\psi}%
(v_{j1}+1)\tilde{\chi}(v_{j2})L_{3\dots n}^{(k)}(\underline{\check{v}})\,.
\]
It has been used that from (\ref{iiik}) and (\ref{ik}) follows
\begin{align*}
&  \prod_{j=3}^{m_{k}}\tilde{\psi}(v_{j1}+1)\tilde{\chi}(v_{j2})L_{3\dots
n}^{(k)}(\underline{\check{v}})\mathbf{\mathring{C}}_{12}\\
&  =\operatorname*{Res}_{v_{12}=1/\nu_{k+1}}L_{12\dots m}^{(k)}(\underline
{v})\\
&  =L_{21\dots m}^{(k)}\operatorname*{Res}_{v_{12}=1/\nu_{k+1}}\tilde{S}%
_{12}^{(k+1)}(v_{12})=L_{21\dots m_{k}}^{(k)}\operatorname*{Res}_{v_{12}%
=1/\nu_{k+1}}\tilde{d}_{k+1}(v_{12})\mathbf{\mathring{K}}_{12}\\
&  =L_{3\dots m_{k}21}^{(k)}(\underline{\check{v}},\underline{\hat{v}}%
)\tilde{S}_{2m_{k}}^{(k+1)}\dots\tilde{S}_{23}^{(k+1)}\tilde{S}_{1m_{k}%
}^{(k+1)}\dots\tilde{S}_{13}^{(k+1)}\operatorname*{Res}_{v_{12}=1/\nu_{k+1}%
}\tilde{d}_{k+1}(v_{12})\mathbf{\mathring{C}}^{21}\mathbf{\mathring{C}}_{12}\\
&  =L_{3\dots m_{k}21}^{(k)}(\underline{\check{v}},\underline{\hat{v}%
})\mathbf{\mathring{C}}^{21}\prod_{j=3}^{m_{k}}\frac{1}{a_{k+1}(v_{1j}%
)a_{k+1}(v_{2j})}\operatorname*{Res}_{v=1/\nu_{k+1}}\tilde{d}_{k+1}%
(v)\mathbf{\mathring{C}}_{12}\,.
\end{align*}
where again unitarity and crossing for $v_{12}=1/\nu_{k+1}$ has been used. We
write%
\begin{align*}
\tilde{h}(\underline{u},\underline{v}) &  =\prod_{i=1}^{n}\prod_{j=1}^{m_{k}%
}\tilde{\phi}_{j}(u_{i}-v_{j})\prod_{1\leq i<j\leq m_{k}}\tau_{ij}(v_{ij})\\
&  =\left(  \prod_{i=1}^{2}\prod_{j=1}^{2}\tilde{\phi}_{j}(u_{i}%
-v_{j})\right)  \left(  \prod_{j=3}^{m_{k}}\tilde{\phi}_{j}(u_{1}-v_{j}%
)\tilde{\phi}_{j}(u_{2}-v_{j})\right)  \left(  \prod_{i=3}^{n_{k}}\tilde{\psi
}(u_{i}-v_{1})\tilde{\chi}(u_{i}-v_{2})\right)  \\
&  \times\tau_{12}(v_{12})\prod_{j=3}^{m_{k}}\left(  \tau_{1j}(v_{1j}%
)\tau_{2j}(v_{2j})\right)  \tilde{h}(\underline{\check{u}},\underline
{\check{v}})
\end{align*}
and obtain finally%
\begin{align*}
&  \operatorname*{Res}_{u_{12}=1/\nu_{k}}K_{\underline{\alpha}}^{(k)}%
(\underline{u})=\tilde{N}_{m_{k}}^{(k)}\left[  \tfrac{1}{2}m_{k}\right]
\left[  \tfrac{1}{2}m_{k}+\tfrac{1}{2}\right]  \\
&  \times\left(  \operatorname*{Res}_{v=1/\nu_{k+1}}\tilde{d}_{k+1}(v)\right)
^{-1}\operatorname*{Res}_{u_{12}=1/\nu_{k}}\oint_{u_{1}-1}dv_{1}\tilde
{c}(u_{1}-v_{1})\left(  -\oint_{u_{2}}\right)  dv_{2}\\
&  \times\tilde{d}_{k}(u_{12})\left(  \prod_{i=1}^{2}\prod_{j=1}^{2}%
\tilde{\phi}_{j}(u_{i}-v_{j})\right)  \tau_{12}(v_{12})\prod_{i=3}^{n_{k}%
}\tilde{\psi}(u_{i1}+1)\tilde{\chi}(u_{i2})\frac{1}{\tilde{N}_{m_{k}-2}^{(k)}%
}\mathbf{C}_{\hat{\alpha}}K_{\underline{\check{\alpha}}}^{(k)}(\underline
{\check{u}})
\end{align*}
It has been used that for $u_{12}=1/\nu_{k},~v_{12}=1/\nu_{k+1},~u_{2}%
=v_{2},~u_{1}=v_{2}+1/\nu_{k}=v_{1}+1$%
\[
\left(  \frac{a_{k+1}(v_{1j})a_{k+1}(v_{2j})}{a_{k}(u_{1}-v_{j})a_{k}%
(u_{2}-v_{j})}\tilde{\psi}(v_{j1}+1)\tilde{\chi}(v_{j2})\right)  \left(
\tilde{\phi}_{j}(u_{1}-v_{j})\tilde{\phi}_{j}(u_{2}-v_{j})\tau_{1j}%
(v_{1j})\tau_{2j}(v_{2j})\right)  =1
\]
for odd and even $j$, which can be shown by means of (\ref{shiftphi}) and the
formula
\[
a_{k}(u_{1})a_{k}(u_{2})=\tilde{b}(-u_{2})/\tilde{b}(u_{1})\,.
\]
The result is that equation (\ref{iiik}) holds if
\begin{align}
&  \frac{\tilde{N}_{m_{k}}^{(k)}}{\tilde{N}_{m_{k}-2}^{(k)}}\left[  \tfrac
{1}{2}m_{k}\right]  \left[  \tfrac{1}{2}m_{k}+\tfrac{1}{2}\right]  \left(
\operatorname*{Res}_{v=1/\nu_{k+1}}\tilde{d}_{k+1}(v)\right)  ^{-1}\left(
-\left(  2\pi i\right)  ^{2}\right)  \label{NkNk}\\
&  \times\operatorname*{Res}_{v_{1}=u_{1}-1}\tilde{c}(u_{1}-v_{1}%
)\operatorname*{Res}_{u_{12}=1/\nu_{k}}\tilde{d}_{k}(u_{12})\tilde{\psi
}(1)\tilde{\psi}(u_{21}+1)\tilde{\chi}(u_{12})\operatorname*{Res}_{v_{2}%
=u_{2}}\tilde{\chi}(u_{2}-v_{2})\tau_{12}(v_{12})\nonumber\\
&  =\frac{\tilde{N}_{m_{k}}^{(k)}}{\tilde{N}_{m_{k}-2}^{(k)}}\left[  \tfrac
{1}{2}m_{k}\right]  \left[  \tfrac{1}{2}m_{k}+\tfrac{1}{2}\right]  4\pi
^{2}\tilde{\psi}^{2}(1)\frac{\tilde{\chi}(1/\nu_{k}-2/\nu)}{\tilde{\chi}%
(1/\nu_{k+1}-2/\nu)}=1
\end{align}
which follows from the assumption (\ref{NN}).

In particular for $k=0$ we have%
\[
\tilde{N}_{m}^{(0)}=\frac{1}{\left[  \frac{1}{2}m\right]  \left[  \frac{1}%
{2}m+\frac{1}{2}\right]  }\frac{1}{4\pi^{2}\tilde{\psi}^{2}(1)}\frac
{\tilde{\chi}(1/\mathring{\nu}-2/\nu)}{\tilde{\chi}(1/\nu-2/\nu)}\tilde
{N}_{m-2}^{(0)}%
\]
and (\ref{2.15}) with (\ref{2.16}), (\ref{Kk}) and (\ref{RKk}) implies
(\ref{norm}).
\end{proof}

\subsection{Proof of lemma
\protect\ref{L3}%
}

\label{se1}

\begin{proof}
For $N$ odd and $k=M=(N-3)/2$ we have $\nu_{k}=2$ and $\nu_{k+1}=-2$ as for
$O(3)$, however, the shift in (\ref{iik}) is not that of $O(3)$ but that of
$O(N)$. We proceed as above for general $k$, for $N>3$ there is for all $j$
even \textquotedblleft pinching\textquotedblright\ at $v_{j}=u_{2}\approx
u_{1}-1/2$ and we take $j=1,2$ as a pair of an odd and an even $j$ and
multiply the result with $\left[  \frac{1}{2}m_{M}\right]  \left[  \frac{1}%
{2}m_{M}+\frac{1}{2}\right]  $. The L-function $L_{1\dots m_{M}}%
^{(M)}(\underline{v})=K_{1\dots m_{M}}^{(M+1)}(\underline{v})$ is a c-number
satisfying (\ref{ik}) with
\[
\tilde{S}_{0}^{(M+1)}(v)=\frac{v+1/\nu_{M+1}}{v-1/\nu_{M+1}}\frac{v+1}%
{v-1}=\frac{v-1/2}{v+1/2}\frac{v+1}{v-1}%
\]
and (\ref{iik}) with the solution (as in subsection \ref{s42})%
\[
L^{(M)}(\underline{v})=\prod\limits_{1\leq i<j\leq m_{M}}L^{(M)}%
(v_{ij})\,,~~L^{(M)}(u)=\frac{\Gamma\left(  \frac{1}{4}\nu+\frac{1}{2}\nu
u\right)  \Gamma\left(  1+\frac{1}{4}\nu-\frac{1}{2}\nu u\right)  }%
{\Gamma\left(  \frac{1}{2}\nu+\frac{1}{2}\nu u\right)  \Gamma\left(
1+\frac{1}{2}\nu-\frac{1}{2}\nu u\right)  }\,.
\]
We have again (\ref{Psik'}) where here%
\[
L_{3\dots m_{k}21}^{(M)}(\underline{v}^{\prime})\mathbf{\mathring{C}}%
^{21}\rightarrow L^{(M)}(\underline{\check{v}})\prod\limits_{j=3}^{m_{M}%
}\left(  L^{(M)}(v_{j1})L^{(M)}(v_{j2})\right)  L^{(M)}(v_{21})
\]
and instead of (\ref{NkNk})%
\begin{align*}
&  \frac{\tilde{N}_{m_{M}}^{(M)}}{\tilde{N}_{m_{M}-2}^{(M)}}\left[  \tfrac
{1}{2}m_{M}\right]  \left[  \tfrac{1}{2}m_{M}+\tfrac{1}{2}\right]
L^{(M)}(v_{21})\left(  -\left(  2\pi i\right)  ^{2}\right)  \\
&  \times\operatorname*{Res}_{v_{1}=u_{1}-1}\tilde{c}(u_{1}-v_{1}%
)\operatorname*{Res}_{u_{12}=1/\nu_{M}}\tilde{d}_{M}(u_{12})\tilde{\psi
}(1)\tilde{\psi}(u_{21}+1)\tilde{\chi}(u_{12})\operatorname*{Res}_{v_{2}%
=u_{2}}\tilde{\chi}(u_{2}-v_{2})\tau_{12}(v_{12})\\
&  =\frac{\tilde{N}_{m_{M}}^{(M)}}{\tilde{N}_{m_{M}-2}^{(M)}}\left[  \tfrac
{1}{2}m_{M}\right]  \left[  \tfrac{1}{2}m_{M}+\tfrac{1}{2}\right]  \frac
{2\pi^{2}\Gamma\left(  1-\frac{1}{4}\nu\right)  }{\Gamma^{3}\left(  1+\frac
{1}{4}\nu\right)  \Gamma\left(  1+\frac{1}{2}\nu\right)  }=1\,,
\end{align*}
which is (\ref{NM}). For $N=3$ there is for all $j$ even and odd
\textquotedblleft pinching\textquotedblright\ at $v_{j}=u_{2}\approx
u_{1}-1/2$ and $\left[  \frac{1}{2}m_{M}\right]  \left[  \frac{1}{2}%
m_{M}+\frac{1}{2}\right]  $ has to be replaced $m\left(  m-1\right)  $. It was
used that for $u_{12}=1/2$ and $v_{12}=-1/2$%
\[
\frac{L^{(M)}(v_{j1})L^{(M)}(v_{j2})}{a_{M}(u_{1}-v_{j})a_{M}(u_{2}-v_{j}%
)}\left(  \tilde{\phi}_{j}(u_{1}-v_{j})\tilde{\phi}_{j}(u_{2}-v_{j})\tau
_{1j}(v_{1j})\tau_{2j}(v_{2j})\right)  =1
\]
which follows as above because%
\[
L^{(M)}(-v_{1})L^{(M)}(-v_{2})=\tilde{\psi}(-v_{1}+1)\tilde{\chi}%
(-v_{2})\tilde{b}(-v_{2})/\tilde{b}(v_{1})\,.
\]

\end{proof}

\subsection{Proof of lemma
\protect\ref{L4}%
}

\label{se2}\textbf{ }

\begin{proof}
For $N$ even and $k=M=(N-4)/2$ we have $\nu_{k}=1$ as for $O(4)$, however, the
shift in (\ref{iik}) is not that of $O(4)$ but that of $O(N)$. We use the
technique of subsection \ref{s2.3} with the $SU(2)$ S-matrix (see
\cite{BFK0,BFK1,BFK3})
\[
S^{SU(2)}(u)=a^{SU(2)}(u)\left(  \frac{u}{u-1}\mathbf{1-}\frac{1}%
{u-1}\mathbf{P}\right)  \,.
\]
First we calculate $\overset{(1)}{\operatorname*{Res}\limits_{u_{12}=1}%
}\,^{\pm}K_{\underline{A}}(\underline{u})$ which is due to the pinching at
$v_{1}=u_{2}\rightarrow u_{1}-1$ and gives the first term in (\ref{1.14ax}).
The contribution is given by $v_{1}$-integration $\oint_{u_{2}}dv_{1}\cdots$
along small circles around $v_{1}=u_{2}$ . For $u_{12}\approx1,~v_{1}\approx
u_{2}$ we may replace the S-matrices inside of $\tilde{\Psi}_{\underline{A}%
}^{SU(2)}$%
\begin{align*}
\tilde{S}(u_{2}-v_{1})  &  \rightarrow\mathbf{P}\\
\tilde{S}(u_{1}-v_{1})  &  \rightarrow\tilde{S}(u_{12})\rightarrow\frac
{1}{u_{12}-1}\left(  \mathbf{1-P}\right)
\end{align*}
such that we may replace%
\[
\tilde{\Psi}_{\underline{A}}^{SU(2)}(\underline{u},\underline{v}%
)\rightarrow\frac{1}{u_{12}-1}\mathbf{C}_{A_{1}A_{2}}\prod_{j=2}^{m}\tilde
{b}(u_{1}-v_{j})\left(  \Omega\tilde{C}(\underline{\check{u}},v_{m}%
)\cdots\tilde{C}(\underline{\check{u}},v_{N})\right)  _{\underline{\check{A}}%
}\,.
\]
because $\left(  \mathbf{1-P}\right)  _{A_{1}A_{2}}^{21}=\mathbf{C}%
_{A_{1}A_{2}}$. We write%
\[
\tilde{h}(\underline{u},\underline{v})=\left(  \tilde{\phi}_{\nu}(u_{1}%
-v_{1})\tilde{\phi}_{\nu}(u_{2}-v_{1})\right)  \prod_{j=2}^{m_{\pm}}\left(
\tilde{\phi}_{\nu}(u_{1}-v_{j})\tilde{\phi}_{\nu}(u_{2}-v_{j})\tau_{\nu
}(v_{1j})\right)  \prod_{i=3}^{n_{M}}\tilde{\phi}_{\nu}(u_{i2})\tilde
{h}(\underline{\check{u}},\underline{\check{v}})
\]
and get%
\begin{align*}
&  \overset{(1)}{\operatorname*{Res}\limits_{u_{12}=1}}\,^{\pm}K_{\underline
{A}}(\underline{u})\\
&  =\frac{\,^{\pm}\tilde{N}_{m_{\pm}}}{\,^{\pm}\tilde{N}_{m_{\pm}-1}}m_{\pm
}\operatorname*{Res}_{u_{12}=1}\frac{1}{u_{12}-1}\left(  \tilde{\phi}_{\nu
}(u_{12})\oint_{u_{2}}dv_{1}\tilde{\phi}_{\nu}(u_{2}-v_{1})\right)
\prod_{i=3}^{n_{M}}\tilde{\phi}_{\nu}(u_{i2})\mathbf{C}_{A_{1}A_{2}}\,^{\pm
}K_{\underline{\check{A}}}(\underline{\check{u}})\\
&  =\frac{\,^{\pm}\tilde{N}_{m_{\pm}}}{\,^{\pm}\tilde{N}_{m_{\pm}-1}}m_{\pm
}2\pi i(-1)^{m_{\pm}}\Gamma^{2}\left(  -\tfrac{1}{2}\nu\right)  \prod
_{i=3}^{n_{M}}\tilde{\phi}_{\nu}(u_{i2})\mathbf{C}_{A_{1}A_{2}}\,^{\pm
}K_{\underline{\check{A}}}(\underline{\check{u}})
\end{align*}
where we have used the identity
\[
\tilde{\phi}_{\nu}(u_{1})\tilde{\phi}_{\nu}(u_{2})\tau_{\nu}(u_{2})\tilde
{b}(u_{1})=-1
\]
because of $\tilde{\phi}_{\nu}(u_{1})\tilde{b}(u_{1})=-\tilde{\phi}_{\nu
}(-u_{2})$. Finally we have
\[
\overset{(1)}{\operatorname*{Res}\limits_{u_{12}=1}}\,^{\pm}K_{\underline{A}%
}(\underline{u})=\prod_{i=3}^{n_{M}}\tilde{\phi}_{\nu}(u_{i2})\mathbf{C}%
_{A_{1}A_{2}}\,^{\pm}K_{\underline{\check{A}}}(\underline{\check{u}})
\]
if
\[
\,^{\pm}\tilde{N}_{m_{\pm}}=\frac{(-1)^{m_{\pm}}}{m_{\pm}2\pi i\Gamma
^{2}\left(  -\tfrac{1}{2}\nu\right)  }\,^{\pm}\tilde{N}_{m_{\pm}-1}.
\]
Now we take the residue of (\ref{KKK})%
\begin{multline*}
\overset{(1)}{\operatorname*{Res}\limits_{u_{12}=1}}K_{\underline{\alpha}%
}^{(M)}(\underline{u})=d_{n_{M}}\left(  \tfrac{1}{2}\pi\nu\right)
\prod_{2<i<j}\sin\tfrac{1}{2}\pi\nu\left(  u_{ij}-1\right) \\
\times\prod_{j=3}^{n_{M}}\left(  \sin\tfrac{1}{2}\pi\nu\left(  u_{1j}%
-1\right)  \sin\tfrac{1}{2}\pi\nu\left(  u_{2j}-1\right)  \right)  \sum
_{l}\overset{(1)}{\operatorname*{Res}\limits_{u_{12}=1}}\,^{+}K_{\underline
{A}}^{\,(l)}(\underline{u})\overset{(1)}{\operatorname*{Res}\limits_{u_{12}%
=1}}\,^{-}K_{\underline{B}}^{(l)\,}(\underline{u})\Gamma_{\underline{\alpha}%
}^{\underline{A}\underline{B}}\\
=-\frac{d_{m_{M}}}{d_{m_{M}-2}}\left(  \tfrac{1}{2}\pi\nu\right)  \pi^{2}%
\prod_{j=3}^{m_{M}}\tilde{\psi}(u_{i1}+1)\tilde{\chi}(u_{i2})\mathbf{C}%
_{\underline{\hat{\alpha}}}K_{\underline{\check{\alpha}}}^{(k)}(\underline
{\check{u}})=\prod_{j=3}^{m_{M}}\tilde{\psi}(u_{i1}+1)\tilde{\chi}%
(u_{i2})\mathbf{C}_{\underline{\hat{\alpha}}}K_{\underline{\check{\alpha}}%
}^{(k)}(\underline{\check{u}})
\end{multline*}
if%
\[
d_{n_{M}}=-\frac{2}{\nu\pi^{3}}d_{n_{M}-2}\,.
\]
We used that $\mathbf{C}_{A_{1}A_{2}}\mathbf{C}_{B_{1}B_{2}}\Gamma_{\alpha
_{1}}^{A_{1}B_{1}}\Gamma_{\alpha_{2}}^{A_{2}B_{2}}=-\mathbf{C}_{\alpha
_{1}\alpha_{2}}$ and
\[
\left(  \tilde{\phi}_{\nu}(u)\right)  ^{2}=\frac{\pi^{2}}{\sin\frac{1}{2}%
\pi\nu\left(  u-1\right)  \,\sin\frac{1}{2}\pi\nu u}\tilde{\psi}(u)\tilde
{\chi}(u)\,.
\]

\end{proof}

\subsection{Two-particle higher level K-functions}

\label{sf}For the examples of section \ref{s5} we need higher level
K-functions. In particular $K_{\alpha_{1}\alpha_{2}}^{(k)}(\theta_{1}%
,\theta_{2})$ (level $k=0,1,2,\dots$) belonging to $O(N-2k)$ in the iso-scalar
two-particle channel (with weights $w=(0,\dots,0)$).

\begin{lemma}
\label{l2}The vector valued functions%
\[
K_{\alpha_{1}\alpha_{2}}^{(k)}(\theta_{1},\theta_{2})=\mathbf{C}_{\alpha
_{1}\alpha_{2}}^{(N-2k)}K_{k}(\theta_{12})
\]
with%
\begin{equation}
K_{k}(\theta)=\frac{\Gamma\left(  -\frac{1}{2}\left(  1-k\nu\right)  +\frac
{1}{2}\left(  1-k\nu\right)  \frac{\theta}{\pi i}\right)  \Gamma\left(
1-\frac{1}{2}\left(  1-k\nu\right)  -\frac{1}{2}\left(  1-k\nu\right)
\frac{\theta}{\pi i}\right)  }{\Gamma\left(  1+\frac{1}{2}\nu-\frac{1}%
{2}\left(  1-k\nu\right)  \frac{\theta}{\pi i}\right)  \Gamma\left(  \frac
{1}{2}\nu+\frac{1}{2}\left(  1-k\nu\right)  \frac{\theta}{\pi i}\right)  }
\label{K}%
\end{equation}
$\allowbreak$satisfy for $k=0,1,2,\dots<N/2-2$ the recursion relation (for a
suitable normalization)%
\begin{align*}
K_{\underline{\alpha}}^{(k)}\left(  \underline{\theta}\right)   &
=N_{n}^{(k)}\int_{\mathcal{C}_{\underline{\theta}}^{o}}dz_{1}\int
_{\mathcal{C}_{\underline{\theta}}^{e}}dz_{2}\,\tilde{h}(\underline{\theta
},\underline{z})\,L_{\underline{\mathring{\beta}}}^{(k)}(\underline{z}%
)\tilde{\Phi}^{(k)}\,_{\underline{\alpha}}^{\underline{\mathring{\beta}}%
}\,(\underline{\theta},\underline{z})\\
L_{\underline{\mathring{\beta}}}^{(k)}(\underline{z})  &  =K_{\underline
{\mathring{\beta}}}^{(k+1)}(\underline{z}\nu_{k+1}/\nu_{k})\,,~~\nu
_{k}=2/(N-2k-2)
\end{align*}
with%
\begin{align*}
\tilde{h}(\underline{\theta},\underline{z})  &  =\prod_{i=1}^{2}\left(
\tilde{\psi}(\theta_{i}-z_{1})\tilde{\chi}(\theta_{i}-z_{2})\right)  \frac
{1}{\tilde{\chi}(z_{12})\tilde{\psi}(-z_{12})}\\
\tilde{\Phi}^{(k)}\,_{\underline{\alpha}}^{\underline{\mathring{\beta}}%
}\,(\underline{\theta},\underline{z})  &  =\left(  \Pi_{\underline{\beta}%
}^{\underline{\mathring{\beta}}}(\underline{z})\Omega\tilde{T}_{1}^{\beta_{2}%
}(\underline{\theta},z_{2})\tilde{T}_{1}^{\beta_{1}}(\underline{\theta}%
,z_{1})\right)  _{\underline{\alpha}}^{(k)}%
\end{align*}

\end{lemma}

The K-function $K_{\underline{\alpha}}^{(0)}\left(  \underline{\theta}\right)
$ belongs to an iso-scalar, spin zero operator (with p-function $p=1$). This
means it belongs to the energy momentum potential (see example \ref{s5.1}
formula (\ref{EM2}))%
\[
F_{\underline{\alpha}}^{T}(\theta)=K_{\underline{\alpha}}^{(0)}\left(
\underline{\theta}\right)  F\left(  \theta\right)  \,.
\]
The L-function $L_{\underline{\mathring{\beta}}}^{(0)}(\underline
{z})=K_{\underline{\mathring{\beta}}}^{(1)}(\underline{z}\nu_{k+1}/\nu_{k})$
is for $n=2$ that of (\ref{Lem}).

\begin{proof}
We do not calculate the integrals in (\ref{Kk}) for general $N$, but we use
arguments of \cite{KW} to prove the claim. In addition we calculate integrals
explicitly for $N\rightarrow\infty,~N=3$ and $N=4$ (see below). That for all
levels $n_{k}=2$ is follows from%
\begin{align*}
w  &  =(w_{1},\dots,w_{\left[  N/2\right]  })=(0,\dots,0)\\
&  =\left\{
\begin{array}
[c]{lll}%
\left(  n-n_{1},\dots,n_{\left[  N/2\right]  -1}-n_{\left[  N/2\right]
}\right)  & \text{for} & N~\text{odd}\\
\left(  n-n_{1},\dots,n_{\left[  N/2\right]  -2}-n_{-}-n_{+},n_{-}%
-n_{+}\right)  & \text{for} & N~\text{even\thinspace.}%
\end{array}
\right.
\end{align*}
Note that for $N~$even we have $n_{-}=n_{+}=1$ (see below). For convenience we
use here the parameterization $\theta=i\pi\nu_{k}u$ and $z=i\pi\nu_{k}v$,
($\nu_{k}=2/(N-2k-2)$) and the S-matrix $S^{(k)}(u)=S^{O(N-2k)}(u)$ (see
(\ref{Su}).

Theorem \ref{TN} (with the proof in appendix \ref{sd}) implies that
$K_{k}(\underline{u})$ defined by (\ref{Kk}) satisfies%
\begin{align*}
\text{(i)}  &  :~~~K_{k}(u)=K_{k}(-u)\tilde{S}_{0}^{(k)}(u)\\
\text{(ii)}  &  :~~~K_{k}(1/\nu-u)=K_{k}(1/\nu+u)
\end{align*}
with (see (\ref{EV}))%
\[
\tilde{S}_{0}^{(k)}(u)=S_{0}^{(k)}(u)/S_{+}^{(k)}(u)=\frac{u+1/\nu_{k}%
}{u-1/\nu_{k}}\frac{u+1}{u-1}=\frac{u+\left(  1/\nu-k\right)  }{u-\left(
1/\nu-k\right)  }\frac{u+1}{u-1}\,.
\]
The minimal solution of (i) and (ii) is
\[
K_{k}^{\min}(u)=\frac{1}{\Gamma\left(  1+\frac{1}{2}\nu-\frac{1}{2}\nu
u\right)  \Gamma\left(  \frac{1}{2}\nu+\frac{1}{2}\nu u\right)  \Gamma\left(
1+\frac{1}{2}\left(  1-k\nu\right)  -\frac{1}{2}\nu u\right)  \Gamma\left(
\frac{1}{2}\left(  1-k\nu\right)  +\frac{1}{2}\nu u\right)  }%
\]
The proof of (iii) in appendix \ref{sd} shows that $K_{k}(u)$ has a pole at
$u=1/\nu_{k}=1/\nu-k$ if $K_{k+1}(u)$ has a pole at $u=1/\nu_{k+1}%
=1/\nu-\left(  k+1\right)  $. If there are no other poles in $0\leq\nu u\leq1$
following \cite{KW} we conclude (up to normalization)%
\[
K_{k}(u)=\frac{1}{\sin\frac{1}{2}\pi\nu\left(  u-\left(  1/\nu-k\right)
\right)  \sin\frac{1}{2}\pi\nu\left(  u+1/\nu-k\right)  }K_{k}^{\min}(u)
\]
which proves (\ref{K}).
\end{proof}

\paragraph{$N\rightarrow\infty:$}

For $\nu\rightarrow0$ with $\theta$ fixed we get from (\ref{K})
\begin{equation}
K_{k}(\theta)=2\pi\frac{\tanh\frac{1}{2}\theta}{\theta-i\pi}+O(\nu)
\label{Kinf}%
\end{equation}
for all $k$. We prove that this result agrees with the recursion relation
(\ref{Kk}).

\begin{proof}
For convenience we use the notation $x_{i}=\nu u_{i}=\theta_{i}/(i\pi
),~y_{i}=\nu v_{i}=z_{i}/(i\pi)$. We calculate the rhs of (\ref{Kk}) for the
component with $\underline{\alpha}=\overline{k+1},k+1$ (up to normalization)%
\[
\int_{\mathcal{C}_{\underline{x}}^{o}}dy_{1}\int_{\mathcal{C}_{\underline{x}%
}^{e}}dy_{2}\,\tilde{h}(\underline{x},\underline{y})K_{k+1}(y_{1}-y_{2}%
)\tilde{\Phi}_{k}(\underline{x},\underline{y})
\]
with $K_{k+1}$ given by (\ref{K}) and
\begin{align}
&  \tilde{\Phi}_{k}(\underline{x},\underline{y})=\mathbf{C}_{\mathring{\beta
}_{1}\mathring{\beta}_{2}}^{(N-2k-2)}\tilde{\Phi}^{(k)}\,_{\overline{k+1}%
,k+1}^{\mathring{\beta}_{1}\mathring{\beta}_{2}}\,(\underline{x},\underline
{y})\label{Phik}\\
&  =\mathbf{C}_{\mathring{\beta}_{1}\mathring{\beta}_{2}}^{(N-2k-2)}%
\mathbf{C}_{(N-2k-2)}^{\mathring{\beta}_{1}\mathring{\beta}_{2}}\left(
\tilde{c}(x_{1}-y_{2})\tilde{d}_{k}(x_{1}-y_{1})+f_{k}(y_{12})\left(
\tilde{c}(x_{1}-y_{1})+\tilde{d}_{k}(x_{1}-y_{1})\right)  \right) \nonumber\\
&  =(N-2k-2)\frac{-\nu^{2}\left(  y_{1}-y_{2}-\nu\right)  }{\left(
y_{1}-y_{2}+\nu/\nu_{k}-\nu\right)  \left(  y_{1}-x_{1}+\nu\right)  \left(
y_{2}-x_{1}+\nu\right)  }\nonumber
\end{align}
where (\ref{1.2}) and (\ref{Pi2}) have been used. We get after a lengthy
calculation the result%
\begin{equation}
\int_{\mathcal{C}_{\underline{x}}^{o}}dy_{1}\int_{\mathcal{C}_{\underline{x}%
}^{e}}dy_{2}\,\tilde{h}(\underline{x},\underline{y})K_{k+1}(y_{1}-y_{2}%
)\tilde{\Phi}_{k}(\underline{x},\underline{y})=-16\pi^{2}\nu^{2}\frac
{1}{x_{12}-1}\tan\frac{1}{2}\pi\left(  x_{12}\right)  +O(\nu^{3}) \label{Krhs}%
\end{equation}
which agrees with (\ref{Kinf}) (up to const.).
\end{proof}

\paragraph{N = 3:}

For $\nu=2$ and $k=0,1$ we get from (\ref{K})%

\begin{align}
K(\theta)  &  =K_{0}(\theta)=8\pi^{3}\frac{\tanh\frac{1}{2}\theta}%
{\theta\left(  \theta-i\pi\right)  \left(  \theta-2\pi i\right)  }\label{K3}\\
L(z)  &  =K_{1}(-z)=2\pi\left(  z-i\pi\right)  \frac{\tanh\frac{1}{2}%
z}{z\left(  z-2\pi i\right)  }\nonumber
\end{align}
The result (\ref{KT4}) in appendix \ref{sh} proves that these functions
satisfy the recursion relation (\ref{Kk}) (for suitable normalization).

\paragraph{N = 4:}

For $\nu=1$ and $k=0$ we get from (\ref{K}) (up to const.)%

\begin{equation}
K(x)=\left(  \frac{1}{\theta_{12}-i\pi}\right)  ^{2} \label{K4}%
\end{equation}

We use the general formulas (\ref{OOO}) and (\ref{FFF}) (for details see
appendix \ref{so4}) with
\begin{align*}
^{+}\mathcal{O}_{1} &  =\,^{-}\mathcal{O}_{2}=\,\mathcal{O}_{1}\\
^{+}\mathcal{O}_{2} &  =\,^{-}\mathcal{O}_{1}=\,\mathcal{O}_{2}%
\end{align*}
and we propose the p-functions $\ $%
\begin{align*}
p^{\mathcal{O}_{1}}(\underline{\theta},\underline{z}) &  ={\textstyle\sum
\limits_{i=1}^{m}}\exp z_{i}\,/{\textstyle\sum\limits_{i=1}^{n}}\exp\theta\\
p^{\mathcal{O}_{2}}(\underline{\theta},\underline{z}) &  =\exp\left(
\tfrac{1}{2}{\textstyle\sum\limits_{i=1}^{n}}\left(  \theta_{i}-\tfrac{1}%
{2}i\pi\right)  -{\textstyle\sum\limits_{i=1}^{m}}z_{i}\right)  \,
\end{align*}
with weights $w_{SU(2)}^{\mathcal{O}_{i}}=(0,0)\Rightarrow w_{O(4)}%
^{\mathcal{O}}=(0,0)\Rightarrow n=2m=$ even and statistics factors $\sigma
_{1}=-\sigma_{2}=1$ which satisfy the condition of appendix \ref{so4}%
\[
\sigma_{1}\sigma_{2}=(-1)^{n-1}\,.
\]
The results of \cite{BFK1,BFK3} imply for $n=2$ (up to constants)%
\begin{align*}
K_{A_{1}A_{2}}^{\mathcal{O}_{1}}(\underline{\theta}) &  =\left(  \delta
_{A_{1}}^{1}\delta_{A_{2}}^{2}-\delta_{A_{1}}^{2}\delta_{A_{2}}^{1}\right)
\frac{1}{\cosh\frac{1}{2}\theta_{12}}\frac{1}{\theta_{12}-i\pi}\\
K_{B_{1}B_{2}}^{\mathcal{O}_{2}}(\underline{\theta}) &  =\left(  \delta
_{B_{1}}^{1}\delta_{B_{2}}^{2}-\delta_{B_{1}}^{2}\delta_{B_{2}}^{1}\right)
\frac{1}{\theta_{12}-i\pi}%
\end{align*}
and therefore%
\begin{align*}
K_{\alpha_{1}\alpha_{2}}^{T}(\underline{\theta}) &  =\cosh\tfrac{1}{2}%
\theta_{12}K_{A_{1}A_{2}}^{\mathcal{O}_{+}}(\underline{\theta})K_{B_{1}B_{2}%
}^{\mathcal{O}_{-}}(\underline{\theta})\Gamma_{\alpha_{1}}^{A_{1}A_{2}}%
\Gamma_{\alpha_{2}}^{B_{1}B_{2}}\\&
=\mathbf{C}_{\alpha_{1}\alpha_{2}}\left(  \frac{1}{\theta_{12}-i\pi
}\right)  ^{2}%
\end{align*}
which agrees with (\ref{K4}) and (\ref{EMN}).

\section{1/N expansion}

\subsection{1/N expansion of the exact 3-particle field form factor}

\label{sg}

For ${\mathcal{O}(}x{)}=i(\square+m^{2})\varphi(x)$ we derive for the
component ${F^{\mathcal{O}}}_{\bar{1}11}(\underline{\theta})$%
\[
F^{\mathcal{O}}_{\bar{1}11}(\underline{\theta})=-\frac{8\pi}{N}m^{2}\left(
\frac{\sinh\theta_{12}}{\theta_{12}-i\pi}+\frac{\sinh\theta_{13}}{\theta
_{13}-i\pi}\right)  +O(N^{-2})
\]

\begin{proof}
The p-function of ${\mathcal{O}(}x{)}$ and three particles for $\nu=0$ is%
\[
p^{\mathcal{O}}=\left(  e^{z_{1}}+e^{z_{2}}\right)  \left(  e^{-z_{1}%
}+e^{-z_{2}}\right)  =4\cosh^{2}\tfrac{1}{2}\left(  z_{1}-z_{2}\right)
\]
We have to consider (up to const.)
\[
K_{\bar{1}11}^{\mathcal{O}}(\underline{\theta})=\int_{\mathcal{C}%
_{\underline{\theta}}^{o}}dz_{1}\int_{\mathcal{C}_{\underline{\theta}}^{e}%
}dz_{2}\,\prod_{i=1}^{3}\left(  \tilde{\psi}(\theta_{i}-z_{1})\tilde{\chi
}(\theta_{i}-z_{2})\right)  \frac{\cosh^{2}\tfrac{1}{2}\left(  z_{1}%
-z_{2}\right)  }{\tilde{\chi}(z_{12})\tilde{\psi}(-z_{12})}\,\tilde{\Psi
}_{\bar{1}11}(\underline{\theta},\underline{z})
\]
with
\begin{align*}
\tilde{\Psi}_{\bar{1}11}(\underline{\theta},\underline{z})  &  =L(z_{12}%
)\mathbf{\mathring{C}}_{\mathring{\beta}_{1}\mathring{\beta}_{2}%
}\mathbf{\mathring{C}}^{\mathring{\beta}_{1}\mathring{\beta}_{2}}\left(
\tilde{c}(\theta_{1}-z_{2})\tilde{d}(\theta_{1}-z_{1})+f(z_{12})\left(
\tilde{c}(\theta_{1}-z_{1})+\tilde{d}(\theta_{1}-z_{1})\right)  \right) \\
&  =L(z_{12})\frac{2}{\nu}\bigg(\frac{\mathbf{-}i\pi\nu}{\theta_{1}-z_{2}%
-i\pi\nu}\frac{\theta_{1}-z_{1}}{\theta_{1}-z_{1}-i\pi\nu}\frac{i\pi\nu
}{\theta_{1}-z_{1}-i\pi}\\
&  +\frac{i\pi\nu}{z_{1}-z_{2}+i\pi-i\pi\nu}\left(  \frac{\mathbf{-}i\pi\nu
}{\theta_{1}-z_{1}-i\pi\nu}+\frac{\theta_{1}-z_{1}}{\theta_{1}-z_{1}-i\pi\nu
}\frac{i\pi\nu}{\theta_{1}-z_{1}-i\pi}\right)  \bigg)\\
&  =\nu\pi^{3}\frac{\left(  z_{1}-z_{2}\right)  \tanh\frac{1}{2}\left(
z_{1}-z_{2}\right)  }{\left(  z_{1}-z_{2}-i\pi\right)  \left(  z_{1}%
-z_{2}+i\pi\right)  \left(  \theta_{1}-z_{1}\right)  \left(  \theta_{1}%
-z_{2}\right)  }+O\left(  \nu^{2}\right)
\end{align*}
for $\nu\rightarrow0$ using $L(z)\overset{\nu\rightarrow0}{\rightarrow}%
\frac{2\pi}{z-i\pi}\tanh\frac{1}{2}z$ (see (\ref{Lem})). The leading terms are
given by the integrals%
\begin{align*}
\int_{\mathcal{C}_{\underline{\theta}}}dz_{1}\int_{\mathcal{C}_{\underline
{\theta}}}dz_{2}\dots &  =\left(  \int_{\theta_{1}-i\pi\nu}+\int_{\theta
_{2}-i\pi\nu}+\int_{\theta_{3}-i\pi\nu}\right)  dz_{1}\left(  \int_{\theta
_{1}}+\int_{\theta_{2}}+\int_{\theta_{3}}\right)  dz_{2}\dots\\
&  =\left(  \int_{\theta_{2}}+\int_{\theta_{3}}\right)  dz_{2}\oint
_{\theta_{1}-i\pi\nu}dz_{1}\dots+O(\nu)=I_{1}+I_{2}+O(\nu)
\end{align*}
where ${\oint_{\theta}}dz\dots$ means an integral along a small circle around
$\theta$ and $\int_{\theta}dz\dots$ an integral around all the poles of the
gamma function according to Figs. \ref{f5.1} and \ref{f5.2}. Up to higher
order terms in $\nu$ and with $\tilde{\psi}\overset{\nu=0}{\rightarrow
}1,~\tilde{\chi}\overset{\nu=0}{\rightarrow}1$ we get (always up to const.)%
\begin{align*}
I_{1}  &  =\int_{\theta_{2}}dz_{2}\tilde{\chi}(\theta_{2}-z_{2})\oint
_{\theta_{1}}dz_{1}\frac{\left(  z_{1}-z_{2}\right)  \tanh\frac{1}{2}\left(
z_{1}-z_{2}\right)  }{\left(  z_{1}-z_{2}-i\pi\right)  \left(  z_{1}%
-z_{2}+i\pi\right)  \left(  \theta_{1}-z_{1}\right)  \left(  \theta_{1}%
-z_{2}\right)  }\cos^{2}\tfrac{1}{2}z_{12}\\
&  =\int_{\theta_{2}}dz_{2}\tilde{\chi}(\theta_{2}-z_{2})\frac{\left(
\theta_{1}-z_{2}\right)  \sin\left(  \theta_{1}-z_{2}\right)  }{\left(
\theta_{1}-z_{2}-i\pi\right)  \left(  \theta_{1}-z_{2}+i\pi\right)  \left(
\theta_{1}-z_{2}\right)  }\\
&  =-\sin\theta_{12}\left(  \oint_{\theta_{1}-i\pi}+\oint_{\theta_{1}+i\pi
}\right)  dz_{2}\frac{\Gamma(-\frac{1}{2\pi i}(\theta_{2}-z_{2}))}%
{\Gamma(\frac{1}{2}\nu-\frac{1}{2\pi i}(\theta_{2}-z_{2}))}\frac{1}{\theta
_{1}-z-i\pi}\frac{1}{\theta_{1}-z+i\pi}\\
&  =\frac{1}{2i\pi}\sinh\theta_{12}\left(  \frac{\Gamma(\frac{1}{2\pi
i}\left(  \theta_{12}-i\pi\right)  )}{\Gamma(\frac{1}{2}\nu+\frac{1}{2\pi
i}\left(  \theta_{12}-i\pi\right)  )}-\frac{\Gamma(\frac{1}{2\pi i}\left(
\theta_{12}+i\pi\right)  )}{\Gamma(\frac{1}{2}\nu+\frac{1}{2\pi i}\left(
\theta_{12}+i\pi\right)  )}\right) \\
&  =\frac{\nu}{2}\sinh\theta_{12}\frac{1}{\theta_{12}-i\pi}+O\left(  \nu
^{2}\right)  \,.
\end{align*}
Similarly%
\[
I_{2}=\frac{\nu}{2}\sinh\theta_{13}\frac{1}{\theta_{13}-i\pi}+O\left(  \nu
^{2}\right)  \,,
\]
which proves (\ref{1overN}).
\end{proof}

\subsection{1/N perturbation theory}

\label{s1/N}

The nonlinear $O(N)$ $\sigma$-model is defined by the Lagrangian and the
constraint
\[
\mathcal{L}=\frac{1}{2}\sum_{\alpha=1}^{N}\left(  \partial_{\mu}%
\varphi_{\alpha}\right)  ^{2}\qquad\text{with}\quad g\sum_{\alpha=1}%
^{N}\varphi_{\alpha}^{2}=1
\]
The Greens's functions may be written as%
\[
\langle\,0\,|\,T\varphi_{\alpha_{1}}(x_{1})\dots\varphi_{\alpha_{n}}%
(x_{n})\,|\,0\,\rangle=i^{-n}\frac{\delta}{\delta J_{\alpha_{1}}(x_{1})}%
\dots\frac{\delta}{\delta J_{\alpha_{1}}(x_{n})}Z(J)
\]
where $\alpha_{i}$ are $O(N)$ labels and $Z(J)$ is the generating functional
of Greens's functions given by the Feynman path integral
\begin{equation}
Z(J)=\int d\varphi\,\delta(\varphi^{2}-1/g)\exp i\left(  \mathcal{A}%
(\varphi)+J\varphi\right)  \label{a.1a}%
\end{equation}
with the action $\mathcal{A}(\varphi)=\int d^{2}x\,\mathcal{L}(\varphi)$. The
fields $\varphi_{\alpha}(x)$ transforms as the vector representation of
$O(N)$. In eq.~(\ref{a.1a}) and in the following we use a matrix notation of
the $x$-integrations e.g.~$J\varphi=\sum_{\alpha=1}^{N}\int d^{2}xJ_{\alpha
}(x)\varphi_{\alpha}(x)$.

For the derivation of the result below the following Feynman integral is
used.
\begin{align*}
I(m,k)  &  =\int\frac{d^{2}p}{(2\pi)^{2}}\,\frac{1}{p^{2}-m^{2}}\,\frac
{1}{(p+k)^{2}-m^{2}}\\
&  =\frac{i}{4\pi}\frac{1}{\sqrt{-k^{2}}\sqrt{4m^{2}-k^{2}}}2\ln\frac
{\sqrt{4m^{2}-k^{2}}+\sqrt{-k^{2}}}{\sqrt{4m^{2}-k^{2}}-\sqrt{-k^{2}}}\\
&  =\frac{i}{4\pi m^{2}}\,\frac{\phi}{\sinh\phi}~,~~~\left(  k^{2}%
=-4m^{2}\sinh^{2}\tfrac{1}{2}\phi\right)  .
\end{align*}

For divergent integrals we use a Pauli-Villars regularization
\[
I(m)\rightarrow I(m)-I(M)\rightarrow\lim_{M\rightarrow\infty}\left(
I(m)-I(M)\right)  ~,
\]
for example
\begin{equation}
I_{\infty}=\int\frac{d^{2}p}{(2\pi)^{2}}\left(  \frac{1}{p^{2}-m^{2}}-\frac
{1}{p^{2}-M^{2}}\right)  =\frac{i}{4\pi}\ln\frac{m^{2}}{M^{2}}~. \label{M}%
\end{equation}
We may introduce the bosonic field $\omega$ and rewrite eq.~(\ref{a.1a}),
equivalently as
\begin{equation}
Z(J)=\int d\varphi\,\,\exp i\left(  \mathcal{A}(\varphi,\omega)+J\varphi
\right)
\end{equation}
with the action $\mathcal{A}(\varphi,\omega)=\int d^{2}x\,\mathcal{L}%
(\varphi,\omega)$ and the Lagrangian
\[
\mathcal{L}(\varphi,\omega)=\frac{1}{2}\left(  \sum_{\alpha=1}^{N}\left(
\partial_{\mu}\varphi_{\alpha}\right)  ^{2}-\omega\left(  \sum_{\alpha=1}%
^{N}\varphi_{\alpha}^{2}-1/g\right)  \right)
\]
Performing the $\varphi$-integrations we obtain
\[
Z(J)=\int d\omega\,\exp\left(  i\mathcal{A}_{eff}(\omega)-\tfrac{1}{2}%
J\Delta(\omega)J\right)
\]
with the propagator $\Delta_{\alpha\beta}(\omega)=i\delta_{\alpha\beta
}(-\square-\omega)^{-1}$ and the effective action
\[
\mathcal{A}_{eff}(\omega)=\tfrac{1}{2}iN\,\operatorname*{Tr}\,\ln(i\Delta
^{-1}(\omega))+\int d^{2}x\,\frac{1}{2g}\omega~.
\]
The symbol $\operatorname*{Tr}$ means the trace with respect to $x$-space, the
trace with respect to $O(N)$-isospin has been taken and given the factor $N.$

We define the vertex functions $\Gamma$ by\footnote{Note that the
$i\Gamma^{(n)}$ are the 1-particle irreducible connected graphs with $n$
external lines.}
\[
\mathcal{A}_{eff}(\omega)=\sum_{n=0}^{\infty}\frac{1}{n!}\int d^{2}x_{1}\dots
d^{2}x_{n}\,\Gamma^{(n)}(x_{1},\dots,x_{n})\,\omega^{\prime}(x_{1})\dots
\omega^{\prime}(x_{n}),
\]
where $\omega=\omega^{\prime}-m^{2}$.

The value $\omega=m^{2}$ is defined by the condition $\mathcal{A}_{eff}%
(\omega)$ being stationary at this point, which means that the one-point
vertex function vanishes $\Gamma^{(1)}(x)=\frac{\delta\mathcal{A}_{eff}%
}{\delta\omega}=0$. Expanding $\mathcal{A}_{eff}$ for small $\omega^{\prime}$%
\begin{align*}
\mathcal{A}_{eff}  &  =-\tfrac{1}{2}iN\,\mathrm{Tr}\,\ln(i\Delta^{-1}%
(\omega))+\int d^{2}x\,\frac{1}{2g}\omega\\
&  =-\tfrac{1}{2}iN\left(  \mathrm{Tr}\,\ln(-\square-m^{2})+\mathrm{Tr}%
\,\left\{  (-\square-m^{2})^{-1}\omega^{\prime}\right.  \right. \\
&  +\left.  \left.  \tfrac{1}{2}(-\square-m^{2})^{-1}\omega^{\prime}%
(-\square-m^{2})^{-1}\omega^{\prime}+\dots\right\}  \right)  +\frac{1}{2g}\int
d^{2}x\,\left(  m^{2}+\omega^{\prime}\right)
\end{align*}
we obtain%
\[
\Gamma^{(1)}(x)=-\tfrac{1}{2}N\Delta(x,x)+\frac{1}{2g}=0
\]
with the propagator$\ \Delta=i(-\square-m^{2})^{-1}$. This equation defines
the mass $m$ by%
\[
m^{2}=M^{2}e^{-\frac{4\pi}{gN}}%
\]
where $M$ is an UV-cutoff (see (\ref{M})). There is the effect of mass
generation and dimensional transmutation: the dimensionless coupling $g$ is
replaced by the mass $m$.

The $1/N$-expansion is obtained by expanding the effective action at this
stationary point. Next we calculate the two point vertex function
\[
\Gamma^{(2)}(x,y)=\frac{\delta^{2}\mathcal{A}_{eff}}{\delta\omega
(x)\delta\omega(y)}=\int\frac{d^{2}k}{(2\pi)^{2}}e^{-i(x-y)k}\,\tilde{\Gamma
}^{(2)}(k)
\]

in momentum space
\[
\tilde{\Gamma}^{(2)}(k)=-\tfrac{1}{2}iN\int\frac{d^{2}p}{(2\pi)^{2}%
}\,\mathrm{tr}\left(  \frac{1}{p^{2}-m^{2}}\,\frac{1}{(p+k)^{2}-m^{2}}\right)
=\frac{N}{8\pi m^{2}}\,\frac{\phi}{\sinh\phi}%
\]
where $k^{2}=-4m^{2}\sinh^{2}(\phi/2)$. The $\omega$-propagator is obtained by
inverting the two-point vertex function $\Delta=i{\Gamma^{(2)}}^{-1}$%
\[
\tilde{\Delta}_{\omega}(k)=i\left(  -\tfrac{1}{2}iNI(m,k)\right)  ^{-1}%
=\frac{8\pi i}{N}m^{2}\,\frac{\sinh\phi}{\phi}~.
\]
This propagator together with the simple vertex $-i\delta_{\alpha}^{\beta}$ of
Fig.~\ref{fa1a} yield the Feynman rules which allow to calculate general
vertex functions in the $1/N$-expansion. For example the four point vertex
function is\begin{figure}[tbh]%
\[%
\begin{array}
[c]{c}%
\unitlength4mm\begin{picture}(3,4) \put(0,1){\line(0,1){2}} \put(-.2,.2){$\alpha$} \put(-.2,3.3){$\beta$} \put(0,2){\dashbox{.2}(2,0){}} \put(.5,1){$k$} \put(.5,2.3){$\leftarrow$} \put(2.2,1.8){$\sigma$} \end{picture}
\end{array}
=-i\delta_{\alpha}^{\beta}%
\]
\caption{The elementary vertex for the $O(N)$ Gross-Neveu model. With respect
to isospin the vertex is proportional to the unit matrix.}%
\label{fa1a}%
\end{figure}%
\begin{equation}
i\tilde{\Gamma}^{(4)}{}_{\alpha\beta}^{\delta\gamma}(-p_{3},-p_{4},p_{1}%
,p_{2})=-\delta_{\alpha}^{\delta}\delta_{\beta}^{\gamma}\,\tilde{\Delta
}_{\omega}(p_{2}-p_{3})-\delta_{\alpha}^{\gamma}\delta_{\beta}^{\delta
}\,\tilde{\Delta}_{\omega}(p_{3}-p_{1})-\delta_{\alpha\beta}\delta
^{\delta\gamma}\,\tilde{\Delta}_{\omega}(p_{1}+p_{2}) \label{a.25}%
\end{equation}
where $\alpha\beta\gamma\delta$ are isospin. We now calculate the three
particle form factor of the fundamental bose field in $1/N$-expansion in
lowest nontrivial order. For convenience we multiply the field with the
Klein-Gordon operator and take
\[
{\mathcal{O}}^{\delta}{(}x{)}=i(\square+m^{2})\varphi^{\delta}(x)
\]
and define
\[
_{out}^{~~\gamma}\langle\,p_{3}\,|\,{\mathcal{O}}^{\delta}(0)\,|\,p_{1}%
,p_{2}\,\rangle_{\alpha\beta}^{in}={F^{\mathcal{O}^{\delta}}\,}_{\alpha\beta
}^{\gamma}(\theta_{3};\theta_{1},\theta_{2}).
\]
By means of LSZ-techniques one can express the connected part in terms of the
4-point vertex function.
\begin{equation}
{F_{conn.}^{\mathcal{O}^{\delta}}}_{\alpha\beta}^{\gamma}(\theta_{3}%
;\theta_{1},\theta_{2})=\,i{{{\tilde{\Gamma}}^{(4)}}{}}_{\alpha\beta}%
^{\delta\gamma}(-p_{3},p_{3}-p_{1}-p_{2},p_{1},p_{2}).
\end{equation}
The lowest order contributions are given by the Feynman graphs of
Fig.~\ref{fa3a}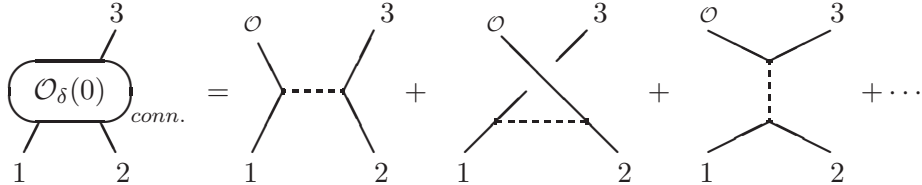
\begin{figure}[tbh]%
\[%
\begin{array}
[c]{c}%
\unitlength4mm\begin{picture}(30,6)\thicklines \put(2,3){\oval(4,2)} \put(2,3){\makebox(0,0){${\cal O}_\delta(0)$}} \put(.5,1){\line(1,2){.5}} \put(3.5,1){\line(-1,2){.5}} \put(3,4){\line(1,2){.5}} \put(4,2){$\scriptstyle conn.$} \put(.1,0){$1$} \put(3.5,0){$2$} \put(3.3,5.3){$3$} \put(6.5,2.7){$=$} \put(8,1){\line(1,2){1}} \put(9,3){\line(-1,2){.8}} \put(7.7,5){$\scriptstyle{\cal O}$} \put(9,3){\dashbox{.2}(2,0){}} \put(12,1){\line(-1,2){1}} \put(11,3){\line(1,2){1}} \put(7.7,0){$1$} \put(12,0){$2$} \put(12.2,5.3){$3$} \put(13,2.7){$+$} \put(15,1){\line(1,1){2}} \put(18,4){\line(1,1){1}} \put(15.7,5){$\scriptstyle{\cal O}$} \put(16,2){\dashbox{.2}(3,0){}} \put(20,1){\line(-1,1){3.8}} \put(14.7,0){$1$} \put(20,0){$2$} \put(19.2,5.3){$3$} \put(22.7,5.3){$\scriptstyle{\cal O}$} \put(25,2){\dashbox{.2}(0,2){}} \put(23,1){\line(2,1){2}} \put(27,1){\line(-2,1){2}} \put(23,5){\line(2,-1){2}} \put(27,5){\line(-2,-1){2}} \put(22.7,0){$1$} \put(27,0){$2$} \put(27,5.3){$3$} \put(21,2.7){$+$} \put(28,2.7){$+\cdots$} \end{picture}
\end{array}
\]
\caption{The connected part of the three particle form factor of the
fundamental fermi field in $1/N$-expansion. }%
\label{fa3a}%
\end{figure}%
\[
{F_{conn.}^{\mathcal{O}^{\delta}}}_{\alpha\beta}^{\gamma}=-\delta_{\alpha
}^{\delta}\delta_{\beta}^{\gamma}\,\tilde{\Delta}_{\omega}(p_{2}-p_{3}%
)-\delta_{\alpha}^{\gamma}\delta_{\beta}^{\delta}\,\tilde{\Delta}_{\omega
}(p_{3}-p_{1})-\delta_{\alpha\beta}\delta^{\delta\gamma}\,\tilde{\Delta
}_{\omega}(p_{1}+p_{2}).
\]

Using $\tilde{\Delta}_{\omega}(k)=\frac{8\pi i}{N}m^{2}\,\frac{\sinh\phi}%
{\phi}$ with $k^{2}=-4m^{2}\sinh^{2}(\phi/2)$ we obtain up to order $1/N$%
\begin{equation}
{F_{conn.}^{\mathcal{O}^{\delta}}}_{\alpha\beta}^{\gamma}=-\frac{8\pi i}%
{N}m^{2}\,\left(  \delta_{\alpha}^{\delta}\delta_{\beta}^{\gamma}\frac
{\sinh\left(  \theta_{23}\right)  }{\theta_{23}}+\delta_{\alpha}^{\gamma
}\delta_{\beta}^{\delta}\frac{\sinh\left(  \theta_{13}\right)  }{\theta_{13}%
}+\delta_{\alpha\beta}\delta^{\delta\gamma}\frac{\sinh\left(  i\pi-\theta
_{12}\right)  }{i\pi-\theta_{12}}\right)  ~.\nonumber
\end{equation}
and by crossing in the complex basis%
\begin{equation}
{F^{\mathcal{O}_{\!}^{\delta}}}_{\alpha\beta\gamma}(\theta_{1},\theta
_{2},\theta_{3})=-\frac{8\pi i}{N}m^{2}\,\left(  \delta_{\alpha}^{\delta
}\mathbf{C}_{\beta\gamma}\frac{\sinh\theta_{23}}{i\pi-\theta_{23}}%
+\delta_{\beta}^{\delta}\mathbf{C}_{\alpha\gamma}\frac{\sinh\theta_{13}}%
{i\pi-\theta_{13}}+\delta_{\gamma}^{\delta}\mathbf{C}_{\alpha\beta}\frac
{\sinh\theta_{12}}{i\pi-\theta_{12}}\right)  \label{a.40b}%
\end{equation}
which agrees with expansion of the exact form factor (\ref{1overN}).

Now we are in the position to check whether this expression is consistent with
the exact S-matrix of Sect.~\ref{s2} (see also \cite{ZZ3}). Using
LSZ-techniques we obtain
\begin{align*}
&  _{out}\langle\gamma,p_{3};\delta,p_{4}\,|\,\alpha,p_{1};\beta,p_{2}%
\rangle^{in}\\
&  =\delta_{14}\delta_{\alpha}^{\delta}\delta_{23}\delta_{\beta}^{\gamma
}+\delta_{13}\delta_{\alpha}^{\gamma}\delta_{24}\delta_{\beta}^{\delta}%
+(2\pi)^{2}\delta^{(2)}(p_{1}+p_{2}-p_{3}-p_{4})\,i{{{\tilde{\Gamma}}^{(4)}}%
{}}_{\alpha\beta}^{\delta\gamma}\,\\
&  =\delta_{14}\delta_{\alpha}^{\delta}\delta_{23}\delta_{\beta}^{\gamma
}+\delta_{13}\delta_{\alpha}^{\gamma}\delta_{24}\delta_{\beta}^{\delta}%
+\frac{1}{4m^{2}\sinh\theta_{12}}(\delta_{14}\delta_{23}+\delta_{13}%
\delta_{24}){F_{conn.}^{\mathcal{O}_{\!\delta}}}_{\alpha\beta}^{\gamma}%
\end{align*}
where $\delta_{14}=2\pi2p_{1}^{0}\delta(p_{1}-p_{4})$ etc. and $\Gamma^{(4)}$
is the 4-point vertex function in (\ref{a.25}). I has been used that
\[
(2\pi)^{2}\delta^{(2)}(p_{1}+p_{2}-p_{3}-p_{4})=\frac{1}{4m^{2}\sinh
\theta_{12}}(\delta_{14}\delta_{23}+\delta_{13}\delta_{24})\,.
\]
By means of the formula%
\begin{align*}
_{out}\langle\gamma,p_{3};\delta,p_{4}\,|\,\alpha,p_{1};\beta,p_{2}%
\rangle^{in}  &  =\,_{out}\langle\gamma,p_{3};\delta,p_{4}\,|\,\beta^{\prime
},p_{2};\alpha^{\prime},p_{1}\rangle^{out}\,S_{\alpha\beta}^{\beta^{\prime
}\alpha^{\prime}}\\
&  =\delta_{13}\delta_{24}S_{\alpha\beta}^{\delta\gamma}+\delta_{14}%
\delta_{23}S_{\alpha\beta}^{\gamma\delta}%
\end{align*}
we obtain
\begin{equation}
S_{\alpha\beta}^{\delta\gamma}(\theta_{12})=\delta_{\alpha}^{\gamma}%
\delta_{\beta}^{\delta}+\frac{1}{4m^{2}\sinh\theta_{12}}{F_{conn.}%
^{\mathcal{O}^{\delta}}}_{\alpha\beta}^{\gamma}(\theta_{3};\theta_{1}%
,\theta_{2})~~~~\mathrm{for}~~~p_{4}\rightarrow p_{2},~p_{3}\rightarrow p_{1}.
\label{FtoS}%
\end{equation}

Equation~(\ref{a.40b}) implies the perturbative result up to order $1/N$ or%
\[
S_{\alpha\beta}^{\delta\gamma}(\theta)=\delta_{\alpha}^{\gamma}\delta_{\beta
}^{\delta}\left(  1-\frac{2\pi i}{N\sinh\theta}\right)  +\delta_{\alpha
}^{\delta}\delta_{\beta}^{\gamma}\left(  -\frac{2\pi i}{N\theta}\right)
+\mathbf{C}^{\delta\gamma}\mathbf{C}_{\alpha\beta}\left(  -\frac{2\pi
i}{N\left(  i\pi-\theta\right)  }\right)  +O(N^{-2})
\]
which agrees with the $1/N$ expansion of (\ref{Smin}). The relation
(\ref{FtoS}) is equivalent to the fact that the form factor of the field (up
to $O(N^{-2})$)%
\[
{F^{\varphi}}_{\alpha\beta\gamma}(\theta_{1},\theta_{2},\theta_{3}%
)=\frac{\frac{\pi}{N}\,\left(  \delta_{\alpha}^{1}\mathbf{C}_{\beta\gamma
}\frac{\sinh\left(  i\pi-\theta_{23}\right)  }{i\pi-\theta_{23}}+\delta
_{\beta}^{1}\mathbf{C}_{\alpha\gamma}\frac{\sinh\left(  i\pi-\theta
_{13}\right)  }{i\pi-\theta_{13}}+\delta_{\gamma}^{1}\mathbf{C}_{\alpha\beta
}\frac{\sinh\left(  i\pi-\theta_{12}\right)  }{i\pi-\theta_{12}}\right)
}{\cosh\frac{1}{2}\theta_{12}\cosh\frac{1}{2}\theta_{13}\cosh\frac{1}{2}%
\theta_{23}}%
\]
satisfies the form factor equation (iii)%

\begin{multline*}
\operatorname*{Res}_{\theta_{12}=i\pi}{F^{\varphi}}_{\alpha\beta\gamma}%
(\theta_{1},\theta_{2},\theta_{3})\\
=-4\frac{\pi}{N}\left(  \delta_{\alpha}^{1}\mathbf{C}_{\beta\gamma}\frac
{1}{i\pi-\theta_{23}}+\delta_{\beta}^{1}\mathbf{C}_{\alpha\gamma}\frac
{1}{\theta_{23}}+\delta_{\gamma}^{1}\mathbf{C}_{\alpha\beta}\frac{1}%
{\sinh\theta_{23}}\right)  +O\left(  N^{-2}\right) \\
=2i\left(  \mathbf{C}_{\alpha\beta}\delta_{\gamma}^{1}-\mathbf{C}_{\alpha
\beta^{\prime}}\delta_{\gamma^{\prime}}^{1}S_{\beta\gamma}^{\gamma^{\prime
}\beta^{\prime}}\left(  \theta_{23}\right)  \right)  .
\end{multline*}

\section{More explicitly calculations}

\label{sh}

\subsection{General formulas}

The maximal number of particles of type $\alpha_{i}=\mathring{\alpha}_{i}%
\neq1,\bar{1}$ in the Bethe ansatz state $\tilde{\Phi}_{\underline{\alpha}%
}^{\underline{\mathring{\beta}}}(\underline{\theta},\underline{z})$ of
(\ref{BS}) is $m$; the other $n-m$ particles are of type $1$. This follows
from the structure of the $\Pi$-matrix (see \cite{BFK5,BFK6}) and the S-matrix
(\ref{S}). We consider the corresponding component of the form factors for $n$
particles
\begin{equation}
F_{\underline{\mathring{\alpha}}\,\underline{1}}(\underline{\theta}%
)=N_{n,m}K_{\underline{\mathring{\alpha}}\,\underline{1}}(\underline{\theta
})\prod_{1\leq i<j\leq n}F(\theta_{i}-\theta_{j})\nonumber
\end{equation}
for $\underline{\mathring{\alpha}}=\mathring{\alpha}_{1},\dots.,\mathring
{\alpha}_{m}$, with $\mathring{\alpha}_{i}\neq1,\bar{1}$ and $\underline
{1}=1,\dots.,1$. For convenience, we use here a different normalization
compared to (\ref{2.10}) and (\ref{2.16}). The K-functions is given by%
\begin{equation}
K_{\underline{\mathring{\alpha}}\,\underline{1}}(\underline{\theta}%
)=\prod_{k=1}^{m}\bigg(\frac{1}{2i\pi}\int_{\mathcal{C}_{\underline{\theta}}%
}dz_{k}\bigg)\tilde{h}(\underline{\theta},\underline{z})p(\underline{\theta
},\underline{z})\sum_{\pi\in S_{m}}k_{\underline{\mathring{\alpha}}%
}(\underline{\theta},\pi\underline{z}) \label{K1}%
\end{equation}
where the sum has to be taken over all permutations of the $z_{i}$ and%
\begin{align*}
\tilde{h}(\underline{\theta},\underline{z})  &  =\prod_{i=1}^{n}\prod
_{j=1}^{m}\tilde{\phi}_{j}\left(  \theta_{i}-z_{j}\right)  \prod_{1\leq
i<j\leq m}\tau_{ij}(z_{ij})\\
k_{\underline{\mathring{\alpha}}}(\underline{\theta},\underline{z})  &
=L_{\underline{\mathring{\alpha}}}(\underline{z})\prod_{1\leq i<j\leq m}%
\frac{1}{\tilde{b}(z_{ij})}\prod_{j=1}^{m}\bigg(\tilde{c}(\theta_{j}%
-z_{j})\prod_{k=j+1}^{m}\tilde{b}(\theta_{j}-z_{k})\bigg).
\end{align*}

\begin{proof}
We use formula (15) of \cite{BFK5}%
\begin{multline*}
\tilde{T}_{a,1\dots m}(\theta,\underline{z})=\prod_{j=1}^{m}\tilde{b}%
(\theta-z_{j})\mathbf{1}_{a}\mathbf{1}_{1}\dots\mathbf{1}_{n}\\
+\sum_{i=1}^{m}\tilde{c}(\theta-z_{i})\tilde{b}(\theta-z_{m})\frac{\tilde
{S}_{am}(z_{im})}{\tilde{b}(z_{im})}\dots\mathbf{P}_{ai}\dots\tilde{b}%
(\theta-z_{1})\frac{\tilde{S}_{a1}(z_{i1})}{\tilde{b}(z_{i1})}\\
+\sum_{i=1}^{m}\tilde{d}(\theta-z_{i})\frac{\tilde{b}(\theta-z_{m})}{\tilde
{b}(\hat{z}_{mi})}\tilde{S}_{am}(\hat{z}_{mi})\dots\mathbf{K}_{ai}\dots
\frac{\tilde{b}(\theta-z_{1})}{\tilde{b}(\hat{z}_{1i})}\tilde{S}_{a1}(\hat
{z}_{1i})\,.
\end{multline*}
which implies for the Bethe state%
\begin{align*}
&  \Psi_{\underline{\mathring{\alpha}}\,\underline{1}}(\underline{\theta
},\underline{z})\\
&  =L_{\underline{\mathring{\beta}}}(\underline{z})\Pi_{\underline
{\mathring{\alpha}}}^{\underline{\mathring{\beta}}}(\underline{z})\prod_{1\leq
i<j\leq m}\frac{1}{\tilde{b}(z_{ij})}\prod_{j=1}^{m}\bigg(\tilde{c}(\theta
_{j}-z_{j})\prod_{k=j+1}^{m}\tilde{b}(\theta_{j}-z_{k}%
)\bigg)+~\text{permutations of the }z_{i}\\
&  =L_{\underline{\mathring{\beta}}}(\underline{z})\prod_{1\leq i<j\leq
m}\frac{1}{\tilde{b}(z_{ij})}~~~%
\begin{array}
[c]{c}%
\unitlength4mm\begin{picture}(15,9.5) \thicklines \put(10.2,1){$\theta_{m+1}$} \put(13.3,1){$\theta_n$} \put(3.2,5.3){$z_m$} \put(6.8,3.5){$z_m$} \put(2,2.3){$z_1$} \put(4,3){$z_1$} \put(5.0,6.2){1} \put(8.0,6.2){1} \put(10.0,6.2){1} \put(13.0,6.2){1} \put(10.0,0){1} \put(13.0,0){1} \put(14,1.8){1} \put(14,4.8){1} \put(5.8,5.5){$\dots$} \put(5.8,1.2){$\dots$} \put(11,5.5){$\dots$} \put(1,5){${\dots}$} \put(10,1){\line(0,1){5}} \put(13,1){\line(0,1){5}} \put(4,6){\oval(8,8)[lb]} \put(4,1){\oval(2,2)[rt]} \put(13.5,6){\oval(17,8)[lb]} \put(7,6){\oval(9,2)[lb]} \put(7,1){\oval(2,8)[rt]} \put(13.5,6){\oval(11,2)[lb]} \put(-.2,6){$\framebox(3,1){$\Pi$}$} \put(0,7){\line(0,1){1}} \put(2.5,7){\line(0,1){1}} \put(4.4,0){$\mathring{\alpha}_1$} \put(7.8,0){$\mathring{\alpha}_m$} \put(-.2,8.3){$\mathring{\beta}_1$} \put(2.2,8.3){$\mathring{\beta}_m$} \end{picture}
\end{array}
+~\text{permutations of the }z_{i}\\
&  =\sum_{\pi\in S_{m}}L_{\underline{\mathring{\alpha}}}(\pi\underline
{z})\prod_{1\leq i<j\leq m}\frac{1}{\tilde{b}(\pi z_{ij})}\prod_{j=1}%
^{m}\bigg(\tilde{c}(\theta_{j}-\pi z_{j})\prod_{k=j+1}^{m}\tilde{b}(\theta
_{j}-\pi z_{k})\bigg)\,.
\end{align*}
because $\Pi_{\underline{\mathring{\alpha}}}^{\underline{\mathring{\beta}}%
}(\underline{z})=\delta_{\underline{\mathring{\alpha}}}^{\underline
{\mathring{\beta}}}$. Note that the $\tilde{d}$-terms do not contribute
because of $\Pi_{\ldots\bar{1}}=0$ (see (\ref{Pi})). It has been used that the
state $\tilde{\Psi}_{\underline{\alpha}}(\underline{\theta},\underline{z})$ is
a symmetric function of the $z_{i}$ (see Remark \ref{r1}).
\end{proof}

Pairs of $\mathring{\alpha}_{i}$ may be replaced by $\bar{1}1$. For example,
we obtain for $\underline{\mathring{\alpha}}=\mathring{\alpha}_{1}%
,\dots.,\mathring{\alpha}_{m-2}$, with $\mathring{\alpha}_{i}\neq1,\bar{1}$
\begin{equation}
K_{\underline{\mathring{\alpha}}\bar{1}1\underline{1}}(\underline{\theta
})=\prod_{k=1}^{m}\left(  \frac{1}{2i\pi}\int_{\mathcal{C}_{\underline{\theta
}}}dz_{k}\right)  \tilde{h}(\underline{\theta},\underline{z})p(\underline
{\theta},\underline{z})\sum_{\pi\in S_{m}}k_{\underline{\mathring{\alpha}}%
}(\underline{\theta},\pi\underline{z}) \label{K2}%
\end{equation}
and
\begin{align*}
k_{\underline{\mathring{\alpha}}}(\underline{\theta},\underline{z})  &
=L_{\underline{\mathring{\alpha}}\mathring{\beta}_{m-1}\mathring{\beta}_{m}%
}(\underline{z})\mathbf{C}^{\mathring{\beta}_{m-1}\mathring{\beta}_{m}}%
\prod_{1\leq i<j\leq m}\frac{1}{\tilde{b}(z_{ij})}\\
&  \times\prod_{j=1}^{m-2}\bigg( \tilde{c}(\theta_{j}-z_{j})\prod
_{k=j+1}^{m-2}\tilde{b}(\theta_{j}-z_{k})\bigg) f(z_{m-1}-z_{m})\tilde
{c}(\theta_{m-1}-z_{m-1})\tilde{b}(\theta_{m-1}-z_{m})
\end{align*}
where (\ref{Pi2}) $\Pi_{\bar{1}1}^{\mathring{\beta}_{1}\mathring{\beta}_{2}%
}(z)=\mathbf{C}^{\mathring{\beta}_{1}\mathring{\beta}_{2}}f(z)$ has been used.
Similar formulas are obtained if more pairs of $\mathring{\alpha}$'s are
replaced by $\bar{1}1$.

\subsection{O(3) form factors}

\label{O3}

For $O(3)$ the general formulas simplify because the integrals reduce to a
finite number of residues%
\[
\frac{1}{2i\pi}\int_{\mathcal{C}_{\underline{\theta}}}dz\,\ldots=\frac
{1}{2i\pi}\sum_{i=1}^{n}\left(  \oint_{\theta_{i}}+\oint_{\theta_{i}-2\pi
i}\right)  dz\,\ldots\,.
\]
and we may replace
\[
\tilde{\phi}_{j}\left(  \theta\right)  \rightarrow\frac{1}{\theta}%
\,,~~\tau_{ij}(z)\rightarrow z^{2}\,.
\]
Furthermore all $\mathring{\alpha}_{i}$ are equal to $0$ and $L_{\underline
{\mathring{\beta}}}(\underline{z})=\prod_{1\leq i<j\leq m}L(z_{ij})$ (see
section \ref{s42}). The example (\ref{K1}) writes as%
\begin{align}
F_{0\cdots0\underline{1}}(\underline{\theta}) &  =N_{m}K_{0\cdots
0\underline{1}}(\underline{\theta})\prod_{1\leq i<j\leq n}F(\theta_{i}%
-\theta_{j})\label{F0}\\
K_{0\cdots0\underline{1}}(\underline{\theta}) &  =\prod_{k=1}^{m}%
\Bigg(\frac{1}{2i\pi}\sum_{i=1}^{n}\bigg(\oint_{\theta_{i}}+\oint_{\theta
_{i}-2\pi i}\bigg)dz_{k}\Bigg)p(\underline{\theta},\underline{z})\sum_{\pi\in
S_{m}}k(\underline{\theta},\pi\underline{z})\label{K0}%
\end{align}
with%
\begin{equation}
k(\underline{\theta},\underline{z})=\prod_{1\leq i<j\leq m}l(z_{i}-z_{j}%
)\prod_{i=1}^{n}\prod_{j=1}^{m}\frac{1}{\theta_{i}-z_{j}}\prod_{j=1}%
^{m}\bigg(\frac{\mathbf{-}2\pi i}{\theta_{j}-z_{j}-2\pi i}\prod_{k=j+1}%
^{m}\frac{\theta_{j}-z_{k}}{\theta_{j}-z_{k}-2\pi i}\bigg).\label{k}%
\end{equation}
We have integrated $\tilde{h}(\underline{\theta},\underline{z})$ into
$k(\underline{\theta},\underline{z})$. We have also used the amplitudes
(\ref{1.2}) and
\[
l(z)=L(z)\tau(z)/\tilde{b}(z)=\left(  z-i\pi\right)  \tanh\frac{1}{2}z\,.
\]
Similarly, we obtain%
\begin{equation}
K_{0\cdots0\bar{1}1\underline{1}}(\underline{\theta})=\prod_{k=1}%
^{m}\bigg(\frac{1}{2i\pi}\int_{\mathcal{C}_{\underline{\theta}}}%
dz_{k}\bigg)p(\underline{\theta},\underline{z})\sum_{\pi\in S_{m}}%
k(\underline{\theta},\pi\underline{z})\label{K1b}%
\end{equation}
with%
\begin{align}
k(\underline{\theta},\underline{z}) &  =\prod_{1\leq i<j\leq m}l(z_{i}%
-z_{j})\prod_{i=1}^{n}\prod_{j=1}^{m}\tilde{\phi}\left(  \theta_{i}%
-z_{j}\right)  \prod_{j=1}^{m-2}\bigg(\tilde{c}(\theta_{j}-z_{j})\prod
_{k=j+1}^{m}\tilde{b}(\theta_{j}-z_{k})\bigg)\nonumber\\
&  \times f(z_{m-1}-z_{m})\tilde{c}(\theta_{m-1}-z_{m-1})\tilde{b}%
(\theta_{m-1}-z_{m})\nonumber
\end{align}
and
\[
K_{0\cdots0\bar{1}\bar{1}11\underline{1}}(\underline{\theta})=\prod_{k=1}%
^{m}\bigg(\frac{1}{2i\pi}\int_{\mathcal{C}_{\underline{\theta}}}%
dz_{k}\bigg)p(\underline{\theta},\underline{z})\sum_{\pi\in S_{m}}%
k(\underline{\theta},\pi\underline{z})
\]
with%
\begin{align*}
k(\underline{\theta},\underline{z}) &  =\prod_{1\leq i<j\leq m}l(z_{i}%
-z_{j})\prod_{i=1}^{n}\prod_{j=1}^{m}\tilde{\phi}\left(  \theta_{i}%
-z_{j}\right)  \prod_{j=1}^{m-4}\bigg(\tilde{c}(\theta_{j}-z_{j})\prod
_{k=j+1}^{m}\tilde{b}(\theta_{j}-z_{k})\bigg)\\
&  \times f(z_{m-3}-z_{m-2})\tilde{c}(\theta_{m-3}-z_{m-3})\tilde{b}%
(\theta_{m-3}-z_{m-2})\tilde{b}(\theta_{m-3}-z_{m-1})\tilde{b}(\theta
_{m-3}-z_{m})\\
&  \times f(z_{m-1}-z_{m})\tilde{c}(\theta_{m-2}-z_{m-1})\tilde{b}%
(\theta_{m-2}-z_{m})
\end{align*}
etc.

It turns out that the form factors for the field and the pseudo-potential of
current are of the form (see also the explicit calculations in subsection
\ref{example})

\begin{equation}
F_{\underline{\alpha}}(\underline{\theta})=g_{\underline{\alpha}}%
(\underline{\theta})G(\underline{\theta}) \label{gG}%
\end{equation}
where $g_{\underline{\alpha}}(\underline{\theta})$ is a polynomial (c.f.
\cite{BN}) and%
\begin{align*}
G(\underline{\theta})  &  =\prod_{1\leq i<j\leq n}G(\theta_{ij})\\
G(\theta)  &  =\frac{\tanh\frac{1}{2}\theta}{\theta\left(  \theta-2\pi
i\right)  }F(\theta)=\frac{\theta-i\pi}{\theta\left(  \theta-2i\pi\right)
}\tanh^{2}\tfrac{1}{2}\theta\,.
\end{align*}

\begin{proof}
A sketch of the proof for (\ref{gG}): The K-functions with $n=m+1$ are of the
form%
\begin{align*}
K(\underline{\theta}) &  =\prod_{k=1}^{m}\left(  \frac{1}{2i\pi}\sum_{i=1}%
^{n}\left(  \oint_{\theta_{i}}+\oint_{\theta_{i}-2\pi i}\right)
dz_{k}\right)  p(\underline{\theta},\underline{z})\sum_{\pi\in S_{m}%
}k(\underline{\theta},\pi\underline{z})\\
&  =\sum I_{i_{1}\dots i_{m}},~~i_{k}\in\{1,\dots,n\}\\
I_{i_{1}\dots i_{m}}(\underline{\theta}) &  =\frac{1}{2i\pi}\bigg(\oint
_{\theta_{i_{1}}}+\oint_{\theta_{i_{1}}-2\pi i}\bigg)dz_{1}\dots\frac{1}%
{2i\pi}\bigg(\oint_{\theta_{i_{m}}}+\oint_{\theta_{i_{m}}-2\pi i}%
\bigg)dz_{m}p(\underline{\theta},\underline{z})\sum_{\pi\in S_{m}}%
k(\underline{\theta},\pi\underline{z})
\end{align*}
where $I_{i_{1}\dots i_{m}}$ is symmetric with respect to the indices and for
two equal indices the integrals vanish $I_{\dots i\dots i\dots}=0$. For
example for (\ref{K0}) we obtain, using (\ref{k})%
\begin{align*}
I_{1\dots n-1}(\underline{\theta}) &  =\prod_{k=1}^{n-1}\left(  \frac{1}%
{2i\pi}\left(  \oint_{\theta_{k}}+\oint_{\theta_{k}-2\pi i}\right)
dz_{k}\right)  p(\underline{\theta},\underline{z})\sum_{\pi\in S_{m}%
}k(\underline{\theta},\pi\underline{z})\\
&  =J(\underline{\theta})\bigg({\textstyle}\prod_{1\leq i<j\leq n-1}%
\tanh\tfrac{1}{2}\theta_{ij}\bigg)p(\underline{\theta},\underline{z}%
=\theta_{1},\dots,\theta_{n-1})\,,
\end{align*}
where%
\begin{align*}
J(\underline{\theta}) &  =\prod_{k=1}^{n-1}\left(  \frac{1}{2i\pi}\left(
\oint_{\theta_{k}}+\oint_{\theta_{k}-2\pi i}\right)  dz_{k}\right)  \sum
_{\pi\in S_{m}}\operatorname*{sign}(\pi)\,j(\underline{\theta},\pi
\underline{z})\,,\\
j(\underline{\theta},\underline{z}) &  =\prod_{1\leq i<j\leq m}(z_{i}%
-z_{j}-i\pi)\prod_{i=1}^{n}\prod_{j=1}^{m}\frac{1}{\theta_{i}-z_{j}}%
\prod_{j=1}^{m}\bigg(\frac{\mathbf{-}2\pi i}{\theta_{j}-z_{j}-2\pi i}%
\prod_{k=j+1}^{m}\frac{\theta_{j}-z_{k}}{\theta_{j}-z_{k}-2\pi i}\bigg)\,.
\end{align*}
Let the p-function satisfy%
\[
\sum_{k=1}^{n}(-1)^{n-k}\Bigg({\textstyle}\prod_{1\leq i<j\leq n,i,j\neq
k}\tanh\tfrac{1}{2}\theta_{ij}\Bigg)\,p(\underline{\theta},\underline
{z}=\theta_{1},\dots,\widehat{\theta_{k}},\dots,\theta_{n})=\prod_{1\leq
i<j\leq n}\tanh\tfrac{1}{2}\theta_{ij}\,
\]
which holds for the p-functions $p^{\varphi}$ and $p^{J}$ of the examples
below. Then, because $I_{1\dots\hat{k}\dots n}=-I_{1\dots\widehat{k+1}\dots
n}$ (where $\hat{k}$ means that the index $k$ is missing) we obtain
\[
K(\underline{\theta})=J(\underline{\theta})\,n!\prod_{1\leq i<j\leq n}%
\tanh\tfrac{1}{2}\theta_{ij}~~~\text{and~~~}F(\underline{\theta}%
)=g(\underline{\theta})G(\underline{\theta})
\]
where $g(\underline{\theta})$ is a polynomial.
\end{proof}

For the energy momentum potential formula (\ref{gG}) holds for $n>2$ (see
(\ref{FT2}) and (\ref{FT4}). However, the prove is more complicated (see the
prove of (\ref{FT2})).

The normalization factor $N_{m}$ is obtained from the form factor equation
(iii), which implies that\footnote{This follows from (\ref{N3}), taking into
account the different normalizations here compared to that of theorem
\ref{TN}.}
\begin{equation}
N_{m}=\frac{2^{2m-3}\pi}{\left(  m-1\right)  m}N_{m-2}\,. \label{N}%
\end{equation}

\subsubsection{Examples:}

\paragraph{Form factors of the field:}

In particular, for the form factors of the field ($n=$ odd) with $m=n-1$ we
have to apply the p-function of (\ref{pphi}) for $\nu=2$%
\[
p^{\varphi}(\underline{\theta},\underline{z})=\frac{({\textstyle}\sum
e^{z_{j}})({\textstyle\sum e^{-z_{j}})}}{({\textstyle}\sum e^{\theta_{j}%
})({\textstyle}\sum e^{-\theta_{j}})-1}\,.
\]
To prove the form (\ref{gG}) of the field form factors one uses the identity
(for $n=$ odd)%
\[
\sum_{i=1}^{n}\Bigg(\bigg({{\textstyle\prod\limits_{\underset{j\neq i}{j=1}%
}^{n}}}\coth\tfrac{1}{2}\theta_{ji}\bigg)\bigg({\textstyle\sum
\limits_{\underset{j\neq i}{j=1}}^{n}}e^{\theta_{j}}%
\bigg)\bigg({\textstyle\sum\limits_{\underset{j\neq i}{j=1}}^{n}}%
e^{-\theta_{j}}\bigg)\Bigg)=1-\left(  {\textstyle\sum\limits_{i=1}^{n}%
}e^{\theta_{i}}\right)  \left(  {\textstyle\sum\limits_{i=1}^{n}}%
e^{-\theta_{i}}\right)  \,.
\]
The normalizations follow from (\ref{N}) as $N_{m}=\dfrac{1}{m!}\pi^{\frac
{1}{2}m}2^{\frac{1}{2}m\left(  m-1\right)  }$\thinspace.

\paragraph{Form factors of the current:}

For the form factors of the pseudo-potential of the current ($n=$ even) with
$m=n-1$ we have to apply the p-function of (\ref{pJ}) for $\nu=2$%
\[
p^{J}(\underline{\theta},\underline{z})=\frac{\exp\left(  {\textstyle\sum
}\theta_{i}-{\textstyle\sum z_{j}}\right)  }{{\textstyle\sum}\exp\theta_{i}%
}\,.
\]
To prove the form (\ref{gG}) of the current form factors one uses the
identity
\[
\sum_{i=1}^{n}e^{\theta_{i}}\prod_{\underset{j\neq i}{j=1}}^{n}\coth\tfrac
{1}{2}\theta_{ij}\,=\sum_{i=1}^{n}e^{\theta_{i}}\,.
\]
The normalizations follow from (\ref{N}) as $N_{m}=\dfrac{1}{m!}i\pi^{\frac
{1}{2}\left(  m+1\right)  }2^{\frac{1}{2}\left(  m\left(  m-1\right)
-2\right)  }$\thinspace.

\paragraph{Form factors of the energy momentum potential:}

For the form factors for the energy momentum potential ($m=n=$ even) we have
to calculate (\ref{K1}) and (\ref{K2}) with the p-function of (\ref{pT})
\[
p^{T}(\underline{\theta},\underline{z})=1.
\]
Again we have to perform the integrations for $O(3)$ by calculating a finite
number of residues. It turns out that the the leading term in the limit
$\nu\rightarrow2$ (i.e. $N\rightarrow3$) vanishes and we have to calculate the
contribution of order $(\nu-2)$.

\subsubsection{Explicit calculations}

\label{example}

\paragraph{n = 1, m = 0:}%

\[
F_{1}^{\varphi}(\underline{\theta})=1
\]

\paragraph{n = 2, m = 1:}%

\[
F_{01}^{J}(\underline{\theta})=\tfrac{1}{2}i\pi K_{01}(\underline{\theta
})=-\tfrac{1}{2}\pi^{2}G(\theta)
\]
which agrees with the result of \cite{KW} (see also (\ref{FJ23}).

\paragraph{n = 3, m = 2:}

After some simple calculations we obtain with $G(\theta)$ given by (\ref{G})
\begin{align*}
F_{001}^{\varphi}(\underline{\theta})  &  =\pi^{3}\theta_{12}\,G(\underline
{\theta})\\
F_{\bar{1}11}^{\varphi}(\underline{\theta})  &  =\pi^{3}\left(  2i\pi
-\theta_{23}\right)  G(\underline{\theta})\,.
\end{align*}
which agree with \cite{BN} which were obtained by different methods. The other
components can be obtained by the form factor equations (i) and (ii)%
\begin{align*}
F_{\alpha\beta\gamma}^{\varphi}(\underline{\theta})  &  =\pi^{3}g_{\alpha
\beta\gamma}^{\varphi}(\underline{\theta})G(\underline{\theta})\\
g_{\alpha\beta\gamma}^{\varphi}(\underline{\theta})  &  =\theta_{23}%
\delta_{\alpha}^{1}\mathbf{C}_{\beta\gamma}+\left(  2\pi i-\theta_{13}\right)
\delta_{\beta}^{1}\mathbf{C}_{\alpha\gamma}+\theta_{12}\delta_{\gamma}%
^{1}\mathbf{C}_{\alpha\beta}\,.
\end{align*}

\paragraph{n = 4, m = 3:}

From (\ref{K0}) and (\ref{K1b}) we get the results%
\begin{align*}
F_{0001}^{J}(\underline{\theta})  &  =\tfrac{1}{2}\pi^{5}\left(  \theta
_{12}\theta_{13}\theta_{23}+2\pi i\theta_{12}\left(  \theta_{32}-2\pi
i\right)  -2i\pi^{3}\right)  G(\underline{\theta})\\
F_{0\bar{1}11}^{J}(\underline{\theta})  &  =\tfrac{1}{2}\pi^{5}\left(
\theta_{32}-2i\pi\right)  \left(  \theta_{23}\theta_{22}-\theta_{12}\left(
\theta_{12}-i\pi\right)  \right)  G(\underline{\theta})\,.
\end{align*}
which again agree with \cite{BN}.

\paragraph{n = 5, m = 4:}

Again from (\ref{K0}) and (\ref{K1b}) we get the results
\[
F_{\underline{\alpha}}^{\varphi}(\underline{\theta})=g_{\underline{\alpha}%
}(\underline{\theta})\,G(\underline{\theta})
\]
with (up to normalizations)\footnote{These results have been obtained by
Mathematica.}

\noindent
\parbox[t]{0.95\textwidth}{$g_{\bar{1}\bar{1}111}(\underline{\theta})=(\theta_{12}-2\pi
i)(\theta_{34}-2\pi i)(\theta_{35}-2\pi i)(\theta_{45}-2\pi i)(-4\pi^{2}-2i\pi\theta_{45}-3i\pi\theta_{1}+2i\pi\theta_{3}-3i\pi\theta_{2}+4\allowbreak
i\pi\theta_{4}-2\theta_{3}\theta_{1}+\theta_{3}^{2}-2\theta_{3}\theta
_{2}+2\theta_{3}\theta_{4}-\theta_{4}\theta_{1}-\allowbreak\theta_{4}\theta_{2}+\theta_{35}\theta_{1}-\theta_{35}\theta_{3}+\theta_{35}\theta
_{2}-\theta_{35}\theta_{4}+3\theta_{2}\theta_{1})$}

\medskip

\noindent\parbox[t]{0.95\textwidth}{$
g_{00\bar{1}11}(\underline{\theta})=-(2\pi+i(\theta_{4}-\theta_{5}))(2\ \pi^{5}+2i\pi^{4}(3\ \theta_{1}+\theta_{2}-2(\theta_{4}+\theta
_{5}))-i(\theta_{1}-\theta_{2})(\theta_{1}^{2}-2\theta_{1}\theta_{3}-
\theta_{4}\theta_{5}+\theta_{3}(\theta_{4}+\theta_{5}))(\theta_{2}^{2}-2\theta_{2}\theta_{3}-\theta_{4}\theta_{5}+\theta_{3}(\theta_{4}+\theta
_{5}))-2\pi^{3}(4\theta_{1}^{2}-\theta_{3}^{2}+\theta_{4}^{2}+3\theta
_{4}\theta_{5}+
\theta_{5}^{2}-\theta_{2}(\theta_{4}+\theta_{5})+\theta_{3}(\theta_{4}+\theta_{5})+\theta_{1}(2\ \theta_{2}-5(\theta_{4}+\theta_{5})))-i\pi
^{2}((-\theta_{2}^{2})\theta_{3}+\theta_{1}^{2}(5\theta_{2}+3\theta_{3}-
4(\theta_{4}+\theta_{5}))+\theta_{1}(\theta_{2}^{2}-4\theta_{3}^{2}+\theta
_{4}(\theta_{3}+4\ \theta_{4})+(\theta_{3}+5\theta_{4})\theta_{5}+4\theta
_{5}^{2}-6\theta_{2}(\theta_{4}+\theta_{5}))+
2(\theta_{3}-\theta_{4}-\theta_{5})(\theta_{4}\theta_{5}+\theta_{3}(\theta
_{4}+\theta_{5}))+\theta_{2}(5\ \theta_{4}\theta_{5}+\theta_{3}(\theta
_{4}+\theta_{5})))-\pi(\theta_{1}^{3}(\theta_{2}-\theta_{3})+
\theta_{2}^{3}\theta_{3}+2\theta_{3}\theta_{4}(2\ \theta_{3}-\theta_{4}-\theta_{5})\theta_{5}-\theta_{2}^{2}(2\ \theta_{3}^{2}-\theta_{4}\theta
_{5}+\theta_{3}(\theta_{4}+\theta_{5}))+\theta_{2}(\theta_{3}^{2}(\theta
_{4}+\theta_{5})+
\theta_{4}\theta_{5}(\theta_{4}+\theta_{5})+\theta_{3}(\theta_{4}^{2}-4\theta_{4}\theta_{5}+\theta_{5}^{2}))+\theta_{1}^{2}(4\theta_{3}^{2}-3\theta_{4}\theta_{5}+\theta_{3}(\theta_{4}+\theta_{5})+\theta_{2}((-7)\theta_{3}+
2(\theta_{4}+\theta_{5})))+\theta_{1}(-\theta_{2}^{3})+3\theta_{2}^{2}\theta_{3}-5\theta_{3}^{2}(\theta_{4}+\theta_{5})+\theta_{4}\theta_{5}(\theta_{4}+\theta_{5})+\theta_{3}(\theta_{4}^{2}+4\theta_{4}\theta_{5}+\theta_{5}^{2})+
2\theta_{2}(\theta_{3}^{2}-\theta_{4}^{2}-\theta_{4}\theta_{5}-\theta_{5}^{2}+\theta_{3}(\theta_{4}+\theta_{5}))))
$}

\medskip

\noindent\parbox[t]{0.95\textwidth}{$g_{00001}(\underline{\theta})=
i (i  ({\theta_1}-{\theta_2}) ({\theta_1}-{\theta_3}) (-{\theta_2}+{\theta_3})
({\theta_1}-{\theta_4}) ({\theta_2}-{\theta_4}) ({\theta_3}-{\theta_4})+2 {{\pi }^5} (3
{\theta_1}+{\theta_2}-5 {\theta_3}+{\theta_4})+
8 i  {{\pi }^4} ({\theta_1}+{\theta_2}-2 {\theta_3}) ({\theta_1}-{\theta_5})+
2 {{\pi }^3} (-{{{\theta_1}}^3}+{{{\theta_2}}^3}-{{{\theta_3}}^3}+{{{\theta_3}}^2} {\theta_4}-3
{\theta_3} {{{\theta_4}}^2}+{{{\theta_4}}^3}-{{{\theta_2}}^2} (3 {\theta_3}+{\theta_4})+2 {\theta_4}
({\theta_3}+{\theta_4}) {\theta_5}+
(3 {\theta_3}+{\theta_4}) {{{\theta_5}}^2}+{{{\theta_1}}^2} (-7 {\theta_2}+{\theta_3}+3
{\theta_4}+6 {\theta_5})+{\theta_2} ({{{\theta_3}}^2}+{\theta_3} {\theta_4}-{{{\theta_4}}^2}-3
{{{\theta_5}}^2})+
{\theta_1} ({{{\theta_2}}^2}+{{{\theta_3}}^2}+{\theta_3} {\theta_4}-{{{\theta_4}}^2}-8
({\theta_3}+{\theta_4}) {\theta_5}-{{{\theta_5}}^2}+3 {\theta_2} ({\theta_3}+{\theta_4}+2
{\theta_5})))-
2 i  {{\pi }^2} (-{\theta_2} {\theta_3} ({\theta_2}-{\theta_4})
{\theta_4}+2 {{{\theta_1}}^3} ({\theta_2}-{\theta_5})+
(2 {{{\theta_2}}^3}-5 {{{\theta_2}}^2} {\theta_3}+3 {\theta_2} {{{\theta_3}}^2}-2
{{{\theta_3}}^3}+{\theta_3} (4 {\theta_2}+3 {\theta_3}) {\theta_4}-(2 {\theta_2}+7
{\theta_3}) {{{\theta_4}}^2}+2 {{{\theta_4}}^3}) {\theta_5}+
2 {\theta_4} (-{\theta_2}+{\theta_3}+{\theta_4}) {{{\theta_5}}^2}-{{{\theta_1}}^2}
({\theta_3} {\theta_4}-2 {\theta_3} {\theta_5}-3 {\theta_4} {\theta_5}-2 {{{\theta_5}}^2}+{\theta_2}
({\theta_3}+2 {\theta_4}+3 {\theta_5}))+
{\theta_1} (-2 {{{\theta_2}}^3}+2 {{{\theta_3}}^3}-3 {{{\theta_3}}^2} {\theta_4}+6
{\theta_3} {{{\theta_4}}^2}-2 {{{\theta_4}}^3}+{{{\theta_2}}^2} (6 {\theta_3}+{\theta_4}-{\theta_5})-{{{\theta_4}}^2}
{\theta_5}-
2 ({\theta_3}+2 {\theta_4}) {{{\theta_5}}^2}+{\theta_2} (-3 {{{\theta_3}}^2}-4
{\theta_3} {\theta_4}+{{{\theta_4}}^2}+4 {\theta_4} {\theta_5}+2 {{{\theta_5}}^2})))+
2 \pi  ({{{\theta_1}}^3} ({\theta_2}-{\theta_5}) ({\theta_4}-{\theta_5})+{{{\theta_2}}^3}
({\theta_4}-{\theta_5}) {\theta_5}-{{{\theta_2}}^2} {\theta_3} ({\theta_4}-{\theta_5}) ({\theta_4}+2
{\theta_5})-
({\theta_3}-{\theta_4}) {\theta_5} ({\theta_3} {\theta_4} ({\theta_3}+{\theta_4})-({{{\theta_3}}^2}+{{{\theta_4}}^2})
{\theta_5})-
{\theta_2} {\theta_5} (-{{{\theta_4}}^2} {\theta_5}+{\theta_3} {\theta_4}
({\theta_4}+{\theta_5})+{{{\theta_3}}^2} (-3 {\theta_4}+2 {\theta_5}))+
{{{\theta_1}}^2} (-{\theta_3} ({\theta_3}-2 {\theta_4}) {\theta_4}+{\theta_3}
{\theta_4} {\theta_5}-({\theta_3}+{\theta_4}) {{{\theta_5}}^2}+
{\theta_2} ({{{\theta_3}}^2}-2 {{{\theta_4}}^2}+4 {\theta_4} {\theta_5}-{{{\theta_5}}^2}+{\theta_3}
(-3 {\theta_4}+{\theta_5})))+
{\theta_1} ({{{\theta_3}}^3} ({\theta_4}-{\theta_5})+{{{\theta_2}}^2} ({\theta_4}-{\theta_5})
(3 {\theta_3}+{\theta_4}-{\theta_5})+2 {{{\theta_3}}^2} {\theta_4} {\theta_5}+{{{\theta_2}}^3}
(-{\theta_4}+{\theta_5})+
{{{\theta_4}}^2} {\theta_5} ({\theta_4}+{\theta_5})-{\theta_3} {\theta_4} ({{{\theta_4}}^2}+4
{\theta_4} {\theta_5}-{{{\theta_5}}^2})+
{\theta_2} (-{\theta_3}+{\theta_4}) ({\theta_3} (3 {\theta_4}-{\theta_5})-{\theta_5}
({\theta_4}+{\theta_5})))))$
}

\paragraph{n = 2, m = 2:}

For the 2-particle form factor we obtain%
\begin{equation}
F_{\underline{\alpha}}^{T}(\underline{\theta})=-\mathbf{C}_{\alpha_{1}%
\alpha_{2}}\tfrac{1}{2}\pi^{2}\frac{1}{\theta_{12}-i\pi}G(\theta_{12})
\label{FT2}%
\end{equation}

\paragraph{n = 4, m = 4:}

For the 4-particle form factor for example a component is%
\begin{equation}
F_{\bar{1}\bar{1}11}^{T}(\underline{\theta})=\tfrac{1}{2}\pi^{5}\left(
\theta_{12}-2\pi i\right)  \left(  \theta_{32}-2\pi i\right)  \prod
_{i<j}G(\theta_{ij})\,. \label{FT4}%
\end{equation}
The other components are obtained by the form factor equations (i) and (ii).
These results agree again with those of \cite{BN} which were obtained by
different methods.

\begin{proof}
We calculate (\ref{K1b}) for $n=m=2$
\begin{align*}
K_{\bar{1}1}(\underline{\theta})  &  =\prod_{k=1}^{2}\left(  \frac{1}{2i\pi
}\int_{\mathcal{C}_{\underline{\theta}}^{k}}dz_{k}\right)  \sum_{\pi\in S_{2}%
}k(\underline{\theta},\pi\underline{z})\\
k(\underline{\theta},\underline{z})  &  =l(z_{1}-z_{2})\prod_{i=1}^{2}%
\prod_{j=1}^{2}\tilde{\phi}_{j}\left(  \theta_{i}-z_{j}\right)  \tilde
{c}(\theta_{1}-z_{1})\tilde{b}(\theta_{1}-z_{2})f(z_{1}-z_{2})
\end{align*}
For convenience we write $\theta=i\pi x$ and $z=i\pi y$
\[
I=\int_{\mathcal{C}_{\underline{x}}^{o}}dy_{1}\int_{\mathcal{C}_{\underline
{x}}^{e}}dy_{2}\phi(\underline{x},\underline{y},\nu)\tau_{12}(y_{1}%
-y_{2})q(y_{1}-y_{2})\varsigma(\underline{x},\underline{y},\nu)
\]
with
\begin{align*}
q(y)  &  =\frac{L(y)f(y)}{\tilde{b}(y)}=\frac{\nu}{y}\frac{\Gamma\left(
-\frac{1}{2}+\frac{1}{2}\nu-\frac{1}{2}y\right)  \Gamma\left(  -\frac{1}%
{2}+\frac{1}{2}\nu+\frac{1}{2}y\right)  }{\Gamma\left(  \frac{1}{2}\nu
-\frac{1}{2}y\right)  \Gamma\left(  \frac{1}{2}\nu+\frac{1}{2}y\right)
}=-q(-y)\\
\phi(\underline{x},\underline{y},\nu)  &  =\tilde{\psi}(x_{1}-y_{1}%
)\tilde{\psi}(x_{2}-y_{1})\tilde{\chi}(x_{1}-y_{2})\tilde{\chi}(x_{2}-y_{2})\\
\varsigma(\underline{x},\underline{y},\nu)  &  =\tilde{c}(x_{1}-y_{1}%
)\tilde{b}(x_{1}-y_{2})-\tilde{c}(x_{1}-y_{2})\tilde{b}(x_{1}-y_{1})\,.
\end{align*}
It turns out that the leading term for $\nu\rightarrow2$ in $I$ vanishes and
we have to calculate all terms in order $O(\nu-2)$. We use
\begin{align*}
q(y)\tau(y)  &  =-\tfrac{1}{2}\left(  \sin\tfrac{1}{2}\pi\left(  y+\nu\right)
\right)  \frac{\nu}{\pi}\Gamma\left(  -\tfrac{1}{2}+\tfrac{1}{2}\nu+\tfrac
{1}{2}y\right)  \Gamma\left(  -\tfrac{1}{2}+\tfrac{1}{2}\nu-\tfrac{1}%
{2}y\right) \\
&  =\left(  \tan\tfrac{1}{2}\pi y\right)  \left(  1+P(y)\right)  +O\left(
\left(  \nu-2\right)  ^{2}\right) \\
P(y)  &  =O\left(  \nu-2\right)
\end{align*}
and%
\[
\int_{\mathcal{C}_{\underline{x}}^{o}}dy_{1}\int_{\mathcal{C}_{\underline{x}%
}^{e}}dy_{2}\phi(\underline{x},\underline{y},\nu)\tan\tfrac{1}{2}\pi
(y_{1}-y_{2})\varsigma(\underline{x},\underline{y},\nu)=O\left(  \left(
\nu-2\right)  ^{2}\right)  .
\]
Therefore we obtain after some calculations up to terms $O\left(  \left(
\nu-2\right)  ^{2}\right)  $%
\begin{align*}
I  &  =\int_{\mathcal{C}_{\underline{x}}^{o}}dy_{1}\int_{\mathcal{C}%
_{\underline{x}}^{e}}dy_{2}\phi(\underline{x},\underline{y},\nu)\tan\tfrac
{1}{2}\pi\left(  y_{1}-y_{2}\right)  P(y_{1}-y_{2})\varsigma(\underline
{x},\underline{y},\nu)\\
&  =I_{11}+I_{12}+I_{21}+I_{22}%
\end{align*}
with
\begin{align}
I_{12}  &  =\left(  \oint_{x_{1}}+\oint_{x_{1}-2}\right)  dy_{1}\left(
\oint_{x_{2}}+\oint_{x_{2}-2}\right)  dy_{2}\phi(\underline{x},\underline
{y},2)\tan\tfrac{1}{2}\pi\left(  y_{1}-y_{2}\right)  P(y_{1}-y_{2}%
)\varsigma(\underline{x},\underline{y},\nu)\nonumber\\
&  =2\left(  \nu-2\right)  \frac{\tan\frac{1}{2}\pi x_{12}}{x_{12}-1}%
\oint_{x_{1}}dy_{1}\oint_{x_{2}}dy_{2}\phi(\underline{x},\underline
{y},2)\varsigma(\underline{x},\underline{y},\nu)\nonumber\\
&  =-32\left(  \nu-2\right)  \frac{1}{x_{12}-1}\frac{1}{x_{12}\left(
x_{12}-2\right)  }\tan\tfrac{1}{2}\pi x_{12}=I_{21} \label{KT4}%
\end{align}
which proves (\ref{FT2}) because $I_{11}=I_{22}=0$. Similarly, one derives
(\ref{FT4}).
\end{proof}

\providecommand{\href}[2]{#2}\begingroup\raggedright\endgroup

\end{document}